\documentclass{SciPost}

\hypersetup{
    colorlinks,
    linkcolor={red!50!black},
    citecolor={blue!50!black},
    urlcolor={blue!80!black}
}

\usepackage[bitstream-charter]{mathdesign}
\DeclareSymbolFont{usualmathcal}{OMS}{cmsy}{m}{n}
\DeclareSymbolFontAlphabet{\mathcal}{usualmathcal}

\fancypagestyle{SPstyle}{
\fancyhf{}
\lhead{\colorbox{scipostblue}{\bf \color{white} ~SciPost Physics }}
\rhead{{\bf \color{scipostdeepblue} ~Submission }}

\fancyfoot[C]{\textbf{\thepage}}
}

\newcommand{\beeq}[1] {\begin{equation}\begin{split}#1\end{split}\end{equation}}
\newcommand{\bracket}[1]{\left\langle #1\right\rangle}

\usepackage{bm}
\usepackage{tikz}
\usetikzlibrary{arrows.meta,decorations.pathreplacing,fit,calc}

\begin{document}

\pagestyle{SPstyle}

\begin{center}{\Large \textbf{\color{scipostdeepblue}{
Dynamical cavity method for continuous-time complex systems on sparse random graphs\\
}}}\end{center}

\begin{center}\textbf{
Fernando L. Metz\textsuperscript{1 $\circ$}
 and
Isaac P\'erez Castillo\textsuperscript{2$\star$}
}\end{center}

\begin{center}
{\bf 1} Physics Institute, Federal University of Rio Grande do Sul, 91501-970 Porto Alegre, Brazil
\\
{\bf 2} Departamento de F\'isica, Universidad Autónoma Metropolitana-Iztapalapa, San Rafael Atlixco 186, Ciudad de México 09340, Mexico
\\[\baselineskip]
$\circ$ \href{mailto:fmetzfmetz@gmail.com}{\small fmetzfmetz@gmail.com}
$\star$ \href{mailto:iperez@izt.uam.mx}{\small iperez@izt.uam.mx}
\end{center}

\section*{\color{scipostdeepblue}{Abstract}}
\textbf{\boldmath{%
Dynamical mean-field theory (DMFT) provides an asymptotic description of broad classes of high-dimensional disordered dynamical systems with dense couplings, reducing the many-body dynamics to a self-consistent effective stochastic process. Many complex systems of current interest, however, interact through sparse and heterogeneous networks, where local fields contain only finitely many strong contributions and the Gaussian closure mechanisms of dense DMFT are not automatically available. We develop a continuous-time cavity derivation of sparse-network DMFT at the level of path measures for stochastic dynamics with general pairwise interactions on sparse random graphs. The resulting cavity equations are exact on trees and give the finite-time thermodynamic cavity description on locally tree-like sparse random graphs. They also make explicit how graph reciprocity changes the structure of dynamical closure: fully directed graphs reduce to the known sparse directed path-probability equation, whereas reciprocal or bidirected edges require conditional path kernels because neighbouring branches are driven by the imposed history of the receiving node. Ensemble averaging then leads to laws over path-probability messages whose barycenters and higher-message moments close by multilinearity of the path update and independence of incoming branches. A causal discrete-time rederivation gives the corresponding finite-time population-dynamics representation, distinguishing ordinary trajectory populations in directed graphs from conditional branch-law or finite-depth tree populations in reciprocal graphs. Since exact finite-time population dynamics carries complete histories, we also formulate finite-memory numerical closures and assess them for an additive-input Recurrent Neural Network (RNN) specialization. Finally, high-connectivity limits are obtained as projections of the sparse path-measure theory under additional scaling assumptions, clarifying when dense effective drift, noise, and response channels reduce to familiar low-dimensional DMFT equations and when path-level descriptions remain necessary.
}}

\vspace{\baselineskip}

\noindent\textcolor{white!90!black}{%
\fbox{\parbox{0.975\linewidth}{%
\textcolor{white!40!black}{\begin{tabular}{lr}%
  \begin{minipage}{0.6\textwidth}%
    {\small Copyright attribution to authors. \newline
    This work is a submission to SciPost Physics. \newline
    License information to appear upon publication. \newline
    Publication information to appear upon publication.}
  \end{minipage} & \begin{minipage}{0.4\textwidth}
    {\small Received Date \newline Accepted Date \newline Published Date}%
  \end{minipage}
\end{tabular}}
}}
}


\vspace{10pt}
\noindent\rule{\textwidth}{1pt}
\tableofcontents
\noindent\rule{\textwidth}{1pt}
\vspace{10pt}


\section{Introduction}
\label{sec:introduction}
Many dynamical phenomena in physics, biology, and information processing arise from the collective evolution of a large number of coupled degrees of freedom with quenched disorder and nonlinear interactions. Canonical examples include random neural-network models exhibiting transitions between fixed-point and chaotic regimes \cite{SompolinskyCrisantiSommers1988,kadmon2015transition,crisanti2018path}, disordered spin systems with nonequilibrium stochastic updates \cite{HatchettEtAl2004,NeriBolle2009,mimura2009parallel}, and random ecological communities described by Lotka--Volterra-type dynamics \cite{Bunin2017,roy2019numerical,altieri2021properties}. In these settings one aims to replace the intractable $N$-body dynamics by a reduced description that remains faithful to the disorder-induced heterogeneity and to the dynamical fluctuations generated by noise and nonlinear feedback. The conceptual difficulty is that disorder and dynamics intertwine: the same random interactions that create complex energy landscapes, metastability, and marginal modes also shape temporal correlations and responses, often making standard perturbative approaches unreliable outside weak-coupling regimes.

For densely connected random systems, the method of dynamical mean-field theory (DMFT) addresses this challenge by converting the original interacting dynamics into an effective single-site stochastic process driven by self-consistent noise and, in general, memory kernels. Historically, this reduction is naturally formulated using the path-integral (generating-functional) representation of stochastic dynamics \cite{MartinSiggiaRose1973,Janssen1976,dedominicis1976techniques,dedominicis1978dynamics}, which makes it possible to average analytically over quenched disorder and to identify the appropriate dynamical order parameters. Modern expositions emphasize how DMFT elevates two-time correlation and response functions to self-consistent objects, thereby producing closed integro-differential equations for effective processes in the thermodynamic limit \cite{HertzRoudiSollich2017,crisanti2018path}. Beyond neural-network models \cite{SompolinskyCrisantiSommers1988,kadmon2015transition}, this perspective has become central in disordered interacting particle systems and in theoretical ecology, where DMFT has been used to obtain phase diagrams and dynamical regimes, and also provides practical numerical schemes for solving the effective stochastic dynamics \cite{EissfellerOpper1992,Bunin2017,roy2019numerical,altieri2021properties}. A crucial structural ingredient in the dense setting is that the effective local field becomes Gaussian in the large-$N$ limit under suitable independence assumptions, which is what ultimately permits closure in terms of low-order dynamical observables.

Sparse random networks violate this simplifying mechanism in a fundamental way. When each node has only finitely many neighbours as $N\to\infty$, the local field is not a sum over $O(N)$ weak contributions but rather a sum over $O(1)$ terms of $O(1)$ magnitude; consequently, non-Gaussianity and sample-to-sample heterogeneity remain intrinsic even in the thermodynamic limit. Sparse random graph ensembles with prescribed degree distributions, including directed configuration-type ensembles, provide the natural probabilistic setting for this finite-connectivity regime \cite{NewmanStrogatzWatts2001,FosdickLarremoreNishimuraUgander2018}. Under the usual sparsity assumptions, their finite-depth neighbourhoods are locally tree-like in the large-system limit, which is the graph-theoretic basis for the branch independence used by cavity constructions \cite{BordenaveLelarge2010}. 

For discrete-spin dynamics on sparse graphs, generating-functional analyses established early that the correct dynamical order parameter is no longer a small set of two-time functions but a disorder-averaged measure over single-site trajectories (path probabilities) \cite{HatchettEtAl2004}. Complementary dynamical-replica analyses treated the dynamics of macroscopic observables such as magnetization and energy in finitely connected disordered Ising systems with random bonds and heterogeneous degree distributions \cite{HatchettPerezCastilloCoolenSkantzos2005}. This insight is closely connected to the cavity method for equilibrium models on locally tree-like graphs \cite{MezardMontanari2009}: on a tree, removing a node disconnects its neighbours and yields exact recursion relations. In the dynamical setting, the analogous recursions propagate probability measures over entire histories along directed edges; they are exact on trees and provide asymptotically exact finite-time descriptions on locally tree-like random graphs \cite{HatchettEtAl2004,NeriBolle2009}. The price is that the message space grows rapidly with time, so obtaining closed macroscopic equations typically requires either additional limits, such as complete asymmetry, or controlled approximations.

The form of these path-measure recursions depends crucially on the reciprocity structure of the graph. In fully directed networks, a neighbour that is driven by a node does not feed back into that same node along the same edge, and the outgoing feedback kernels normalize away. The resulting sparse theory can close at the level of unconditioned single-site path probabilities, with source histories retained as response-generating arguments when needed. This directed closure is the setting of early sparse-graph analyses of kinetic Ising systems and related models \cite{HatchettEtAl2004,derrida1987exactly,mimura2009parallel}, and in continuous-valued nonlinear dynamics it has recently been formulated as a sparse directed-network DMFT for single-site path probabilities \cite{Metz2025}. By contrast, when reciprocal or bidirected edges are present, a neighbouring branch is driven by the imposed history of the receiving node through an inserted drift in its branch equation. This insertion is additive in the branch dynamics, but for a general pairwise interaction it may still depend on the branch trajectory. The corresponding message is therefore an imposed-history conditional path kernel generated by a modified branch dynamics, not merely an unconditioned path law and not, in general, an ordinary source shift. This temporal feedback is the mechanism captured here for retarded self-interactions in reciprocal sparse dynamics, and it explains why directed and reciprocal sparse systems have different closure structures.

This distinction is closely related to, but not exhausted by, existing dynamical cavity and dynamic message-passing approaches. For graphs with reciprocal edges, dynamical cavity equations generally couple messages to the history of the receiving node \cite{NeriBolle2009,mimura2009parallel}. Approximation strategies for dilute kinetic Ising models and related systems retain the local tree-like structure while attempting to manage the resulting time correlations, including dynamic TAP-type expansions \cite{roudi2011dynamicaltap,aurell2012dynamic}, dynamic message-passing schemes with memory effects \cite{delferraro2015dmp,karrer2010message,lokhov2014inferring}, cavity-master-equation approaches for discrete-spin dynamics \cite{aurell2017cavity}, matrix-product representations of dynamical cavity messages \cite{barthel2018matrix,barthel2020matrix}, and backtracking dynamical cavity methods for long-time or attractor-related questions \cite{behrens2023backtracking}. These developments show both the usefulness of path messages on locally tree-like graphs and the difficulty of reducing them to tractable macroscopic equations once reciprocal temporal feedback is present.

For continuous-valued degrees of freedom and general nonlinear stochastic dynamics on sparse locally tree-like graphs, the corresponding model-agnostic path-measure structure is less developed, particularly for reciprocal feedback beyond ordinary source-shift reductions. For linearly coupled stochastic dynamics, dynamic cavity formulations can be made explicit within Gaussian approximations that preserve key temporal correlations; in the linear case with additive noise, this route leads to closed causal equations beyond the discrete-spin setting \cite{TaraboloDallAsta2025}. Recent sparse-interaction studies of generalized Lotka--Volterra models have also emphasized that finite connectivity can generate non-Gaussian and topology-driven effects absent from dense descriptions \cite{tonolo2026generalized,tarabolo2025sparse}. The sparse directed-network DMFT investigated by one the authors provides a fully directed continuous-valued path-probability equation for nonlinear dynamics on sparse random networks and demonstrates its use in neural, ecological, epidemic, and synchronization models \cite{Metz2025}. The present work embeds that directed result in a broader continuous-time cavity derivation for sparse random graphs with general pairwise interactions, where reciprocal feedback is treated through imposed-history conditional path kernels.

In this paper we develop the dynamical cavity method for stochastic dynamics of continuous degrees of freedom on sparse random networks with general pairwise interactions and both directed and reciprocal graph structure. Starting from the continuous-time path measure, we derive fixed-graph cavity messages and then pass to ensemble laws over these messages. The ensemble object is a probability law over path-probability messages; its barycenter is the first moment of that law, and higher-message moments are described by corresponding multi-copy path measures. The barycenter and higher-message-moment equations close because the path update is multilinear in incoming messages and because distinct incoming branches are independent in the locally tree-like finite-time limit. This is a closure at the level of path measures and message-law moments, not a general closure in a small set of observables such as means, correlations, and responses.

A further role of the path-measure formulation is to separate general conditional kernels from special source-shift reductions. For a general pairwise interaction $g(x_i,x_j)$, the history imposed by a reciprocal neighbour enters the branch dynamics through an additive inserted drift, but the inserted term may still depend on the branch trajectory. The corresponding object is therefore an imposed-history conditional kernel. In the source-shift cases, including additive-input models where the interaction depends only on the neighbour history and the linear interaction as a special case, this conditional dependence can be represented as an ordinary shift of the source history. This distinction is essential for keeping the general sparse theory separate from the model-specific reductions in which more familiar response-generating equations can be obtained. Closed Volterra equations for means, correlations, and responses require further structure, such as the linear--Gaussian reciprocal specialization, and do not follow automatically from the existence of a source shift.

The same structure also determines the causal finite-time numerical representation. We introduce a causal discretization as a discrete-time rederivation of the continuous-time path-measure equations, not as a separate theory. Its purpose is to make the time ordering explicit, to define population dynamics unambiguously, and to represent reciprocal feedback without causality ambiguities. In fully directed ensembles, the population-dynamics representation acts on ordinary trajectory populations. In reciprocal or bidirected ensembles, the propagated object is a conditional branch law, or equivalently a finite-depth causal tree of histories. This finite-time construction gives an algorithmic realization of the conditional-kernel structure that appears in the continuous-time cavity derivation. Because exact finite-time population dynamics carries complete histories whose dimension grows with the observation horizon, we also introduce finite-memory closures as numerical approximations to this representation and assess them for an additive-input RNN specialization.

Finally, the sparse path-measure theory provides a systematic starting point for model-specific reductions and high-connectivity projections. In the linear--Gaussian reciprocal case, the source-shift structure, Gaussianity, and linearity allow the conditional path-kernel recursion to be parametrized explicitly and decoded into Volterra equations for means, correlations, and responses. In high-connectivity limits, many sparse incoming contributions combine into effective drift, colored-noise, and, when reciprocal feedback is present, response channels. These dense effective processes are projections of the sparse path-measure theory under additional scaling assumptions. Closure in low-dimensional observables is then a further model-dependent question, not an automatic consequence of sparsity or of the cavity construction itself.

To orient the reader we have organized this work as follows. In Sec.~\ref{sec:model} we define the class of dynamical processes and network ensembles we consider, together with the auxiliary fields and noise conventions used to generate response and correlation functions. In Sec.~\ref{sec:derivations} we develop the continuous-time cavity formulation at the level of path probabilities and derive the corresponding dynamical cavity equations for the general pairwise model, making explicit how the graph structure enters, how reciprocity modifies the dependence of cavity messages on the history of the receiving node, and how the fully directed reduction is obtained. In Sec.~\ref{sec:known} we discuss representative reductions and checks, especially the linear--Gaussian reciprocal case, where the path-kernel recursion can be decoded into Volterra equations for means, correlations, and responses. In Sec.~\ref{sec:causal_discretization_population_dynamics} we introduce a causal discretization of the same path-measure equations and discuss the resulting population-dynamics representations, including directed trajectory populations, finite-depth constructions for bidirected graphs, finite-memory numerical closures, and their additive-input RNN comparison. In Sec.~\ref{sec:high_connectivity_response_channels} we discuss high-connectivity limits, separating dense effective path laws from the stronger problem of obtaining closed equations for low-dimensional dynamical observables. Additional technical and algorithmic details are collected in Apps.~\ref{app:rcopy_moments}, \ref{app:finite_depth_reciprocal_tree}, and \ref{app:directed_population_algorithm}.

\section{Model definitions}
\label{sec:model}
Coupled stochastic dynamical equations on graphs provide a common microscopic description for recurrent neural networks and related disordered dynamical systems \cite{SompolinskyCrisantiSommers1988}, interacting ecological communities \cite{Bunin2017}, spreading processes \cite{PastorSatorrasCastellanoVanMieghemVespignani2015}, and synchronization dynamics \cite{RodriguesPeronJiKurths2016}. We now fix the class of microscopic dynamics, graph ensembles, coupling conventions, noise convention, and source-field notation used in the path-measure cavity analysis below. The formulation includes the sparse directed-network setting of \cite{Metz2025} as a special case and is compatible with the standard MSRJD/generating-functional representation of stochastic dynamics \cite{MartinSiggiaRose1973,Janssen1976,dedominicis1978dynamics,HertzRoudiSollich2017} and with locally tree-like sparse graph ensembles \cite{MezardMontanari2009,BordenaveLelarge2010}.

We consider a system of $N$ interacting degrees of freedom $x_i(t)\in\mathbb{R}$ evolving on a graph. The real-valued coordinate setting is the default one used in the derivation; bounded, positive, or phase-valued examples are understood either through an appropriate coordinate representation or with the corresponding state-space interpretation. The microscopic dynamics is defined by the set of coupled stochastic differential equations
\begin{equation}
\dot x_i(t) = -f\big(x_i(t)\big)+\sum_{j=1}^N c_{ij} J_{ij} g\big(x_i(t),x_j(t)\big)+h_i(t)+\xi_i(t)\,, \qquad i=1,\dots,N\,,
\label{eq:microscopic_dynamics}
\end{equation}
where $f$ is a single-site drift term and $g$ specifies the pairwise interaction kernel. We assume throughout that the functions $f$ and $g$, together with the initial law and the noise convention specified below, define a well-posed normalized finite-time path measure under the causal discretization used in the sequel.

This notation covers several representative models by appropriate choices of the single-site drift and the pairwise kernel. A linear Ornstein--Uhlenbeck network is obtained from $f(x)=\lambda x$, with $\lambda>0$, and $g(x,x')=x'$. A standard rate-based recurrent neural-network model corresponds to $f(x)=x$ and $g(x,x')=\varphi(x')$, for instance $\varphi(x')=\tanh x'$. In a quenched mean-field SIS representation for infection probabilities $x_i\in[0,1]$, one may use $f(x)=x$ and $g(x,x')=(1-x)x'$, with infection-rate factors absorbed into $J_{ij}$. The Kuramoto form is represented by $f(x)=0$ and $g(x,x')=\sin(x'-x)$, with $x$ interpreted as a phase variable and intrinsic frequencies included, when present, as node fields. Lotka--Volterra dynamics in abundance variables $N_i>0$ fits directly into the pairwise formulation: in the normalization
\beeq{
\dot N_i(t)=N_i(t)\bigl(1-N_i(t)\bigr)+\lambda_{\rm im} +\sum_{j=1}^{N}c_{ij}J_{ij}N_i(t)N_j(t)\,,
}
one has $f(N)=N(N-1)-\lambda_{\rm im}$ and $g(N,N')=NN'$. At the level of the displayed deterministic drift in our previous definition, the coordinate change $x_i=\log N_i$ gives
\beeq{
\dot x_i(t)=1-e^{x_i(t)}+\lambda_{\rm im}e^{-x_i(t)}
+\sum_{j=1}^{N}c_{ij}J_{ij}e^{x_j(t)},
}
which has the same microscopic form with $f(x)=e^x-1-\lambda_{\rm im}e^{-x}$ and $g(x,x')=e^{x'}$. Thus the abundance variables give a genuinely pairwise kernel, whereas the log-abundance variables give a transformed representation whose interaction term depends only on the neighbour coordinate. Comparisons of stochastic sources, noise conventions, observables, and responses across such coordinate representations require transforming the full stochastic problem consistently. These examples are illustrative choices of $f$ and $g$, not restrictions of the general pairwise model.

The quenched couplings are encoded by (i) a binary adjacency matrix, whose entries $c_{ij}\in\{0,1\}$ indicate the presence of an edge $j\to i$ (we set $c_{ii}=0$ for simplicity), and (ii) real-valued interaction strengths $J_{ij}$. With this convention, $J_{ij}$ multiplies the contribution of $x_j$ to the equation of \(x_i\). If both $c_{ij}$ and $c_{ji}$ are present, then $J_{ij}$ and $J_{ji}$ are the two directed weights of the reciprocal pair. The functions $h_i(t)$ are external source fields. They are normally used to generate response functions and then set to zero after functional differentiation, although the same notation also allows one to keep actual external perturbations if these are part of the physical model. The stochastic force $\xi_i(t)$ is taken to be Gaussian with mean zero and covariance
\begin{equation}
\bracket{ \xi_i(t) \xi_j(t')} = \sigma^2\delta_{ij}\delta(t-t')\,,
\label{eq:noise_covariance}
\end{equation}
where $\bracket{(\cdots)}$ denotes an average over noise realizations for fixed disorder $(\mathbf c,\mathbf J)$. The deterministic dynamics is included as the degenerate limit $\sigma=0$. The additive nature of the noise makes the It\^{o}/Stratonovich convention immaterial for the present definitions; in the path-integral representation below, consistent with the standard Martin--Siggia--Rose--Janssen--De~Dominicis (MSRJD) construction \cite{MartinSiggiaRose1973,Janssen1976,dedominicis1978dynamics,HertzRoudiSollich2017}, we use the standard causal discretization, for which the Jacobian associated with the change from noises to trajectories is field-independent and can be absorbed into the normalization.

Together with the factorized initial distribution
\beeq{
p_0(\mathbf x(0))=\prod_{i=1}^N p_0(x_i(0))\,,
}
the stochastic dynamics in Eq.~\eqref{eq:microscopic_dynamics} and the noise convention in Eq.~\eqref{eq:noise_covariance} determine a conditional path probability for the trajectories given the prescribed source history. On a finite time interval $[0,T]$, this physical conditional path probability can be written in causal functional-delta form as
\beeq{
P[\mathbf{x}\mid\mathbf{h}] =p_0(\mathbf{x}(0))\int [d\boldsymbol{\xi}]P_\sigma[\boldsymbol{\xi}]\prod_{i=1}^N\delta_{(\mathrm{F})}\Bigg[&\dot x_i(t)+f(x_i(t))\\
&-\sum_{j=1}^N c_{ij}J_{ij}g(x_i(t),x_j(t))-h_i(t)-\xi_i(t)\Bigg]\,,
\label{eq:physical_path_probability}
}
where $[d\boldsymbol{\xi}]=\prod_i[d\xi_i]$, $P_\sigma[\boldsymbol{\xi}]$ denotes the Gaussian noise path weight associated with Eq.~\eqref{eq:noise_covariance}, and $\delta_{(\mathrm{F})}(\cdots)$ is a functional Dirac delta enforcing the stochastic equations of motion pointwise in time. The conditioning on $\mathbf h$ means that the path law is evaluated for a prescribed source history; it is not a posterior conditioning on an observed trajectory.

The graph structure of this dynamical system will enter through the in- and out-neighbourhoods of each node,
\begin{equation}
\partial_i^{\mathrm{in}} \equiv \{j:\,c_{ij}=1\}\,,\qquad  \partial_i^{\mathrm{out}} \equiv \{j:\,c_{ji}=1\}\,,
\label{eq:neighborhoods}
\end{equation}
with in- and out-degrees $k_i^{\mathrm{in}}=|\partial_i^{\mathrm{in}}|$ and $k_i^{\mathrm{out}}=|\partial_i^{\mathrm{out}}|$. The framework does not assume symmetry: undirected networks correspond to the special case $c_{ij}=c_{ji}$ (and, if desired, symmetric weights $J_{ij}=J_{ji}$), while purely directed networks correspond to ensembles in which reciprocal pairs are completely or asymptotically absent. When reciprocal pairs are present, local tree-likeness refers to the underlying sparse marked neighbourhood, with reciprocal pairs treated as bidirected edge marks rather than as long graph cycles that destroy branch independence. Allowing reciprocity is important because it can create rich dynamical feedback loops that render cavity messages history-dependent, an effect that is absent in fully asymmetric cases and which is central in certain dynamical-cavity approaches to nonequilibrium processes on graphs \cite{HatchettEtAl2004,NeriBolle2009,Metz2025}.

We focus on sparse random-graph ensembles, where the typical degrees remain $O(1)$ as $N\to\infty$. A convenient canonical choice is to think about a directed configuration-model ensemble specified by a normalized joint distribution $p_{k,\ell}$ for $(k^{\mathrm{in}},k^{\mathrm{out}})$, with finite and equal in- and out-degree means
\begin{equation}
c\equiv \sum_{k,\ell} k\,p_{k,\ell}=\sum_{k,\ell}\ell\,p_{k,\ell}<\infty\,.
\label{eq:mean_degree}
\end{equation}
Conditioned on a graphical degree sequence drawn from $p_{k,\ell}$ with equal total numbers of incoming and outgoing half-edges, edges are placed uniformly at random among half-edges. One may alternatively use the closely related constrained Erd\H{o}s--R\'enyi representation used in \cite{Metz2025}; in the thermodynamic sparse limit both descriptions lead to locally tree-like directed neighbourhoods under the usual assumptions \cite{NewmanStrogatzWatts2001,FosdickLarremoreNishimuraUgander2018,BordenaveLelarge2010}. The distribution $p_{k,\ell}$ is sufficient for the default non-reciprocal directed ensemble, where reciprocal pairs have vanishing density. Ensembles with a positive density of reciprocal edges require additional information, for instance a reciprocal-degree variable; the ensemble law is therefore refined later in Sec.~\ref{sec:derivations} to a joint law over in-degree, out-degree, and reciprocal degree when reciprocal pairs are treated explicitly. Undirected ensembles are obtained by imposing $c_{ij}=c_{ji}$. The sparsity assumption is empirically well-motivated across domains, since many measured interaction networks have finite mean degree and broad degree heterogeneity even at large system size \cite{Newman2010Networks,Guimaraes2020}, and directedness and non-normality are ubiquitous features with direct dynamical consequences \cite{AsllaniLambiotteCarletti2018}. Throughout, the disorder $(\mathbf c,\mathbf J)$ is quenched: we analyze the stochastic dynamics for a fixed realization and then consider typical behavior under the graph/coupling ensemble.

The weights $\{J_{ij}\}_{i\neq j}$ are taken as independent, or conditionally independent given the adjacency matrix, unless otherwise specified. Unreciprocated directed edges carry weights drawn from a distribution $p_J$, independent of the connectivity unless an explicit dependence is imposed. Reciprocal pairs may instead carry a joint law for the two directed weights, denoted later by $p_{\leftrightarrow}(dJ,d\widetilde J)$, with independent directed weights and symmetric weights as special cases. Thus, if reciprocal edges are present, the two directed weights $J_{ij}$ and $J_{ji}$ are distinct random variables unless a symmetric-weight constraint is imposed explicitly. For later comparisons between finite-connectivity behavior and dense mean-field limits, it is useful to allow for degree-dependent scalings that keep the total input $O(1)$ as connectivity increases, e.g.\ parametrizations where $\mathbb{E}[J_{ij}] = O(c^{-1})$ and $\mathrm{Var}(J_{ij})=O(c^{-1})$ so that the sum of $k^{\mathrm{in}}\sim c$ terms has finite mean and variance, in direct analogy with classical random-network scalings \cite{SompolinskyCrisantiSommers1988}. In the sparse regime with fixed $c$, such scalings amount to a convenient parametrization of the coupling distribution rather than an additional approximation; only in Sec.~\ref{sec:high_connectivity_response_channels} are they used as asymptotic assumptions in a high-connectivity projection.

To define correlation and response observables in a way that interfaces naturally with the cavity method at the level of path measures, we introduce the MSRJD path weight for trajectories $\{x_i(t)\}$ and auxiliary response fields $\{\hat x_i(t)\}$ \cite{MartinSiggiaRose1973,Janssen1976,dedominicis1978dynamics,HertzRoudiSollich2017}. The Fourier representation of the functional delta in Eq.~\eqref{eq:physical_path_probability}, followed by the Gaussian integration over the noises, gives the formal weight
\beeq{
P[\mathbf{x},\hat{\mathbf{x}}\mid\mathbf{h}]= p_0(\mathbf{x}(0))\exp\Bigg[&-\frac{\sigma^2}{2}\sum_{i=1}^N \int_0^T dt \hat x_i(t)^2-i\sum_{i=1}^N \int_0^T dt\hat x_i(t)\Bigg(\dot x_i(t)+f(x_i(t))\\
&-\sum_{j=1}^N c_{ij}J_{ij}g(x_i(t),x_j(t))-h_i(t)\Bigg)\Bigg]\,,
\label{eq:msrjd_weight}
}
up to normalization constants independent of the paths and sources. The physical path probability is obtained by integrating out $\hat{\mathbf{x}}$, i.e.\ $P[\mathbf{x}\mid\mathbf{h}]=\int [d\hat{\mathbf{x}}] P[\mathbf{x},\hat{\mathbf{x}}\mid\mathbf{h}]$. Here $[d\hat{\mathbf{x}}]$ denotes the formal path-integral measure over the auxiliary fields. If a generic object $\boldsymbol{\theta}$ is added additively to the source field, whether or not it can be recast as a source field, one obtains the identity
\begin{equation}
P[\mathbf{x}\mid\mathbf{h}+\boldsymbol{\theta}]= \int [d\hat{\mathbf{x}}] P[\mathbf{x},\hat{\mathbf{x}}\mid\mathbf{h}]\exp\left[i\sum_{i=1}^N\int dt\,\hat x_i(t)\theta_i(t)\right]\,.
\label{eq:source_shift_identity}
\end{equation}
This identity will be used repeatedly in the cavity derivations below. It simply records the fact that any additional contribution entering additively in the dynamical constraint can be represented, in the MSRJD weight, by the corresponding Fourier factor. As we will see in the cavity construction, the effect of a removed neighbour is also an additive insertion at the level of the source field. When this insertion is independent of the branch trajectory, it can be recast as an ordinary shift of the source history. For a general pairwise kernel $g(x_\ell,x_i)$, the same additive insertion may depend on the branch trajectory $x_\ell$, and is then interpreted as part of the imposed-history conditional kernel rather than as an external source history independent of the branch path.

Given this representation, two-time correlations and linear responses can be expressed in the usual way either by functional differentiation with respect to the sources $h_i$ or equivalently by mixed correlations involving $\hat x_i$:
\begin{equation}
C_{ij}(t,t') \equiv \bracket{ x_i(t) x_j(t')}\Big|_{\mathbf{h}=\mathbf{0}}\,,\qquad R_{ij}(t,t') \equiv \left.\frac{\delta}{\delta h_j(t')}\bracket{ x_i(t)}\right|_{\mathbf{h}=\mathbf{0}}= i\bracket{ x_i(t) \hat x_j(t')}\Big|_{\mathbf{h}=\mathbf{0}}\,,
\label{eq:correlation_response_definitions}
\end{equation}
where the last equality follows from Eq.~\eqref{eq:msrjd_weight} and the normalization of the physical path measure. After the MSRJD representation has been introduced, the same bracket denotes expectation with respect to the corresponding physical or auxiliary-field path measure for fixed disorder, with the meaning clear from the inserted variables; when only physical variables are present, it coincides with the noise average defined above. Since we will use the cavity method, disorder-averaged operations are not generally needed at the level of a fixed graph realization. When ensemble averages over the quenched graph and couplings are considered, they will be denoted by an overline $\overline{(\cdots)}$ when needed. These definitions fix the microscopic dynamics, graph ensemble, source convention, and path-measure notation from which the cavity equations are derived in the following section.

\section{Continuous-time path-measure cavity equations}
\label{sec:derivations}
We derive the cavity equations at the level of path measures. The fixed-graph construction is exact on trees, and on locally tree-like sparse random graphs it gives the finite-time cavity description in the thermodynamic limit \cite{MezardMontanari2009,BordenaveLelarge2010}. The starting point is the stochastic network dynamics defined in Sec.~\ref{sec:model}, supplemented with external source fields $h_i(t)$ to generate responses and additive Gaussian noise $\xi_i(t)$ to keep the path measure in the standard MSRJD class. We first derive the fixed-graph cavity equation, then pass to ensemble laws over messages, and finally take barycenters of those laws to obtain closed path-measure equations for averaged single-site dynamics. In the fully directed case the construction recovers the sparse directed path-probability equation derived by generating-functional methods in \cite{Metz2025}, while in the presence of reciprocal edges it requires imposed-history conditional path kernels. Throughout this section we use the physical path law $P[\mathbf{x}\mid\mathbf{h}]$ and the MSRJD weight $P[\mathbf{x},\hat{\mathbf{x}}\mid\mathbf{h}]$ defined in Eqs.~\eqref{eq:physical_path_probability} and \eqref{eq:msrjd_weight}.

The cavity derivation proceeds by constructing the single-node marginal path measure. We use the factorized initial law fixed in Sec.~\ref{sec:model}, so that single-site initial factors appear below. For a given node $i$ we define
\beeq{
P_i[x_i,\hat x_i\mid h_i]\equiv \int [d\mathbf{x}_{\setminus i} d\hat{\mathbf{x}}_{\setminus i}]P[\mathbf{x},\hat{\mathbf{x}}\mid\mathbf{h}]\,,
\label{eq:single_node_marginal}
}
and analogously its physical marginal $P_i[x_i\mid h_i]=\int [d\hat x_i] P_i[x_i,\hat x_i\mid h_i]$. Here the notation $\mathbf{x}_{\setminus i}$ stands for the vector $\mathbf{x}$ without the entry $i$. The dependence of $P_i$ on the full field history $\mathbf{h}$ is implicit in Eq.~\eqref{eq:single_node_marginal}; the local source $h_i$ appears only through the local response field $\hat x_i$ in the MSRJD weight \eqref{eq:msrjd_weight}, which is what allows local response functions to be generated from the marginal.

To expose the graph structure in the action, we introduce the in- and out-neighbourhoods
\beeq{
\partial_i^{\mathrm{in}}=\{j:\,c_{ij}=1\}\,,\qquad \partial_i^{\mathrm{out}}=\{j:\,c_{ji}=1\}\,,\qquad \partial_i=\partial_i^{\mathrm{in}}\cup\partial_i^{\mathrm{out}}\,.
\label{eq:neigh_sets_deriv}
}
Only in-neighbours appear in the equation of motion of $x_i$, whereas out-neighbours are the nodes whose dynamics receives a contribution from $x_i$. Writing the exponent in the MSRJD weight \eqref{eq:msrjd_weight} as an action $S[\mathbf{x},\hat{\mathbf{x}}]$, one can separate the terms involving node $i$ and its immediate neighbours as
\beeq{
S[\mathbf{x},\hat{\mathbf{x}}]&=S_i[x_i,\hat x_i\mid h_i]+i\int_0^T dt \hat{x}_i(t)\sum_{j\in\partial_i^{\mathrm{in}}}J_{ij}g\big(x_i(t),x_j(t)\big)\\
&+i\int_0^Tdt\sum_{j\in\partial_i^{\mathrm{out}}}\hat{x}_j(t)J_{ji}g\big(x_j(t),x_i(t)\big)+S^{(i)}[\mathbf{x}_{\setminus i},\hat{\mathbf{x}}_{\setminus i}\mid\mathbf{h}_{\setminus i}]\,,
\label{eq:action_decomposition}
}
where
\beeq{
S_i[x_i,\hat x_i\mid h_i] = -\frac{\sigma^2}{2}\int_0^T dt \hat x_i(t)^2-i\int_0^T dt \hat x_i(t)\left(\dot x_i(t)+f(x_i(t))-h_i(t)\right)\,,
\label{eq:single_site_action}
}
and $S^{(i)}$ is the action of the cavity system where node $i$ is removed, with all other edges among the remaining nodes kept.

The cavity step uses the fact that, on a tree, removing $i$ disconnects its neighbouring branches. On locally tree-like sparse random graphs, the same separation holds for fixed finite depth in the thermodynamic limit \cite{MezardMontanari2009,BordenaveLelarge2010}. Denoting by $P^{(i)}[\mathbf{x}_{\setminus i},\hat{\mathbf{x}}_{\setminus i}\,|\,\mathbf{h}_{\setminus i}]$ the MSRJD path weight of the system where node $i$ has been removed, we introduce the joint cavity marginal of the neighbour paths as follows,
\beeq{
P_{\partial_i}^{(i)}[x_{\partial_i},\hat x_{\partial_i}\mid h_{\partial_i}]\equiv
\int [d \mathbf{x}_{\setminus (i\cup\partial_i)}d\hat{\mathbf{x}}_{\setminus (i\cup\partial_i)}] P^{(i)}[\mathbf{x}_{\setminus i},\hat{\mathbf{x}}_{\setminus i}\mid \mathbf{h}_{\setminus i}]\,,
\label{eq:neigh_cavity_marginal}
}
where $x_{\partial_i}$ stands for the collection of $x$-variables associated with neighbours of $i$. This gives the exact finite-graph identity
\beeq{
P_i&[x_i,\hat x_i\mid h_i]= p_0\big(x_i(0)\big)\exp\left[-\frac{\sigma^2}{2}\int_0^T dt\hat x_i(t)^2-i\int_0^Tdt\hat x_i(t)\Big(\dot x_i(t)+f(x_i(t))-h_i(t)\Big)\right]\\
&\times\int [dx_{\partial_i} d\hat {x}_{\partial_i}]P^{(i)}_{\partial_i}[x_{\partial_i},\hat x_{\partial_i}\mid h_{\partial_i}]\exp\Bigg[
i\int_0^Tdt\hat x_i(t)\sum_{\ell\in\partial_i^{\mathrm{in}}}J_{i\ell}g\big(x_i(t),x_\ell(t)\big)
\\
&+i\int_0^T dt \sum_{\ell\in\partial_i^{\mathrm{out}}}\hat x_\ell(t)J_{\ell i}g\big(x_\ell(t),x_i(t)\big) \Bigg]\,.
\label{eq:msrjd_marginal_exact}
}
The tree or locally tree-like approximation enters only when this joint cavity marginal is factorized:
\beeq{
P^{(i)}_{\partial_i}[x_{\partial_i},\hat x_{\partial_i}\mid h_{\partial_i}]\simeq \prod_{\ell\in\partial_i} P^{(i)}_\ell[x_\ell,\hat x_\ell\mid h_\ell]\,,
\label{eq:cavity_factorization}
}
where $P^{(i)}_\ell[x_\ell,\hat x_\ell\mid h_\ell]$ --sometimes denoted as $P_{\ell\to i}[x_\ell,\hat x_\ell\mid h_\ell]$-- is the cavity marginal of node $\ell$ in the graph where $i$ is removed, equivalently the dynamical message sent from $\ell$ to $i$. Throughout, \(P_a^{(b)}\) denotes the message carried by node \(a\) in the cavity graph where \(b\) has been removed; along an edge \(a\to b\), the same object may be read as the message from \(a\) to \(b\). The factorization in Eq.~\eqref{eq:cavity_factorization} is exact on trees and is the standard finite-time closure underlying dynamical cavity methods on locally tree-like sparse graphs --sometimes also called the Bethe-Peierls approximation-- \cite{HatchettEtAl2004,NeriBolle2009}. Substituting \eqref{eq:cavity_factorization} into \eqref{eq:msrjd_marginal_exact} expresses the single-node path weight $P_i[x_i,\hat x_i\mid h_i]$ in terms of single-node cavity path weights $P^{(i)}_{\ell}[x_\ell,\hat x_\ell\mid h_\ell]$:
\beeq{
P_i[x_i,\hat x_i&\mid h_i]\simeq p_0\big(x_i(0)\big) \exp\left[-\frac{\sigma^2}{2}\int_0^T dt\hat x_i(t)^2-i\int_0^Tdt\hat x_i(t)\Big(\dot x_i(t)+f(x_i(t))-h_i(t)\Big)\right]\\
&\times\int \left[\prod_{\ell\in\partial_i}[d x_\ell d\hat x_\ell]P^{(i)}_{\ell}[x_\ell,\hat x_\ell\mid h_\ell]\right]\exp\Bigg[
i\int_0^Tdt\hat x_i(t)\sum_{\ell\in\partial_i^{\mathrm{in}}}J_{i\ell}g\big(x_i(t),x_\ell(t)\big)\\
&+i\int_0^Tdt\sum_{\ell\in\partial_i^{\mathrm{out}}}\hat{x}_\ell(t)J_{\ell i}g\big(x_\ell(t),x_i(t)\big)\Bigg]\,.
\label{eq:msrjd_marginal_message}
}

Equation~\eqref{eq:msrjd_marginal_message} is not yet a closed system of message-paths, because the full single-node path weight and the cavity path weight are different objects. The cavity path message is obtained by removing a neighbour of $i$, say $j\in\partial_i$, and repeating the same construction in the graph where $j$ is absent. With the convention that $A\setminus j$ means $A\setminus\{j\}$, this gives the unified cavity expression
\beeq{
P^{(j)}_i[x_i,\hat x_i&\mid h_i]\simeq p_0\big(x_i(0)\big) \exp\left[-\frac{\sigma^2}{2}\int_0^T dt\hat x_i(t)^2-i\int_0^Tdt\hat x_i(t)\Big(\dot x_i(t)+f(x_i(t))-h_i(t)\Big)\right]\\
&\times\int \left[\prod_{\ell\in\partial_i\setminus j}[d x_\ell d\hat x_\ell] P^{(i)}_{\ell}[x_\ell,\hat x_\ell\mid h_\ell]\right]\exp\Bigg[
i\int_0^Tdt\hat x_i(t)\sum_{\ell\in\partial_i^{\mathrm{in}}\setminus j}J_{i\ell}g\big(x_i(t),x_\ell(t)\big)\\
&+i\int_0^Tdt\sum_{\ell\in\partial_i^{\mathrm{out}}\setminus j}\hat{x}_\ell(t)J_{\ell i}g\big(x_\ell(t),x_i(t)\big)\Bigg]\,.
\label{eq:cavity_msrjd_message_unified}
}
This form covers the three possibilities in a general graph: $j$ may be only an incoming neighbour, only an outgoing neighbour, or a reciprocal neighbour in $\partial_i^{\mathrm{in}}\cap\partial_i^{\mathrm{out}}$. In the reciprocal case, removing $j$ removes it from both the incoming driving term and the outgoing feedback term.

We now pass from the auxiliary-field expression to physical path probabilities. For each neighbour $\ell\in\partial_i\setminus j$ and for a fixed admissible trajectory $x_i$, define the history-dependent drift
\begin{equation}
\Theta_{\ell\leftarrow i}[x_\ell,x_i](t)\equiv
\begin{cases}
J_{\ell i}g\big(x_\ell(t),x_i(t)\big)\,, & \ell\in\partial_i^{\mathrm{out}}\,,\\[1mm]
0\,, & \ell\notin\partial_i^{\mathrm{out}}\,.
\end{cases}
\label{eq:theta_feedback_drift}
\end{equation}
The neighbour factor generated by the Fourier term defines the imposed-history path law
\begin{equation}
P_\ell^{(i)}[x_\ell\Vert x_i;h_\ell]\equiv\int[d\hat x_\ell]P_\ell^{(i)}[x_\ell,\hat x_\ell\mid h_\ell]\exp\left[i\int_0^Tdt\,\hat x_\ell(t)\Theta_{\ell\leftarrow i}[x_\ell,x_i](t)\right]\,.
\label{eq:conditional_kernel_density}
\end{equation}
For each fixed admissible history $x_i$, the measure $[dx_\ell]P_\ell^{(i)}[x_\ell\Vert x_i;h_\ell]$ is a probability measure over histories $x_\ell$. It is the path law of the cavity branch of $\ell$ when $x_i$ is held fixed as an imposed input history. This should be distinguished from the regular conditional law obtained by conditioning the full coupled process on the event $X_i=x_i$; in that posterior object the observed trajectory $x_i$ also carries information about $x_\ell$ through the equation of motion of $i$. The cavity construction uses instead the path law in Eq.~\eqref{eq:conditional_kernel_density}, generated by the branch dynamics with node $i$ removed and then driven by the prescribed history $x_i$. For the general pairwise kernel, the imposed history $x_i$ enters the branch dynamics through the inserted drift $\Theta_{\ell\leftarrow i}[x_\ell,x_i]$. This contribution is additive in the branch equation and in the corresponding MSRJD Fourier factor, but it need not be recast as an external source history independent of the branch trajectory $x_\ell$. The notation $P_\ell^{(i)}[x_\ell\Vert x_i;h_\ell]$ therefore denotes the path law of the modified cavity branch driven by the imposed history $x_i$, with the possible $x_\ell$-dependence of the inserted drift kept inside the branch dynamics.

This path-law interpretation follows directly from the functional-delta representation. In the cavity branch of $\ell$ with node $i$ removed, the local functional constraint at $\ell$ has the form
\beeq{
\dot x_\ell(t)+f(x_\ell(t))-\sum_{m\in\partial_\ell^{\mathrm{in}}\setminus i}J_{\ell m}g(x_\ell(t),x_m(t))-h_\ell(t)-\xi_\ell(t)=0\,.
}
Multiplication by the Fourier factor in \eqref{eq:conditional_kernel_density} changes this local constraint to
\beeq{
\dot x_\ell(t)+f(x_\ell(t))-\sum_{m\in\partial_\ell^{\mathrm{in}}\setminus i}J_{\ell m}g(x_\ell(t),x_m(t))-h_\ell(t)-\Theta_{\ell\leftarrow i}[x_\ell,x_i](t)-\xi_\ell(t)=0\,.
\label{eq:abc}
}
Thus the expression in \eqref{eq:conditional_kernel_density} is the physical path probability of the modified cavity stochastic dynamics of node $\ell$ with the additional drift $\Theta_{\ell\leftarrow i}[x_\ell,x_i]$. Consequently,
\begin{equation}
\int [dx_\ell]\,P_\ell^{(i)}[x_\ell\Vert x_i;h_\ell]=1
\label{eq:conditional_kernel_normalization}
\end{equation}
for each fixed admissible $x_i$, under the same causal discretization and well-posedness assumptions as in the original stochastic dynamics. When no ambiguity is possible, we will expand the compact notation as follows
\beeq{
P_\ell^{(i)}[x_\ell\Vert x_i;h_\ell]\equiv P_\ell^{(i)}\left[x_\ell\mid h_\ell+\Theta_{\ell\leftarrow i}[x_\ell,x_i]\right]\,.
\label{eq:dfdf}
}
Here the notation $h_\ell+\Theta_{\ell\leftarrow i}[x_\ell,x_i]$ records the additive insertion produced by the Fourier factor in Eq.~\eqref{eq:conditional_kernel_density} and by the modified constraint in Eq.~\eqref{eq:abc}. For a general pairwise kernel, this inserted object may depend on the branch path $x_\ell$, so Eq.~\eqref{eq:dfdf} should be read as a definition of the normalized imposed-history path law rather than as an ordinary source-field representation. When $g(u,v)$ depends only on its second argument, we write $g(u,v)=g_{\rm add}(v)$. The inserted drift then becomes $\Theta_{\ell\leftarrow i}[x_\ell,x_i](t)=J_{\ell i}g_{\rm add}(x_i(t))$, which is independent of $x_\ell$. In this case the same notation can be recast as the usual source-history shift $h_\ell+J_{\ell i}g_{\rm add}(x_i)$. The linear case $g(u,v)=v$ gives the special source-history shift $h_\ell+J_{\ell i}x_i$.

After the neighbour auxiliary fields have been integrated into normalized imposed-history path laws, the remaining auxiliary field $\hat x_i$ enforces the stochastic equation for $x_i$ with branch histories sampled from those path laws. Integrating Eq.~\eqref{eq:cavity_msrjd_message_unified} over $\hat x_i$, representing the remaining local Fourier integral by the noise variable $\xi_i$, and using Eq.~\eqref{eq:conditional_kernel_density} gives the general cavity equation in physical path space:
\beeq{
P^{(j)}_{i}[x_i\mid h_i]&\simeq p_0\big(x_i(0)\big)\int [d\xi_i]P_\sigma[\xi_i]\int\left[\prod_{\ell\in\partial_i\setminus j}[dx_\ell] P^{(i)}_{\ell}\left[x_\ell\Vert x_i;h_\ell\right]\right]\\
&\times\delta_{(\mathrm{F})}\left[\dot{x}_i(t)+f\big(x_i(t)\big)-h_i(t)-\sum_{\ell\in\partial_i^{\mathrm{in}}\setminus j}J_{i\ell}g\big(x_i(t),x_\ell(t)\big)-\xi_i(t)\right]\,.
\label{eq:cavity_equation_general_density}
}
Equation \eqref{eq:cavity_equation_general_density} is the central fixed-graph dynamical cavity equation in physical path space after the tree or locally tree-like factorization. It is exact on trees and has the usual finite-time locally tree-like status in the thermodynamic sparse limit. It expresses the trajectory measure of $x_i$ in a cavity graph in terms of the imposed-history path laws of its remaining neighbours, with the effect of $x_i$ on a neighbour $\ell$ entering through $P_\ell^{(i)}[x_\ell\Vert x_i;h_\ell]$. This is the mechanism by which bidirectional couplings generate history-dependent feedback in sparse dynamics, a feature already emphasized in the kinetic Ising literature \cite{HatchettEtAl2004,NeriBolle2009}. The distinction is therefore not between an additive and a non-additive modification of the branch equation: the modification is additive in the local constraint in both cases. The distinction is whether the added object can be recast as an ordinary source history. For general $g(x_\ell,x_i)$, the added drift $J_{\ell i}g(x_\ell,x_i)$ remains part of the conditional branch dynamics, whereas for $g(x_\ell,x_i)=g_{\rm add}(x_i)$ it becomes an ordinary source history imposed on the branch. The full single-node marginal is recovered by replacing $\partial_i\setminus j$ with $\partial_i$ and $\partial_i^{\mathrm{in}}\setminus j$ with $\partial_i^{\mathrm{in}}$ in Eq.~\eqref{eq:cavity_equation_general_density}, with the same tree or finite-time locally tree-like status as the cavity equation itself. Thus, once the single-node cavity path probabilities are known, the single-node path probabilities are given by
\beeq{
P_{i}[x_i\mid h_i]&\simeq p_0\big(x_i(0)\big)\int [d\xi_i]P_\sigma[\xi_i]\int\left[\prod_{\ell\in\partial_i}[dx_\ell]P_\ell^{(i)}[x_\ell\Vert x_i;h_\ell]\right]\\
&\times\delta_{(\mathrm{F})}\left[\dot{x}_i(t)+f\big(x_i(t)\big)-h_i(t)-\sum_{\ell\in\partial_i^{\mathrm{in}}}J_{i\ell}g\big(x_i(t),x_\ell(t)\big)-\xi_i(t)\right]\,.
\label{eq:physical_equation_general}
}

Looking at Eq.~\eqref{eq:cavity_equation_general_density}, the role of reciprocity becomes transparent. If $\ell\in\partial_i^{\mathrm{out}}$ but $\ell\notin\partial_i^{\mathrm{in}}$, then $x_\ell$ does not appear in the functional constraint for $x_i$, and the path law $P_\ell^{(i)}[x_\ell\Vert x_i;h_\ell]$ integrates to one by normalization. Thus, on purely directed graphs without bidirected edges, so that $\partial_i^{\mathrm{in}}\cap\partial_i^{\mathrm{out}}=\varnothing$ for all $i$, the recursion collapses to a simpler form in which only incoming neighbours remain. For the directed cavity message sent along an oriented edge $i\to j$, the removed node $j$ is an outgoing neighbour of $i$ and hence $j\notin\partial_i^{\mathrm{in}}$. Therefore $\partial_i^{\mathrm{in}}\setminus j=\partial_i^{\mathrm{in}}$, and the directed cavity equation can be written as
\begin{align}
P^{(j)}_{i}[x_i\mid h_i]
&=p_0\big(x_i(0)\big)\int [d\xi_i]\,P_\sigma[\xi_i]\int \left[\prod_{\ell\in\partial_i^{\mathrm{in}}}[dx_\ell]
P^{(i)}_{\ell}\left[x_\ell\mid h_\ell\right]\right]\nonumber\\
&\hspace{2.0cm}\times\delta_{(\mathrm{F})}\left[\dot x_i(t)+f\big(x_i(t)\big)-h_i(t)-\sum_{\ell\in\partial_i^{\mathrm{in}}}J_{i\ell}g\big(x_i(t),x_\ell(t)\big)-\xi_i(t)\right].
\label{eq:cavity_equation_directed}
\end{align}
This absence of back-action is the dynamical simplification underlying the sparse directed-network DMFT of \cite{Metz2025}. When bidirected edges are present, the neighbours in $\partial_i^{\mathrm{in}}\cap\partial_i^{\mathrm{out}}$ appear both in the driving term of $x_i$ and through imposed-history path laws depending on $x_i$, so the recursion \eqref{eq:cavity_equation_general_density} remains explicitly history-dependent. The general formulation remains valid and provides the natural starting point to analyze model-specific closures.

At fixed directed tree depth, deleting the downstream node $i$ does not change the branch law of an incoming source node $\ell$. Thus one may identify $P_\ell^{(i)}$ with the corresponding full marginal path law $P_\ell$ of that source node. With this fixed-graph identification, Eq.~\eqref{eq:cavity_equation_directed} becomes
\begin{align}
P^{(j)}_{i}[x_i\mid h_i]
&=p_0\big(x_i(0)\big)\int [d\xi_i]\,P_\sigma[\xi_i]\int \left[\prod_{\ell\in\partial_i^{\mathrm{in}}}[dx_\ell]
P_{\ell}\left[x_\ell\mid h_\ell\right]\right]\nonumber\\
&\hspace{2.0cm}\times\delta_{(\mathrm{F})}\left[\dot x_i(t)+f\big(x_i(t)\big)-h_i(t)-\sum_{\ell\in\partial_i^{\mathrm{in}}}J_{i\ell}g\big(x_i(t),x_\ell(t)\big)-\xi_i(t)\right].
\label{eq:cavity_equation_directed_rewritten}
\end{align}
The corresponding expression for the \emph{full} single-site marginal $P_i[x_i\mid h_i]$ has the same local form,
\begin{align}
P_{i}[x_i\mid h_i]
&=p_0\big(x_i(0)\big)\int [d\xi_i]\,P_\sigma[\xi_i]\int \left[\prod_{\ell\in\partial_i^{\mathrm{in}}}[dx_\ell]\,
P_{\ell}\left[x_\ell\mid h_\ell\right]\right]\nonumber\\
&\hspace{2.0cm}\times\delta_{(\mathrm{F})}\left[\dot x_i(t)+f\big(x_i(t)\big)-h_i(t)-\sum_{\ell\in\partial_i^{\mathrm{in}}}J_{i\ell}g\big(x_i(t),x_\ell(t)\big)-\xi_i(t)\right].
\label{eq:node_marginal_directed_from_messages}
\end{align}
The equality of the local fixed-graph formulas should not be confused with the ensemble average. The cavity messages are sampled along directed edges and therefore use the source-biased edge law, whereas the full single-site marginals are sampled over nodes. This is why the directed ensemble equations distinguish the source-sampled path law from the node-uniform path law.

We now turn to the ensemble average of the finite-graph cavity equations. Up to this point, the cavity messages $P_i^{(j)}[\cdot\mid\cdot]$ have been defined on a fixed graph with fixed couplings; for such a quenched instance they are path-probability functionals attached to cavity edges. After averaging over a sparse random graph ensemble, the direct object is therefore a probability law over messages, i.e.\ a probability law over path probabilities. This object is one level higher than a single averaged path law. The point of the derivation below is to make this layer of randomness explicit and then show that the first moment of this law---its barycenter---satisfies a closed path-measure equation by multilinearity of the cavity update and independence of distinct incoming branches in the locally tree-like limit.

To make this explicit, we first define the message space and the relevant empirical laws. Let
\beeq{
\mathcal{M}\equiv \Big\{\eta:\ (x,h)\mapsto \eta[x\mid h] \text{ such that }\eta[\cdot\mid h] \text{ is a normalized path-probability on } \mathcal{X}\Big\}\,,
}
where $\mathcal{X}$ denotes the single-site path space on the chosen time interval, and the dependence on $h$ is retained to generate response functions. When a reciprocal edge transmits the history $x$ back into a neighbour message, we use the compact notation
\beeq{
\eta[y\Vert x;\widetilde J]\equiv \eta\left[y\mid \widetilde J g(y,x)\right]\,,
}
with the same imposed-history conditional-kernel interpretation established in the finite-graph cavity derivation. The expression $\widetilde J g(y,x)$ records an additive insertion in the branch equation. In the general pairwise case it may depend on the branch path \(y\), and need not be recast as an ordinary source field. Thus $\eta[y\Vert x;\widetilde J]$ denotes the path density of the neighbour branch driven by the imposed history $x$ through this inserted drift, not an independent new message space and not a collapse of the general kernel to a source shift. In the additive-input specialization $g(u,v)=g_{\rm add}(v)$, the inserted object is independent of \(y\), and the same notation reduces to the ordinary source-history form
\beeq{
\eta[y\Vert x;\widetilde J]=\eta\left[y\mid \widetilde J g_{\rm add}(x)\right]\,.
}

We first formulate the ensemble average for a sparse locally tree-like ensemble with both unreciprocated directed edges and reciprocal pairs. To make this explicit, decompose the neighbourhood of node $i$ into reciprocal, purely incoming, and purely outgoing parts:
\beeq{
\partial_i^{\leftrightarrow}\equiv \partial_i^{\mathrm{in}}\cap \partial_i^{\mathrm{out}}\,,\qquad
\partial_i^{\mathrm{in},0}\equiv \partial_i^{\mathrm{in}}\setminus \partial_i^{\mathrm{out}}\,,\qquad
\partial_i^{\mathrm{out},0}\equiv \partial_i^{\mathrm{out}}\setminus \partial_i^{\mathrm{in}}\,.
}
The associated counts are
\beeq{
K_i\equiv|\partial_i^{\mathrm{in}}|\,,\qquad
L_i\equiv|\partial_i^{\mathrm{out}}|\,,\qquad
B_i\equiv|\partial_i^{\leftrightarrow}|\,,
}
so that, by disjointness, $|\partial_i^{\mathrm{in},0}|=K_i-B_i$ and $|\partial_i^{\mathrm{out},0}|=L_i-B_i$. We assume a sparse locally tree-like configuration-type ensemble specified by a normalized limiting joint distribution
\begin{equation}
p_{k,\ell,b}\equiv \lim_{N\to\infty}\frac{1}{N}\sum_{i=1}^N \delta_{K_i,k}\,\delta_{L_i,\ell}\,\delta_{B_i,b}\,, \qquad 0\le b\le \min(k,\ell),
\label{eq:pkellb_def}
\end{equation}
with $\sum_{k,\ell,b}p_{k,\ell,b}=1$. For every finite directed graph, the total number of incoming half-edges equals the total number of outgoing half-edges, $\sum_{i}K_i=\sum_{i}L_i$, since each directed edge contributes once to the indegree of its target and once to the outdegree of its source. Hence any limiting degree law arising from graphical realizations must satisfy
\beeq{
\sum_{k,\ell,b} k\,p_{k,\ell,b}=\sum_{k,\ell,b}\ell\,p_{k,\ell,b}=c\,.
}
Here $c$ is the common finite mean in/out-degree. The mean reciprocal degree is
\beeq{
c_{\leftrightarrow}\equiv \sum_{k,\ell,b} b\,p_{k,\ell,b}\,.
}
Since $B_i$ contributes equally to the in- and out-degree of node $i$, the mean purely directed incoming and outgoing degrees are also equal. We denote this common mean by
\begin{equation}
c_{\rightarrow} \equiv\sum_{k,\ell,b}(k-b)\,p_{k,\ell,b}=\sum_{k,\ell,b}(\ell-b)\,p_{k,\ell,b}=c-c_{\leftrightarrow}\,.
\label{eq:c_split}
\end{equation}
Unreciprocated directed edges carry couplings drawn from $p_J$. For reciprocal pairs we write the joint law of the two directed couplings as $p_{\leftrightarrow}(dJ,d\tilde J)$, where $J$ denotes the coupling from the neighbour into the root and $\tilde J$ denotes the feedback coupling from the root into that neighbour. The independent directed-weight convention corresponds to $p_{\leftrightarrow}(dJ,d\tilde J)=p_J(dJ)p_J(d\tilde J)$, while symmetric undirected weights correspond to $p_{\leftrightarrow}(dJ,d\tilde J)=p_J(dJ)\delta_J(d\tilde J)$,  where $\delta_J(d\widetilde J)$ denotes the Dirac measure at $J$ in the variable $\widetilde J$.

As in the directed case, the quenched cavity description is encoded in empirical measures on $\mathcal{M}$, but now one must distinguish whether an oriented edge is part of a reciprocal pair. For a finite instance, define the node-uniform empirical measure of full marginals by
\begin{equation}
\mathcal{Q}_{N}(\mathrm{d}\eta)\equiv\frac{1}{N}\sum_{i=1}^N\delta_{P_i}\big(\mathrm{d}\eta\big)\,,
\label{eq:empirical_Q_node}
\end{equation}
the unreciprocated-edge source empirical measure by
\begin{equation}
\mathcal{Q}^{\rightarrow}_{N,s}(\mathrm{d}\eta)\equiv 
\frac{1}{\sum_{i=1}^N(L_i-B_i)}\sum_{i=1}^N (L_i-B_i)\,\delta_{\,P_i}\big(\mathrm{d}\eta\big)\,,
\label{eq:empirical_Q_source_unidir}
\end{equation}
and the reciprocal-edge empirical measure on oriented reciprocal edges by
\begin{equation}
\mathcal{Q}^{\leftrightarrow}_{N,s}(\mathrm{d}\eta)\equiv 
\frac{1}{\sum_{i=1}^N B_i}\sum_{i=1}^N\sum_{j\in \partial_i^{\leftrightarrow}}
\delta_{\,P_i^{(j)}}\big(\mathrm{d}\eta\big)\,.
\label{eq:empirical_Q_source_bidir}
\end{equation}
In \eqref{eq:empirical_Q_source_unidir} one samples a uniformly random unreciprocated directed edge $i\to j$, so $j\in\partial_i^{\mathrm{out},0}$, and records the message on that edge. Since removing such an outgoing neighbour does not affect the incoming field of $i$, the edge message coincides with the full marginal $P_i$. In \eqref{eq:empirical_Q_source_bidir} one samples a uniformly random oriented reciprocal edge $i\to j$, so $j\in\partial_i^{\leftrightarrow}$, and records the corresponding cavity message $P_i^{(j)}$, which differs from $P_i$ because $j$ is also an incoming neighbour and is removed from the driving term.

At this point it is useful to separate the different layers of randomness and the sense in which independence is invoked below. For fixed \(N\) and fixed disorder, the messages are deterministic objects and are generally correlated when viewed as functions of the full disorder realization. The cavity limit concerns a local weak limit statement: when one samples a typical node or a typical oriented edge from the ensemble and inspects a finite-depth neighbourhood, this neighbourhood converges in distribution to a random rooted tree. Distinct incoming branches of that limiting tree are independent because they depend on disjoint sets of quenched random variables. This branch-level independence, together with the independence of edge couplings from the cavity branch messages, will be used as a source of statistical independence that yields factorization identities.

Under the standard finite-time locally tree-like cavity/self-averaging assumption, the empirical measures \eqref{eq:empirical_Q_node}--\eqref{eq:empirical_Q_source_bidir} are represented in the thermodynamic limit by deterministic laws $\mathcal{Q}$, $\mathcal{Q}^{\rightarrow}_s$, and $\mathcal{Q}^{\leftrightarrow}_s$ on $\mathcal M$. The edge-sampled limits are characterized by distributional fixed-point equations. To write these equations, introduce an update operator that encodes the general physical cavity recursion after purely outgoing neighbours have been discarded by normalization. Let $u\ge 0$ denote the number of purely incoming neighbours and $v\ge 0$ the number of reciprocal neighbours included in the driving sum. Let $\mathbf{J}=(J_1,\dots,J_u)$ denote the incoming couplings from purely incoming neighbours, and let $(J^\leftrightarrow_s,\tilde J_s)_{s=1}^v$ denote the incoming and feedback couplings along reciprocal pairs. Define $\mathfrak{G}_{u,v,\mathbf{J},\mathbf{J}^{\leftrightarrow},\tilde{\mathbf{J}}}$ by
\begin{align}
&\big(\mathfrak{G}_{u,v,\mathbf{J},\mathbf{J}^{\leftrightarrow},\tilde{\mathbf{J}}}(\eta_1,\dots,\eta_u;\tilde\eta_1,\dots,\tilde\eta_v)\big)[x\mid h]=p_0(x(0))\int [d\xi] P_\sigma[\xi]\int \left[\prod_{r=1}^{u}[dx_r] \eta_r[x_r\mid 0]\right]\nonumber\\
&\quad\times \int\left[\prod_{s=1}^{v}[dy_s] \tilde\eta_s\left[y_s\Vert x;\tilde J_s\right]\right] \delta_{(\mathrm{F})}\Bigg[\dot x(t)+f(x(t))-h(t)-\sum_{r=1}^{u}J_rg\big(x(t),x_r(t)\big)\nonumber\\
&\hspace{7cm}-\sum_{s=1}^{v}J^\leftrightarrow_s g\big(x(t),y_s(t)\big)-\xi(t)\Bigg]\,.
\label{eq:G_operator_def}
\end{align}

Two properties of this update will be used repeatedly. The first is multilinearity in the incoming messages. Although the functions $f$ and $g$ may be nonlinear in the dynamical variables, each unreciprocated message $\eta_r$ appears in Eq.~\eqref{eq:G_operator_def} only through the factor $\eta_r[x_r\mid 0]$, and each reciprocal message $\tilde\eta_s$ appears only through the factor $\tilde\eta_s[y_s\Vert x;\tilde J_s]$. The integrations over paths, the average over the local noise, and the functional constraint act linearly on these factors. Reciprocal-edge insertions change the branch dynamics at which a reciprocal message is evaluated. They enter additively in the branch dynamical constraint; in the additive-input specialization this additive insertion can be recast as an ordinary source-history shift, while in the general pairwise case it may depend on the branch trajectory. In either case, these reciprocal factors do not introduce a normalization factor depending on the collection of incoming messages. Thus $\mathfrak{G}_{u,v,\mathbf{J},\mathbf{J}^{\leftrightarrow},\tilde{\mathbf{J}}}$ is multilinear in $(\eta_1,\dots,\eta_u;\tilde\eta_1,\dots,\tilde\eta_v)$. More explicitly, for any $\alpha\in[0,1]$ and any two unreciprocated input messages $\eta_r,\eta'_r\in\mathcal{M}$, one has
\begin{align}
&\mathfrak{G}_{u,v,\mathbf{J},\mathbf{J}^{\leftrightarrow},\tilde{\mathbf{J}}}(\eta_1,\dots,\alpha\eta_r+(1-\alpha)\eta'_r,\dots,\eta_u;\tilde\eta_1,\dots,\tilde\eta_v) \nonumber\\
&\qquad =\alpha\mathfrak{G}_{u,v,\mathbf{J},\mathbf{J}^{\leftrightarrow},\tilde{\mathbf{J}}}(\eta_1,\dots,\eta_r,\dots,\eta_u;\tilde\eta_1,\dots,\tilde\eta_v)\nonumber\\
&\qquad\quad+(1-\alpha)\mathfrak{G}_{u,v,\mathbf{J},\mathbf{J}^{\leftrightarrow},\tilde{\mathbf{J}}}(\eta_1,\dots,\eta'_r,\dots,\eta_u;\tilde\eta_1,\dots,\tilde\eta_v)\,,
\label{eq:G_multilinear_unreciprocated}
\end{align}
and similarly, for any two reciprocal input messages $\tilde\eta_s,\tilde\eta'_s\in\mathcal{M}$,
\begin{align}
&\mathfrak{G}_{u,v,\mathbf{J},\mathbf{J}^{\leftrightarrow},\tilde{\mathbf{J}}}(\eta_1,\dots,\eta_u;\tilde\eta_1,\dots,\alpha\tilde\eta_s+(1-\alpha)\tilde\eta'_s,\dots,\tilde\eta_v)\nonumber\\
&\qquad =\alpha\mathfrak{G}_{u,v,\mathbf{J},\mathbf{J}^{\leftrightarrow},\tilde{\mathbf{J}}}(\eta_1,\dots,\eta_u;\tilde\eta_1,\dots,\tilde\eta_s,\dots,\tilde\eta_v)\nonumber\\
&\qquad+(1-\alpha)\mathfrak{G}_{u,v,\mathbf{J},\mathbf{J}^{\leftrightarrow},\tilde{\mathbf{J}}}(\eta_1,\dots,\eta_u;\tilde\eta_1,\dots,\tilde\eta'_s,\dots,\tilde\eta_v)\,.
\label{eq:G_multilinear_reciprocal}
\end{align}
The normalization of the output path law is supplied by the stochastic kernel defined by the noise measure and the functional constraint in Eq.~\eqref{eq:G_operator_def}, together with the normalization of the input path-probability messages. It is not supplied by a denominator of the form $Z(\eta_1,\dots,\eta_u;\tilde\eta_1,\dots,\tilde\eta_v)$. This is important because, if such a denominator were present, the update would be a ratio of message-dependent functionals rather than a multilinear map.

The second property is the branch independence supplied by the locally tree-like ensemble. Distinct incoming branches of the limiting random tree depend on disjoint sets of quenched variables. Therefore the messages occupying different input slots of $\mathfrak G$ are independent random messages. More concretely, after writing the distributional fixed point for the message law, substituting the explicit formula for $\mathfrak G$ produces random-message factors of the form
\beeq{
\prod_{r=1}^{u}\eta_r[x_r\mid 0]\prod_{s=1}^{v}\tilde\eta_s[y_s\Vert x;\tilde J_s]\,.
}
Because the messages in different slots are independent, the ensemble average of this product factorizes into the product of ensemble averages:
\beeq{
\mathbb{E}\left[\prod_{r=1}^{u}\eta_r[x_r\mid 0]\prod_{s=1}^{v}\tilde\eta_s[y_s\Vert x;\tilde J_s]\right]=\prod_{r=1}^{u}\mathbb{E}\left[\eta_r[x_r\mid 0]\right]\prod_{s=1}^{v}\mathbb{E}\left[\tilde\eta_s[y_s\Vert x;\tilde J_s]\right]\,.
}
Each expectation on the right-hand side is the corresponding mean message. Thus, after taking the first moment of the message-law fixed point, every random incoming message can be replaced by its mean message, giving closed equations for the mean messages, or barycenters, of the message laws at the path-measure level. The same mechanism has a higher-moment analogue. The $r$-th moment of a message law is the path probability of $r$ copies of the same cavity process evolving in the same quenched cavity environment, with independent dynamical noises conditional on that environment. Equivalently, if $\eta$ is a random path-probability message with law $\mathcal Q$, the corresponding $r$-copy path measure is
\beeq{
S^{(r)}_{\mathcal{Q}}[dx^1,\dots,dx^r\mid h^1,\dots,h^r]\equiv\int \mathcal Q(d\eta)\eta[dx^1\mid h^1]\cdots \eta[dx^r\mid h^r]\,.
}
For $r=1$ this gives the mean message, while for $r=2$ it gives the joint path probability of two copies evolving in the same random cavity environment and therefore contains the covariance of the random path probabilities themselves. More generally, multilinearity of the update and independence of distinct incoming branches imply a closed hierarchy for these $r$-copy path measures. In the main text we focus on the first member of this hierarchy; the $r$-copy formulation is developed in App.~\ref{app:rcopy_moments}. We now proceed to the ensemble equations for the laws of path-probability messages.

For an unreciprocated oriented edge, the source degree triple $(K,L,B)$ is sampled with size-bias proportional to $L-B$, giving weight $(\ell-b)p_{k,\ell,b}/c_{\rightarrow}$. Conditional on $(k,\ell,b)$, the source has $u=k-b$ purely incoming neighbours and $v=b$ reciprocal neighbours contributing to its dynamics. For a reciprocal oriented edge, the source is sampled with size-bias proportional to $B$, giving weight $b\,p_{k,\ell,b}/c_{\leftrightarrow}$. The message on that edge is a cavity message with the corresponding reciprocal neighbour removed, so $u=k-b$ and $v=b-1$. Thus, whenever $c_{\rightarrow}>0$ and $c_{\leftrightarrow}>0$, the coupled distributional fixed point is
\begin{align}
\mathcal{Q}^{\rightarrow}_s(\mathrm{d}\eta)&=\sum_{\substack{k,\ell,b\ge 0\\0\le b\le \min(k,\ell)}}\frac{(\ell-b)\,p_{k,\ell,b}}{c_{\rightarrow}}\int\Bigg[\prod_{r=1}^{k-b}\mathcal{Q}^{\rightarrow}_s(\mathrm{d}\eta_r)\,p_J(\mathrm{d}J_r)\Bigg]\nonumber\\
&\hspace{-.5cm}\times\int\Bigg[\prod_{s=1}^{b}\mathcal{Q}^{\leftrightarrow}_s(\mathrm{d}\tilde\eta_s)\,p_{\leftrightarrow}(\mathrm{d}J^\leftrightarrow_s,\mathrm{d}\tilde J_s)\Bigg]\delta_{\mathcal{M}}\Big(\eta-\mathfrak{G}_{k-b,b,\mathbf{J},\mathbf{J}^{\leftrightarrow},\tilde{\mathbf{J}}}(\eta_1,\dots,\eta_{k-b};\tilde\eta_1,\dots,\tilde\eta_b)\Big)\,,\label{eq:Q_fixed_point_unidir_general}\\
\mathcal{Q}^{\leftrightarrow}_s(\mathrm{d}\eta)&=\sum_{\substack{k,\ell,b\ge 0\\1\le b\le \min(k,\ell)}}\frac{b\,p_{k,\ell,b}}{c_{\leftrightarrow}}\int\Bigg[\prod_{r=1}^{k-b}\mathcal{Q}^{\rightarrow}_s(\mathrm{d}\eta_r)\,p_J(\mathrm{d}J_r)\Bigg]\nonumber\\
&\hspace{-.5cm}\times\int\Bigg[\prod_{s=1}^{b-1}\mathcal{Q}^{\leftrightarrow}_s(\mathrm{d}\tilde\eta_s)\,p_{\leftrightarrow}(\mathrm{d}J^\leftrightarrow_s,\mathrm{d}\tilde J_s)\Bigg]\delta_{\mathcal{M}}\Big(\eta-\mathfrak{G}_{k-b,b-1,\mathbf{J},\mathbf{J}^{\leftrightarrow},\tilde{\mathbf{J}}}(\eta_1,\dots,\eta_{k-b};\tilde\eta_1,\dots,\tilde\eta_{b-1})\Big)\,.
\label{eq:Q_fixed_point_bidir_general}
\end{align}
The convention is that empty products are equal to one. If $c_{\rightarrow}=0$, the first equation is absent; if $c_{\leftrightarrow}=0$, the second equation is absent.

Equations \eqref{eq:Q_fixed_point_unidir_general}--\eqref{eq:Q_fixed_point_bidir_general} are the ensemble-level cavity equations for the probability laws of path-probability messages. Their structure is one level higher than an ordinary self-consistency equation for a path probability. A draw from $\mathcal{Q}^{\rightarrow}_s$ or $\mathcal{Q}^{\leftrightarrow}_s$ is itself a normalized path measure, and the fixed point determines the probability distribution of such measures under the random graph and coupling ensemble. From a formal point of view this is the natural analogue of the distribution of cavity messages in sparse disordered systems. From a practical point of view, however, solving \eqref{eq:Q_fixed_point_unidir_general}--\eqref{eq:Q_fixed_point_bidir_general} directly would amount to representing a population whose elements are themselves path probabilities, or equivalently a population of stochastic processes rather than a population of trajectories.

For the averaged single-site dynamics, one does not need to solve the full probability law over path probabilities. The distributional fixed point is the correct ensemble-level object, but its first moment has a closed path-measure equation. Thus, instead of attempting to propagate the full laws $\mathcal{Q}^{\rightarrow}_s$ and $\mathcal{Q}^{\leftrightarrow}_s$, we can extract their mean path-probability messages. These first moments are the barycenters of the message laws:
\begin{equation}
P^{\rightarrow}_s[x\mid h]\equiv \int \mathcal{Q}^{\rightarrow}_s(\mathrm{d}\eta)\,\eta[x\mid h]\,,
\qquad P^{\leftrightarrow}_s[x\mid h]\equiv \int \mathcal{Q}^{\leftrightarrow}_s(\mathrm{d}\eta)\,\eta[x\mid h]\,.
\label{eq:Ps_barycenters_general_def}
\end{equation}
We also write
\beeq{
P^\leftrightarrow_s[y\Vert x;\widetilde J] \equiv P^\leftrightarrow_s\!\left[y\mid \widetilde J g(y,x)\right]\,,
}
with the imposed-history conditional-kernel interpretation introduced above. In this notation $\widetilde J g(y,x)$ is the additive drift inserted in the branch equation. For a general pairwise kernel it may depend on $y$, while for $g(u,v)=g_{\rm add}(v)$ it can be recast as the ordinary source history $\widetilde J g_{\rm add}(x)$.

Multiplying \eqref{eq:Q_fixed_point_unidir_general}--\eqref{eq:Q_fixed_point_bidir_general} by $\eta[x\mid h]$, integrating over $\eta$, and substituting the explicit form of $\mathfrak G$, the only dependence on the random incoming messages is through products of the form
\beeq{
\prod_{r=1}^{u}\eta_r[x_r\mid 0]\;\prod_{s=1}^{v}\tilde\eta_s\!\left[y_s\Vert x;\tilde J_s\right]\,.
}
For fixed paths $(x_1,\dots,x_u,y_1,\dots,y_v,x)$ and fixed couplings, the product measures in \eqref{eq:Q_fixed_point_unidir_general}--\eqref{eq:Q_fixed_point_bidir_general} give the factorization identities
\begin{align}
\int\Bigg[\prod_{r=1}^{u}\mathcal{Q}^{\rightarrow}_s(\mathrm{d}\eta_r)\Bigg]\prod_{r=1}^{u}\eta_r[x_r\mid 0]&=\prod_{r=1}^{u}P^{\rightarrow}_s[x_r\mid 0]\,,\label{eq:factorization_unidir_general}\\
\int\Bigg[\prod_{s=1}^{v}\mathcal{Q}^{\leftrightarrow}_s(\mathrm{d}\tilde\eta_s)\Bigg]\prod_{s=1}^{v}\tilde\eta_s\!\left[y_s\Vert x;\tilde J_s\right]&=\prod_{s=1}^{v}P^{\leftrightarrow}_s\!\left[y_s\Vert x;\tilde J_s\right]\,.
\label{eq:factorization_bidir_general}
\end{align}
These identities are the precise point at which branch independence enters the first-moment equations: the random messages being averaged occupy distinct incoming slots of the update operator.

Inserting \eqref{eq:factorization_unidir_general}--\eqref{eq:factorization_bidir_general} yields the closed barycenter equations. For the unreciprocated-edge barycenter,
\begin{align}
P^{\rightarrow}_s[x\mid h]&=\sum_{\substack{k,\ell,b\ge 0\\0\le b\le \min(k,\ell)}}\frac{(\ell-b)\,p_{k,\ell,b}}{c_{\rightarrow}} p_0(x(0))\int [d\xi] P_\sigma[\xi]\int \left[\prod_{r=1}^{k-b}[dx_r] P^{\rightarrow}_s[x_r\mid 0]\right]\nonumber\\
&\times \int\Bigg[\prod_{r=1}^{k-b}p_J(\mathrm{d}J_r)\Bigg]\int\Bigg[\prod_{s=1}^{b}p_{\leftrightarrow}(\mathrm{d}J^\leftrightarrow_s,\mathrm{d}\tilde J_s)\Bigg]\int \left[\prod_{s=1}^{b}[dy_s]\,P^{\leftrightarrow}_s \left[y_s\Vert x;\tilde J_s\right]\right] \nonumber\\
&\times\delta_{(\mathrm{F})}\left[\dot x(t)+f(x(t))-h(t)-\sum_{r=1}^{k-b}J_r g\big(x(t),x_r(t)\big)-\sum_{s=1}^{b}J^\leftrightarrow_s g\big(x(t),y_s(t)\big)-\xi(t)\right]\,,
\label{eq:Ps_unidir_closed_general}
\end{align}
provided $c_{\rightarrow}>0$. For the reciprocal-edge barycenter,
\begin{align}
P^{\leftrightarrow}_s[x\mid h]&=\sum_{\substack{k,\ell,b\ge 0\\1\le b\le\min(k,\ell)}}\frac{b\,p_{k,\ell,b}}{c_{\leftrightarrow}}p_0(x(0))
\int [d\xi] P_\sigma[\xi]\int\left[\prod_{r=1}^{k-b}[dx_r] P^{\rightarrow}_s[x_r\mid 0]\right]\nonumber\\
&\times \int\Bigg[\prod_{r=1}^{k-b}p_J(\mathrm{d}J_r)\Bigg]\int\Bigg[\prod_{s=1}^{b-1}p_{\leftrightarrow}(\mathrm{d}J^\leftrightarrow_s,\mathrm{d}\tilde J_s)\Bigg]\int \left[\prod_{s=1}^{b-1}[dy_s]\,P^{\leftrightarrow}_s \left[y_s\Vert x;\tilde J_s\right]\right]\nonumber\\
&\times\delta_{(\mathrm{F})}\left[\dot x(t)+f(x(t))-h(t)-\sum_{r=1}^{k-b}J_r g\big(x(t),x_r(t)\big)-\sum_{s=1}^{b-1}J^\leftrightarrow_s g\big(x(t),y_s(t)\big)-\xi(t)\right]\,,
\label{eq:Ps_bidir_closed_general}
\end{align}
provided $c_{\leftrightarrow}>0$. The node-uniform averaged path-probability for a uniformly sampled node is obtained by sampling $(k,\ell,b)$ from $p_{k,\ell,b}$ rather than from an edge-biased law:
\begin{align}
P[x\mid h]&=\sum_{\substack{k,\ell,b\ge 0\\0\le b\le \min(k,\ell)}}p_{k,\ell,b}p_0(x(0))\int [d\xi] P_\sigma[\xi]\int \left[\prod_{r=1}^{k-b}[dx_r] P^{\rightarrow}_s[x_r\mid 0]\right]\nonumber\\
&\quad\times\int\Bigg[\prod_{r=1}^{k-b}p_J(\mathrm{d}J_r)\Bigg]\int\Bigg[\prod_{s=1}^{b}p_{\leftrightarrow}(\mathrm{d}J^\leftrightarrow_s,\mathrm{d}\tilde J_s)\Bigg]\int \left[\prod_{s=1}^{b}[dy_s] P^{\leftrightarrow}_s\left[y_s\Vert x;\tilde J_s\right]\right] \nonumber\\
&\quad\times\delta_{(\mathrm{F})}\left[\dot x(t)+f(x(t))-h(t)-\sum_{r=1}^{k-b}J_rg\big(x(t),x_r(t)\big)-\sum_{s=1}^{b}J^\leftrightarrow_s g \big(x(t),y_s(t)\big)-\xi(t)\right]\,.
\label{eq:P_node_uniform_general}
\end{align}
Equations \eqref{eq:Q_fixed_point_unidir_general}--\eqref{eq:Q_fixed_point_bidir_general} are the ensemble-level cavity statements at the level of message distributions. Equations \eqref{eq:Ps_unidir_closed_general}--\eqref{eq:P_node_uniform_general} are the corresponding first-moment identities at the path-measure level: they follow from multilinearity of the update and independence of distinct incoming branches.

It is useful to emphasize, in view of the equilibrium cavity analogy, why this barycenter closure is special. In a standard normalized belief-propagation update one has schematically
\beeq{
\eta(\sigma)=\frac{1}{Z(\eta_1,\dots,\eta_k)}\Phi(\sigma)\prod_{r=1}^k\sum_{\sigma_r}\Psi_r(\sigma,\sigma_r)\,\eta_r(\sigma_r)\,,
}
so the updated message is a ratio of multilinear functionals of the incoming messages. Averaging such an update gives an expectation of a ratio and cannot generally be rewritten only in terms of the barycenter $\bar\eta=\int \mathcal{Q}(d\eta)\eta$. In the present dynamical path-measure update, the noise integral and functional constraint define a stochastic kernel, and no additional incoming-message-dependent normalization appears in \eqref{eq:G_operator_def}. This is the reason why the first moment of the message law closes.

We now recover the fully directed non-reciprocal case. This corresponds to $p_{k,\ell,b}=p_{k,\ell}\delta_{b,0}$, hence $c_{\leftrightarrow}=0$, $c_{\rightarrow}=c$, and $P^{\rightarrow}_s=P_s$. The reciprocal-edge law is absent. The update operator reduces to
\begin{align}
\big(\mathfrak{F}_{k,\mathbf{J}}(\eta_1,\dots,\eta_k)\big)[x\mid h]&=p_0(x(0))\int [d\xi] P_\sigma[\xi]\int \left[\prod_{r=1}^{k}[dx_r]\eta_r[x_r\mid 0]\right]\nonumber\\
&\hspace{1.0cm}\times\delta_{(\mathrm{F})}\left[\dot x(t)+f(x(t))-h(t)-\sum_{r=1}^{k}J_r g\big(x(t),x_r(t)\big)-\xi(t)\right]\,.
\label{eq:F_operator_def}
\end{align}
The source-sampled message law satisfies
\begin{align}
\mathcal{Q}_s(\mathrm{d}\eta)&=\sum_{k,\ell\ge 0}\frac{\ell p_{k,\ell}}{c}\int\left[\prod_{r=1}^{k}\mathcal{Q}_s(\mathrm{d}\eta_r) p_J(\mathrm{d}J_r)\right]\delta_{\mathcal{M}}\Big(\eta-\mathfrak{F}_{k,\mathbf{J}}(\eta_1,\dots,\eta_k)\Big)\,.
\label{eq:Q_fixed_point}
\end{align}
The barycenter
\begin{equation}
P_s[x\mid h]\equiv \int \mathcal{Q}_s(\mathrm{d}\eta)\eta[x\mid h]
\label{eq:Ps_barycenter_def}
\end{equation}
therefore satisfies
\begin{align}
P_s[x\mid h]&=\sum_{k,\ell\ge 0}\frac{\ell p_{k,\ell}}{c}p_0(x(0))\int [d\xi] P_\sigma[\xi]\int \left[\prod_{r=1}^{k} [dx_r] P_s[x_r\mid 0]\right]\int\left[\prod_{r=1}^{k} p_J(\mathrm{d}J_r)\right]\nonumber\\
&\hspace{2.0cm}\times\delta_{(\mathrm{F})}\left[\dot x(t)+f(x(t))-h(t)-\sum_{r=1}^{k}J_r g\big(x(t),x_r(t)\big)-\xi(t)\right]\,.
\label{eq:Ps_closed}
\end{align}
The node-uniform averaged path-probability is obtained by sampling the root degree pair $(k,\ell)$ from $p_{k,\ell}$:
\begin{align}
P[x\mid h]&=\sum_{k,\ell\ge 0}p_{k,\ell}\ p_0(x(0))\int [d\xi] P_\sigma[\xi]\int \left[\prod_{r=1}^{k} [dx_r] P_s[x_r\mid 0]\right]\int \left[\prod_{r=1}^{k}p_J(\mathrm{d}J_r)\right]\nonumber\\
&\hspace{2.0cm}\times\delta_{(\mathrm{F})}\left[\dot x(t)+f(x(t))-h(t)-\sum_{r=1}^{k}J_r g\big(x(t),x_r(t)\big)-\xi(t)\right]\,.
\label{eq:P_from_Ps}
\end{align}
Equations \eqref{eq:Ps_closed}--\eqref{eq:P_from_Ps} are the recovered directed sparse-network path-probability equations. If indegrees and outdegrees are independent, $p_{k,\ell}=p_{\mathrm{in}}(k)p_{\mathrm{out}}(\ell)$, then
\beeq{
\sum_{\ell\ge 0}\frac{\ell\,p_{k,\ell}}{c}=p_{\mathrm{in}}(k)\,,
}
and the source-sampled and node-sampled barycenters coincide. In that case $P_s=P$ and the directed equation reduces to the single self-consistency
\begin{align}
P[x\mid h]&=\sum_{k\ge 0}p_{\mathrm{in}}(k)\ p_0(x(0)) \int [d\xi] P_\sigma[\xi]\int \left[\prod_{r=1}^{k}[dx_r] P[x_r\mid 0]\right]\int\left[\prod_{r=1}^{k}p_J(\mathrm{d}J_r)\right]\nonumber\\
&\hspace{1.0cm}\times\delta_{(\mathrm{F})}\left[\dot x(t)+f(x(t))-h(t)-\sum_{r=1}^{k}J_r g\big(x(t),x_r(t)\big)-\xi(t)\right]\,.
\label{eq:Metz_single_equation_rederived}
\end{align}
This is the single path-probability equation obtained for uncorrelated sparse directed ensembles. For correlated in/out degrees, the two averaged path laws $P_s$ and $P$ need not coincide.

Finally, consider the undirected specialization. In this case every edge is reciprocal, so $K_i=L_i=B_i=D_i$, and the degree law reduces to a single distribution $p_d$:
\beeq{
p_{k,\ell,b}=\sum_{d\ge 0}p_d\,\delta_{k,d}\delta_{\ell,d}\delta_{b,d}\,.
}
Then $c_{\rightarrow}=0$, $c_{\leftrightarrow}=c=\sum_d d\,p_d$, and the only edge-sampled law is the reciprocal one. Writing $P_e\equiv P^{\leftrightarrow}_s$, the edge-message equation becomes
\begin{align}
P_e[x\mid h]&=\sum_{d\ge 1}\frac{d\,p_d}{c}\,p_0(x(0))\int [d\xi] P_\sigma[\xi]\int\Bigg[\prod_{s=1}^{d-1}p_{\leftrightarrow}(\mathrm{d}J_s,\mathrm{d}\tilde J_s)\Bigg]\int \left[\prod_{s=1}^{d-1} [dy_s] P_e \left[y_s\Vert x;\tilde J_s\right]\right]\nonumber\\
&\hspace{2.0cm}\times\delta_{(\mathrm{F})}\left[\dot x(t)+f(x(t))-h(t)-\sum_{s=1}^{d-1}J_s g\big(x(t),y_s(t)\big)-\xi(t)\right]\,,
\label{eq:undirected_edge_barycenter}
\end{align}
while the node-uniform averaged path-probability is
\begin{align}
P[x\mid h]&=\sum_{d\ge 0}p_d\,p_0(x(0))\int [d\xi] P_\sigma[\xi]\int\Bigg[\prod_{s=1}^{d}p_{\leftrightarrow}(\mathrm{d}J_s,\mathrm{d}\tilde J_s)\Bigg]\int \left[\prod_{s=1}^{d}[dy_s] P_e\left[y_s\Vert x;\tilde J_s\right]\right]\nonumber\\
&\hspace{2.0cm}\times\delta_{(\mathrm{F})}\left[\dot x(t)+f(x(t))-h(t)-\sum_{s=1}^{d}J_s g\big(x(t),y_s(t)\big)-\xi(t)\right]\,.
\label{eq:undirected_node_barycenter}
\end{align}
Equations \eqref{eq:undirected_edge_barycenter}--\eqref{eq:undirected_node_barycenter} have the same first-moment logic as the generic equations, but not the same closure structure as the fully directed case. The reciprocal neighbour messages are evaluated as conditional kernels depending on the receiver history $x$. For symmetric undirected weights one sets $p_{\leftrightarrow}(\mathrm{d}J,\mathrm{d}\tilde J)=p_J(\mathrm{d}J)\delta_J(\mathrm{d}\tilde J)$; for bidirected graphs with independent directed weights one sets $p_{\leftrightarrow}(\mathrm{d}J,\mathrm{d}\tilde J)=p_J(\mathrm{d}J)p_J(\mathrm{d}\tilde J)$.

The derivation above focuses on the first moment of the ensemble-level message laws, because this is the object that yields the averaged single-site effective dynamics. The same probabilistic structure, however, extends beyond the barycenter. Higher moments of the message laws can be interpreted as path probabilities for several copies of the same cavity process evolving in a common quenched cavity environment, with independent dynamical noises conditional on that environment. This gives a natural hierarchy of $r$-copy path-measure equations: the case $r=1$ is the barycenter equation derived here, while $r\ge 2$ probes the fluctuations of the random path-probability messages themselves. We develop this $r$-copy formulation in App.~\ref{app:rcopy_moments}, where it is used to make precise the distinction between closed equations for averaged path probabilities and possible concentration properties of the underlying probability law on path probabilities.

\section{Representative reductions and linear--Gaussian reciprocal case}
\label{sec:known}
The fully directed non-reciprocal reduction is already contained in the path-measure derivation of Sec.~\ref{sec:derivations}. In that case the conditional kernels generated by outgoing neighbours integrate to one, incoming neighbours do not receive an imposed-history insertion from the root along the same edge, and the cavity recursion closes directly at the level of unshifted single-site path probabilities. We will not rederive that result here. Instead, this section uses the linear--Gaussian reciprocal case as a verification bridge between the general imposed-history conditional-kernel formalism and the more familiar Volterra equations for means, responses, and correlations. The point is not that reciprocal conditional kernels are intrinsically linear or Gaussian. Rather, the linear interaction is a source-shift specialization of the general theory: the imposed history enters a neighbouring branch as an additive source history independent of the branch trajectory. When this source-shift structure is combined with Gaussian initial laws, linear dynamics, and additive Gaussian noise, the path-kernel recursion closes exactly within Gaussian path measures. The Volterra equations are then obtained by decoding that Gaussian kernel/source recursion at the observable level.

To compare our results with the linear--Gaussian dynamic-cavity setting of Ref.~\cite{TaraboloDallAsta2025}, we specialize the dynamics in this section. We allow the single-site decay rate to depend on the node by replacing the general drift term $f(x_i)$ with $\lambda_i x_i$, and we take the interaction kernel to be linear:
\begin{equation}
f(x_i)\longmapsto \lambda_i x_i,\qquad g(x,x')=x'\,.
\label{eq:linear_choice}
\end{equation}
The microscopic equations of motion are then
\begin{equation}
\dot x_i(t)=-\lambda_i x_i(t)+\sum_{j=1}^N c_{ij}J_{ij}x_j(t)+h_i(t)+\xi_i(t)\,, \qquad \bracket{\xi_i(t)\xi_j(t')}=\delta_{ij}\Delta_i(t,t')\,,
\label{eq:OU_network}
\end{equation}
where $\Delta_i(t,t')$ denotes an additive Gaussian noise covariance. The white-noise convention of Sec.~\ref{sec:model} is recovered by taking $\Delta_i(t,t')=\sigma^2\delta(t-t')$; the additive thermal-noise setting of Ref.~\cite{TaraboloDallAsta2025} corresponds to the same white-in-time form, with the appropriate thermal-noise amplitude. For the exact Gaussian closure stated below, we assume Gaussian initial cavity laws, with deterministic initial conditions included as degenerate Gaussian initial laws. Under this assumption, the linear dynamics, additive Gaussian noise, and Gaussian conditional kernels imply that every cavity message is a Gaussian path measure.

For each $\ell\in\partial_i^{\mathrm{in}}\setminus j$, the coupling $J_{i\ell}$ is the coupling from $\ell$ into the root $i$. If $j$ is not an incoming neighbour of $i$, then $\partial_i^{\mathrm{in}}\setminus j=\partial_i^{\mathrm{in}}$. The coupling through which the root feeds back into $\ell$ is instead
\begin{equation}
\widetilde J_{\ell i}\equiv
\begin{cases}
J_{\ell i}, & \ell\in\partial_i^{\mathrm{out}},\\
0, & \ell\notin\partial_i^{\mathrm{out}}\,.
\end{cases}
\label{eq:feedback_coupling_def}
\end{equation}
Thus no symmetry of the two directed weights is assumed. In a genuinely bidirected graph, $J_{i\ell}$ and $\widetilde J_{\ell i}$ are two directed couplings, while in the symmetric undirected specialization one sets $\widetilde J_{\ell i}=J_{\ell i}=J_{i\ell}$.

The cavity equation \eqref{eq:cavity_equation_general_density}, applied to \eqref{eq:linear_choice}, gives the shifted linear cavity path equation
\begin{align}
P_i^{(j)}[x_i\mid h_i]&=p_0\big(x_i(0)\big)\int[d\xi_i]P_{\Delta_i}[\xi_i]\int\left[\prod_{\ell\in\partial_i^{\mathrm{in}}\setminus j}[dx_\ell] P_\ell^{(i)}[x_\ell\Vert x_i;h_\ell]\right]\nonumber\\
&\quad\times\delta_{(F)}\left[\dot x_i(t)+\lambda_i x_i(t)-h_i(t)-\sum_{\ell\in\partial_i^{\mathrm{in}}\setminus j}J_{i\ell}x_\ell(t)-\xi_i(t)\right]\,.
\label{eq:linear_cavity_shifted}
\end{align}
The functional delta enforces the linear equation of motion of the root trajectory after its remaining incoming neighbours have been sampled from their cavity branch laws. For the linear interaction, the conditional-kernel notation introduced in Sec.~\ref{sec:derivations} reduces to the source-shift identity
\begin{equation}
P_\ell^{(i)}[x_\ell\Vert x_i;h_\ell]=P_\ell^{(i)}[x_\ell\mid h_\ell+\widetilde J_{\ell i}x_i]\,,
\label{eq:linear_shifted_neighbour_kernel}
\end{equation}
where the right-hand side denotes the normalized cavity branch law of $\ell$ driven by the imposed trajectory $x_i$. This equality is a specialization of the general imposed-history kernel: it holds because, for $g(x,x')=x'$, the additive insertion is independent of the branch trajectory $x_\ell$ and can be recast as an ordinary additive source history. It is still a conditional path-probability kernel generated by the branch dynamics with $i$ removed and then driven by the prescribed history $x_i$, not the posterior conditional law of $x_\ell$ given the event $X_i=x_i$ in the full coupled process. In the remainder of this section, kernels such as $P_\ell^{(i)}[x_\ell\Vert x_i;h_\ell]$ are probability kernels over paths, whereas two-time objects such as $K_i^{(j)}$, $R_i^{(j)}$, $\Gamma_i^{(j)}$, and $\Omega_i^{(j)}$ are operator kernels acting on histories.

We use the following operator notation. If $A$ is a two-time kernel, then
\begin{equation}
[Au](t)\equiv \int_0^T ds A(t,s)u(s)\,, \qquad (AB)(t,t')\equiv \int_0^T ds A(t,s)B(s,t')\,.
\label{eq:linear_operator_conventions}
\end{equation}
The identity operator has kernel $\mathbb I(t,t')=\delta(t-t')$. The adjoint $A^\dagger$ is defined by the time-pairing
\begin{equation}
\langle u,Av\rangle=\langle A^\dagger u,v\rangle,
\qquad
\langle u,v\rangle\equiv \int_0^Tdt\,u(t)v(t),
\label{eq:linear_adjoint_convention}
\end{equation}
with the boundary convention fixed by the causal initial-value problem. A response kernel is causal when it vanishes for perturbation times later than observation times. Thus $R(t,t')=0$ for $t<t'$. When an operator such as $\Lambda$ has a causal inverse, $\Lambda^{-1}$ denotes the retarded Green operator $R$ satisfying $\Lambda R=\mathbb I$ together with this causal support.

Because the message is Gaussian, we parametrize it as the full path measure
\begin{equation}
P_i^{(j)}[x_i\mid h_i]=\frac{1}{Z_i^{(j)}[h_i]}\exp\left[-\frac{1}{2}\left\langle x_i,[K_i^{(j)}]^{-1}x_i\right\rangle+\left\langle \mathfrak j_i^{(j)}[h_i],x_i\right\rangle\right]\,.
\label{eq:Gaussian_message_param}
\end{equation}
Here $K_i^{(j)}$ is the centered covariance kernel of the message, $[K_i^{(j)}]^{-1}$ is its precision kernel, and $\mathfrak j_i^{(j)}[h_i]$ is the Gaussian source in the exponent. The dependence on the additive field history is contained in $\mathfrak j_i^{(j)}[h_i]$, while $K_i^{(j)}$ is independent of $h_i$ in the linear additive-source problem. Boundary contributions from a nondegenerate Gaussian initial law are absorbed into the same precision/source representation; deterministic initial data are treated as the corresponding degenerate boundary condition on the causal path space. All inverses are understood on this causal path space with the stated boundary convention.

The normalization of \eqref{eq:Gaussian_message_param} is obtained by completing the square:
\begin{equation}
Z_i^{(j)}[h_i]=Z_{i,0}^{(j)}\exp\left[\frac{1}{2}\left\langle\mathfrak j_i^{(j)}[h_i],K_i^{(j)}\mathfrak j_i^{(j)}[h_i]\right\rangle\right]\,,
\label{eq:Gaussian_message_normalisation}
\end{equation}
where $Z_{i,0}^{(j)}$ is independent of $h_i$. Differentiating $\log Z_i^{(j)}$ with respect to the Gaussian source gives the mean,
\begin{equation}
m_i^{(j)}(t)\equiv\int[dx_i] x_i(t)P_i^{(j)}[x_i\mid h_i]=\frac{\delta\log Z_i^{(j)}[h_i]}{\delta\mathfrak j_i^{(j)}[h_i](t)}=\int_0^Tdt' K_i^{(j)}(t,t')\mathfrak j_i^{(j)}[h_i](t')\,.
\label{eq:linear_mean_from_kernel}
\end{equation}
A second derivative gives the centered covariance,
\begin{align}
C_i^{(j)}(t,t')&\equiv\int[dx_i]\big(x_i(t)-m_i^{(j)}(t)\big)\big(x_i(t')-m_i^{(j)}(t')\big)P_i^{(j)}[x_i\mid h_i]\nonumber\\
&=\frac{\delta^2\log Z_i^{(j)}[h_i]}{\delta\mathfrak j_i^{(j)}[h_i](t)\delta\mathfrak j_i^{(j)}[h_i](t')}=K_i^{(j)}(t,t')\,.
\label{eq:linear_covariance_from_kernel}
\end{align}
Finally, define the source-response operator
\begin{equation}
\mathcal B_i^{(j)}(s,t')\equiv\frac{\delta\mathfrak j_i^{(j)}[h_i](s)}{\delta h_i(t')}\,.
\label{eq:linear_source_response_operator}
\end{equation}
Since $K_i^{(j)}$ is independent of the additive source $h_i$, differentiating \eqref{eq:linear_mean_from_kernel} gives the physical causal response
\begin{equation}
R_i^{(j)}(t,t')\equiv\frac{\delta m_i^{(j)}(t)}{\delta h_i(t')}=\int_0^T ds\,K_i^{(j)}(t,s)\mathcal B_i^{(j)}(s,t')\,.
\label{eq:linear_response_from_kernel}
\end{equation}
Thus $\mathcal B_i^{(j)}$ is the response of the Gaussian source in the exponent, whereas $R_i^{(j)}=K_i^{(j)}\mathcal B_i^{(j)}$ is the physical response of the mean trajectory.

For the linear interaction in \eqref{eq:linear_shifted_neighbour_kernel}, the effect of the imposed source history follows from the same Gaussian identities. For a fixed cavity branch $\ell\to i$, consider an arbitrary additive source increment $\theta_\ell$. In the linear--Gaussian message, the quadratic precision kernel is independent of the source and the Gaussian source is affine in the additive field. Therefore
\begin{equation}
\mathfrak j_\ell^{(i)}[h_\ell+\theta_\ell](u)=\mathfrak j_\ell^{(i)}[h_\ell](u)+\int_0^Tds \mathcal B_\ell^{(i)}(u,s)\theta_\ell(s)\,,\qquad\mathcal B_\ell^{(i)}(u,s)=\frac{\delta\mathfrak j_\ell^{(i)}[h_\ell](u)}{\delta h_\ell(s)}\,.
\label{eq:neighbour_source_affine_shift}
\end{equation}
Substituting \eqref{eq:neighbour_source_affine_shift} into the Gaussian mean formula gives
\begin{align}
m_\ell^{(i)}[h_\ell+\theta_\ell](t)&=\int_0^Tdu K_\ell^{(i)}(t,u)\mathfrak j_\ell^{(i)}[h_\ell+\theta_\ell](u)\nonumber\\
&=\int_0^Tdu K_\ell^{(i)}(t,u)\mathfrak j_\ell^{(i)}[h_\ell](u)+\int_0^Tdu\int_0^Tds K_\ell^{(i)}(t,u)\mathcal B_\ell^{(i)}(u,s)\theta_\ell(s)\nonumber\\
&=m_\ell^{(i)}[h_\ell](t)+\int_0^Tds R_\ell^{(i)}(t,s)\theta_\ell(s)\,,
\label{eq:neighbour_affine_mean_shift_general}
\end{align}
where
\begin{equation}
R_\ell^{(i)}(t,s)= \int_0^Tdu K_\ell^{(i)}(t,u)\mathcal B_\ell^{(i)}(u,s)\,.
\label{eq:neighbour_response_from_source_operator}
\end{equation}
This identity is exact in the linear--Gaussian setting; it is not a small-perturbation approximation. The functional expansion terminates after first order because the map from additive source history to mean trajectory is affine.

For the reciprocal cavity shift, the imposed root trajectory enters the branch $\ell\to i$ as the source increment
\begin{equation}
\theta_\ell(s)=\widetilde J_{\ell i}x_i(s)\,.
\label{eq:reciprocal_shift_as_source_increment}
\end{equation}
Equation \eqref{eq:neighbour_affine_mean_shift_general} therefore gives
\begin{equation}
m_\ell^{(i)}[h_\ell+\widetilde J_{\ell i}x_i](t)=m_\ell^{(i)}[h_\ell](t)+\int_0^Tds R_\ell^{(i)}(t,s)\widetilde J_{\ell i}x_i(s)\,.
\label{eq:shifted_neighbour_mean}
\end{equation}
This is the mean of the branch law driven by the imposed history $x_i$, not a posterior conditional expectation in the full coupled process. From this point on, except when a source shift is displayed explicitly, the dependence of cavity means on their local source histories is left implicit; thus $m_\ell^{(i)}(t)$ denotes $m_\ell^{(i)}[h_\ell](t)$.

The centered covariance is unchanged by the same source shift. By Eq.~\eqref{eq:linear_covariance_from_kernel}, applied to the branch message of $\ell$, and because $K_\ell^{(i)}$ does not depend on the additive source, one has
\begin{equation}
\operatorname{Cov}_{h_\ell+\widetilde J_{\ell i}x_i}\big(x_\ell(t),x_\ell(t')\big)=K_\ell^{(i)}(t,t')\,.
\label{eq:shifted_neighbour_covariance_unchanged}
\end{equation}
The shifted neighbour law is therefore the unshifted centered Gaussian fluctuation law translated by a deterministic mean shift.

It is useful to make this statement into notation before performing the Gaussian integrations. For fixed imposed root trajectory $x_i$, define the branch-kernel expectation
\begin{equation}
\bracket{F}_{\mathrm{br}\Vert x_i}^{(j)}\equiv\int\left[\prod_{\ell\in\partial_i^{\mathrm{in}}\setminus j}[dx_\ell] P_\ell^{(i)}[x_\ell\Vert x_i;h_\ell]\right]F\left(\{x_\ell\}_{\ell\in\partial_i^{\mathrm{in}}\setminus j}\right)\,.
\label{eq:branch_expectation_def}
\end{equation}
The subscript $\mathrm{br}\Vert x_i$ indicates an average over the product of shifted cavity branch kernels with $x_i$ held fixed as an imposed driving history. It does not denote a posterior conditional average in the full process. For each branch, define the centered fluctuation
\begin{equation}
\zeta_{\ell\to i}(t)\equiv x_\ell(t)-m_\ell^{(i)}(t)-\int_0^Tds R_\ell^{(i)}(t,s)\widetilde J_{\ell i}x_i(s)\,.
\label{eq:branch_fluctuation_def}
\end{equation}
Using \eqref{eq:shifted_neighbour_mean}, the branch mean of this residual is zero,
\begin{equation}
\bracket{\zeta_{\ell\to i}(t)}_{\mathrm{br}\Vert x_i}^{(j)}=0\,.
\label{eq:branch_fluctuation_zero_mean}
\end{equation}
Using \eqref{eq:shifted_neighbour_covariance_unchanged} and the product structure of the cavity branches at fixed imposed $x_i$, its covariance is
\begin{equation}
\bracket{\zeta_{\ell\to i}(t)\zeta_{\ell'\to i}(t')}_{\mathrm{br}\Vert x_i}^{(j)}=\delta_{\ell\ell'}K_\ell^{(i)}(t,t')\,.
\label{eq:direct_neighbour_fluctuation_covariance}
\end{equation}
Here $\delta_{\ell\ell'}$ is a Kronecker delta over neighbour labels. It appears because different shifted branches are independent under the product measure in \eqref{eq:branch_expectation_def}; the time dependence is carried by the covariance kernel $K_\ell^{(i)}(t,t')$.

We now insert the Gaussian parametrization into the shifted cavity equation. Use the Fourier representation
\begin{equation}
\delta_{(F)}[Y]\propto \int[d\hat x_i]\exp\left[-i\int_0^Tdt\,\hat x_i(t)Y(t)\right]\,.
\label{eq:functional_delta_fourier_linear}
\end{equation}
For a fixed neighbour $\ell$, the contribution of the branch law to the Fourier-transformed root constraint is
\begin{equation}
I_{\ell\to i}[x_i,\hat x_i]\equiv\int[dx_\ell]P_\ell^{(i)}[x_\ell\Vert x_i;h_\ell]\exp\left[
iJ_{i\ell}\int_0^Tdt\,\hat x_i(t)x_\ell(t)\right]\,.
\label{eq:one_neighbour_integral_def}
\end{equation}
Substituting the decomposition implied by \eqref{eq:branch_fluctuation_def},
\begin{equation}
x_\ell(t)=m_\ell^{(i)}(t)+\int_0^Tds R_\ell^{(i)}(t,s)\widetilde J_{\ell i}x_i(s)+\zeta_{\ell\to i}(t)\,,
\label{eq:shifted_neighbour_decomposition}
\end{equation}
gives
\begin{align}
I_{\ell\to i}[x_i,\hat x_i] &=\exp\left[iJ_{i\ell}\int_0^Tdt\,\hat x_i(t)m_\ell^{(i)}(t)\right]\nonumber\\
&\quad\times\exp\left[iJ_{i\ell}\widetilde J_{\ell i}\int_0^Tdt\int_0^Tds\hat x_i(t)R_\ell^{(i)}(t,s)x_i(s)\right]\nonumber\\
&\quad\times\bracket{\exp\left[iJ_{i\ell}\int_0^Tdt\,\hat x_i(t)\zeta_{\ell\to i}(t)\right]}_{\ell\Vert x_i}\,,
\label{eq:one_neighbour_integral_decomposed}
\end{align}
where, after the linear change of variables in Eq.~\eqref{eq:shifted_neighbour_decomposition}, the last average is over the centered Gaussian fluctuation $\zeta_{\ell\to i}$ under the shifted branch law. Since $\zeta_{\ell\to i}$ has zero mean and covariance $K_\ell^{(i)}$, its characteristic functional is
\begin{equation}
\bracket{\exp\left[iJ_{i\ell}\int_0^Tdt\,\hat x_i(t)\zeta_{\ell\to i}(t)\right]}_{\ell\Vert x_i}=\exp\left[-\frac12J_{i\ell}^2\int_0^Tdt\int_0^Tdt'\hat x_i(t)K_\ell^{(i)}(t,t')\hat x_i(t')\right]\,.
\label{eq:branch_fluctuation_characteristic}
\end{equation}
Combining \eqref{eq:one_neighbour_integral_decomposed} and \eqref{eq:branch_fluctuation_characteristic} gives
\begin{align}
I_{\ell\to i}[x_i,\hat x_i]&=\exp\left[iJ_{i\ell}\int_0^Tdt\,\hat x_i(t)m_\ell^{(i)}(t)+iJ_{i\ell}\widetilde J_{\ell i}\int_0^Tdt\int_0^Tds\hat x_i(t)R_\ell^{(i)}(t,s)x_i(s)\right]\nonumber\\
&\quad\times\exp\left[-\frac{1}{2}J_{i\ell}^2\int_0^Tdt\int_0^Tdt'\hat x_i(t)K_\ell^{(i)}(t,t')\hat x_i(t')\right]\,.
\label{eq:normalised_neighbour_gaussian_integral}
\end{align}
The three factors have different meanings. The first is the direct mean input from the neighbour. The second is the retarded response loop $i\to\ell\to i$ generated by reciprocal feedback. The last is the covariance injected into the root by the centered fluctuations of the neighbour. The two coupling products are therefore distinct: the memory term contains $J_{i\ell}\widetilde J_{\ell i}$, whereas the covariance-injection term contains two factors of the incoming coupling $J_{i\ell}$.

Fourier-transforming the root constraint in Eq.~\eqref{eq:linear_cavity_shifted}, substituting Eq.~\eqref{eq:normalised_neighbour_gaussian_integral} for each incoming neighbour, and integrating over the local Gaussian noise gives the effective Gaussian MSRJD form
\begin{align}
P_i^{(j)}[x_i,\hat x_i\mid h_i]&\propto p_0\big(x_i(0)\big)\exp\left[-\frac{1}{2}\int_0^Tdt\int_0^Tdt' \hat x_i(t)\Omega_i^{(j)}(t,t')\hat x_i(t')\right]\nonumber\\
&\quad\times\exp\left[-i\int_0^Tdt\,\hat x_i(t)\left([\Lambda_i^{(j)}x_i](t)-h_i(t)-M_i^{(j)}(t)\right)\right]\,.
\label{eq:effective_linear_msrjd_message}
\end{align}
The induced objects appearing here collect, respectively, deterministic neighbour means, reciprocal memory, effective innovation covariance, and the residual causal operator. The deterministic mean field from incoming neighbour means is
\begin{equation}
M_i^{(j)}(t)\equiv\sum_{\ell\in\partial_i^{\mathrm{in}}\setminus j}J_{i\ell}m_\ell^{(i)}(t)\,.
\label{eq:linear_mean_field_kernel}
\end{equation}
The causal memory kernel generated by reciprocal response loops is
\begin{equation}
\Gamma_i^{(j)}(t,s)\equiv\sum_{\ell\in\partial_i^{\mathrm{in}}\setminus j}J_{i\ell}\widetilde J_{\ell i}R_\ell^{(i)}(t,s)=\sum_{\ell\in\partial_i^{\mathrm{in}}\setminus j}J_{i\ell}\widetilde J_{\ell i}\int_0^Tdu K_\ell^{(i)}(t,u)\mathcal B_\ell^{(i)}(u,s)\,.
\label{eq:linear_memory_kernel}
\end{equation}
The effective innovation covariance is
\begin{equation}
\Omega_i^{(j)}(t,t')\equiv\Delta_i(t,t')+\sum_{\ell\in\partial_i^{\mathrm{in}}\setminus j}J_{i\ell}^2K_\ell^{(i)}(t,t')\,.
\label{eq:linear_noise_kernel}
\end{equation}
Finally, with $[D_i x](t)=\dot x(t)+\lambda_i x(t)$, the causal residual operator is
\begin{equation}
[\Lambda_i^{(j)}x](t)\equiv[D_i x](t)-\int_0^Tds\,\Gamma_i^{(j)}(t,s)x(s)\,.
\label{eq:linear_Lambda_def}
\end{equation}
Thus $M_i^{(j)}$ is a deterministic neighbour mean input, $\Gamma_i^{(j)}$ is a generally non-symmetric retarded memory kernel, $\Omega_i^{(j)}$ is a symmetric covariance kernel for the effective innovation, and $\Lambda_i^{(j)}$ maps a root trajectory to its effective dynamical residual. Causality of the neighbour responses implies $\Gamma_i^{(j)}(t,s)=0$ for $s>t$, so the convolution in \eqref{eq:linear_Lambda_def} is Volterra even though it is written over the full interval.

Integrating out $\hat x_i$ in \eqref{eq:effective_linear_msrjd_message} gives the $x_i$-only Gaussian path measure
\begin{equation}
P_i^{(j)}[x_i\mid h_i]\propto p_0\big(x_i(0)\big)\exp\left[-\frac{1}{2}\left\langle
\Lambda_i^{(j)}x_i-h_i-M_i^{(j)},[\Omega_i^{(j)}]^{-1}(\Lambda_i^{(j)}x_i-h_i-M_i^{(j)})
\right\rangle\right]\,.
\label{eq:linear_effective_x_only_gaussian}
\end{equation}
Expanding the quadratic form in \eqref{eq:linear_effective_x_only_gaussian} gives
\begin{align}
&-\frac{1}{2}\left\langle\Lambda_i^{(j)}x_i-h_i-M_i^{(j)},[\Omega_i^{(j)}]^{-1}(\Lambda_i^{(j)}x_i-h_i-M_i^{(j)})\right\rangle\nonumber\\
&\quad =-\frac{1}{2}\left\langle x_i,[\Lambda_i^{(j)}]^\dagger[\Omega_i^{(j)}]^{-1}\Lambda_i^{(j)}x_i\right\rangle+\left\langle
[\Lambda_i^{(j)}]^\dagger[\Omega_i^{(j)}]^{-1}\big(h_i+M_i^{(j)}\big),x_i\right\rangle+\text{constant}\,.
\label{eq:linear_effective_quadratic_expansion}
\end{align}
Comparing \eqref{eq:linear_effective_quadratic_expansion} with the Gaussian parametrization \eqref{eq:Gaussian_message_param} yields the closed kernel/source equations
\begin{align}
[K_i^{(j)}]^{-1}&= [\Lambda_i^{(j)}]^\dagger[\Omega_i^{(j)}]^{-1}\Lambda_i^{(j)}\,,\label{eq:Gaussian_kernel_equation_K}\\
\mathfrak j_i^{(j)}[h_i]&=[\Lambda_i^{(j)}]^\dagger[\Omega_i^{(j)}]^{-1}\big(h_i+M_i^{(j)}\big)\,,\label{eq:Gaussian_kernel_equation_j}\\
\mathcal B_i^{(j)}&=[\Lambda_i^{(j)}]^\dagger[\Omega_i^{(j)}]^{-1}\,.
\label{eq:Gaussian_kernel_equation_B}
\end{align}
The first equation says that the precision of the root path measure is obtained by pulling back the effective innovation precision through the residual operator $\Lambda_i^{(j)}$. The second says that the Gaussian source is the corresponding pullback of the deterministic drive $h_i+M_i^{(j)}$. The third follows because $h_i$ enters the residual additively and $M_i^{(j)}$, $\Gamma_i^{(j)}$, and $\Omega_i^{(j)}$ are independent of the local source $h_i$. Together with \eqref{eq:linear_mean_from_kernel}, \eqref{eq:linear_response_from_kernel}, and \eqref{eq:linear_mean_field_kernel}--\eqref{eq:linear_Lambda_def}, these equations form a closed recursion for Gaussian path kernels and sources. The full single-node marginal is obtained by replacing every occurrence of $\partial_i^{\mathrm{in}}\setminus j$ in the definitions above by $\partial_i^{\mathrm{in}}$.

The central object produced by the reciprocal linear--Gaussian cavity formalism is therefore the Gaussian path law itself, encoded by the operator triple $(K_i^{(j)},\mathfrak j_i^{(j)},\mathcal B_i^{(j)})$ and by the induced kernels $(M_i^{(j)},\Gamma_i^{(j)},\Omega_i^{(j)})$. The Volterra equations for means, responses, and correlations are not the starting point of the cavity calculation; they are observable-level consequences of the Gaussian kernel/source recursion. On a time grid, \eqref{eq:linear_mean_field_kernel}--\eqref{eq:Gaussian_kernel_equation_B} become finite-dimensional matrix recursions for Gaussian path measures. This is a structural observation about the representation and does not assert numerical performance.

We now decode the Volterra equations from the kernel/source equations. The mean is obtained from $[K_i^{(j)}]^{-1}m_i^{(j)}=\mathfrak j_i^{(j)}$. The response is obtained from $R_i^{(j)}=K_i^{(j)}\mathcal B_i^{(j)}$, equivalently $[K_i^{(j)}]^{-1}R_i^{(j)}=\mathcal B_i^{(j)}$. The centered covariance is $C_i^{(j)}=K_i^{(j)}$, and the precision equation implies the factorization
\begin{equation}
K_i^{(j)}=[\Lambda_i^{(j)}]^{-1}\Omega_i^{(j)}[\Lambda_i^{(j)\dagger}]^{-1}\,.
\label{eq:kernel_decoding_factorisation}
\end{equation}
Each of these identities follows from the Gaussian representation already derived; no moment equation is imposed independently.

For the mean, start with
\begin{equation}
[K_i^{(j)}]^{-1}m_i^{(j)}=\mathfrak j_i^{(j)}\,.
\label{eq:mean_decoding_start}
\end{equation}
Substituting \eqref{eq:Gaussian_kernel_equation_K} and \eqref{eq:Gaussian_kernel_equation_j} gives
\begin{equation}
[\Lambda_i^{(j)}]^\dagger[\Omega_i^{(j)}]^{-1}\Lambda_i^{(j)}m_i^{(j)}=[\Lambda_i^{(j)}]^\dagger[\Omega_i^{(j)}]^{-1}\big(h_i+M_i^{(j)}\big)\,,
\label{eq:mean_decoding_substituted}
\end{equation}
or
\begin{equation}
[\Lambda_i^{(j)}]^\dagger[\Omega_i^{(j)}]^{-1}\left(\Lambda_i^{(j)}m_i^{(j)}-h_i-M_i^{(j)}
\right)=0\,.
\label{eq:kernel_to_mean_operator}
\end{equation}
On the causal path space on which the Gaussian innovation covariance is nondegenerate, the operator multiplying the parentheses has no null direction relevant to the path measure. Therefore
\begin{equation}
\Lambda_i^{(j)}m_i^{(j)}=h_i+M_i^{(j)}\,.
\label{eq:mean_operator_volterra}
\end{equation}
Expanding $\Lambda_i^{(j)}=D_i-\Gamma_i^{(j)}$ gives
\begin{equation}
\partial_t m_i^{(j)}(t)+\lambda_i m_i^{(j)}(t)-\int_0^Tds\,\Gamma_i^{(j)}(t,s)m_i^{(j)}(s)=h_i(t)+M_i^{(j)}(t)\,.
\label{eq:mean_operator_expanded}
\end{equation}
Using the causal support of $\Gamma_i^{(j)}$ and substituting \eqref{eq:linear_mean_field_kernel} and \eqref{eq:linear_memory_kernel}, one obtains
\begin{align}
\partial_t m_i^{(j)}(t)+\lambda_i m_i^{(j)}(t)&=h_i(t)+\sum_{\ell\in\partial_i^{\mathrm{in}}\setminus j}J_{i\ell}m_\ell^{(i)}(t)+\int_0^t ds \Gamma_i^{(j)}(t,s)m_i^{(j)}(s)\nonumber\\
&=h_i(t)+\sum_{\ell\in\partial_i^{\mathrm{in}}\setminus j}J_{i\ell}m_\ell^{(i)}(t)+\sum_{\ell\in\partial_i^{\mathrm{in}}\setminus j}\int_0^t ds J_{i\ell}\widetilde J_{\ell i}R_\ell^{(i)}(t,s)m_i^{(j)}(s)\,.
\label{eq:mean_Volterra}
\end{align}
This is the causal mean equation. The retarded self-interaction is present only through reciprocal feedback. If an incoming edge $\ell\to i$ is not accompanied by the feedback edge $i\to\ell$, then $\widetilde J_{\ell i}=0$ and that neighbour contributes no memory term.

For the response, begin with the Gaussian identity
\begin{equation}
R_i^{(j)}=K_i^{(j)}\mathcal B_i^{(j)}\,.
\label{eq:response_decoding_start}
\end{equation}
Applying $[K_i^{(j)}]^{-1}$ gives
\begin{equation}
[K_i^{(j)}]^{-1}R_i^{(j)}=\mathcal B_i^{(j)}\,.
\label{eq:response_decoding_Kinv}
\end{equation}
Substituting \eqref{eq:Gaussian_kernel_equation_K} and \eqref{eq:Gaussian_kernel_equation_B} gives
\begin{equation}
[\Lambda_i^{(j)}]^\dagger[\Omega_i^{(j)}]^{-1}\Lambda_i^{(j)}R_i^{(j)}=[\Lambda_i^{(j)}]^\dagger[\Omega_i^{(j)}]^{-1}\,.
\label{eq:response_decoding_substituted}
\end{equation}
Using the same nondegeneracy on the causal path space gives
\begin{equation}
\Lambda_i^{(j)}R_i^{(j)}=\mathbb I\,.
\label{eq:response_operator_volterra}
\end{equation}
In time variables, this means
\begin{equation}
\partial_t R_i^{(j)}(t,t')+\lambda_i R_i^{(j)}(t,t')-\int_0^Tds \Gamma_i^{(j)}(t,s)R_i^{(j)}(s,t')=\delta(t-t')\,.
\label{eq:response_operator_expanded}
\end{equation}
The factor $\delta(t-t')$ is the kernel of the identity operator. Because $\Gamma_i^{(j)}(t,s)=0$ for $s>t$ and $R_i^{(j)}(s,t')=0$ for $s<t'$, the convolution is restricted to $t'\le s\le t$. Hence
\begin{equation}
\partial_t R_i^{(j)}(t,t')+\lambda_i R_i^{(j)}(t,t')=\delta(t-t')+\int_{t'}^t ds \Gamma_i^{(j)}(t,s)R_i^{(j)}(s,t')\,.
\label{eq:response_Volterra}
\end{equation}
Substituting \eqref{eq:linear_memory_kernel} gives the equivalent expanded response equation
\begin{equation}
\partial_t R_i^{(j)}(t,t')+\lambda_i R_i^{(j)}(t,t')=\delta(t-t')+\sum_{\ell\in\partial_i^{\mathrm{in}}\setminus j}\int_{t'}^t ds J_{i\ell}\widetilde J_{\ell i}R_\ell^{(i)}(t,s)R_i^{(j)}(s,t')\,.
\label{eq:response_Volterra_expanded}
\end{equation}

For the covariance, use $C_i^{(j)}=K_i^{(j)}$ and \eqref{eq:kernel_decoding_factorisation}. Since $R_i^{(j)}$ is the causal inverse of $\Lambda_i^{(j)}$, \eqref{eq:response_operator_volterra} implies
\begin{equation}
C_i^{(j)} =K_i^{(j)}=R_i^{(j)}\Omega_i^{(j)}R_i^{(j)\dagger}\,.
\label{eq:covariance_kernel_factorisation}
\end{equation}
Writing this factorization in time variables gives the double convolution
\begin{equation}
C_i^{(j)}(t,t')=\int_0^Tdu\int_0^Tdv R_i^{(j)}(t,u)\Omega_i^{(j)}(u,v)R_i^{(j)}(t',v)\,.
\label{eq:covariance_double_convolution}
\end{equation}
The causal support of the two response kernels restricts the effective integration domain to $u\le t$ and $v\le t'$, but the full interval notation keeps the operator product transparent.

Act with $\Lambda_i^{(j)}$ on the first time argument of \eqref{eq:covariance_double_convolution}. Using
\begin{equation}
\int_0^Tdu \Lambda_i^{(j)}(t,u)R_i^{(j)}(u,s)=\delta(t-s)\,,
\label{eq:Lambda_R_collapse_for_covariance}
\end{equation}
one obtains
\begin{equation}
[\Lambda_i^{(j)}C_i^{(j)}](t,t')=\int_0^Tdv \Omega_i^{(j)}(t,v)R_i^{(j)}(t',v)\,.
\label{eq:covariance_after_Lambda}
\end{equation}
The causal support of $R_i^{(j)}(t',v)$ reduces the right-hand side to $0\le v\le t'$. Expanding $\Lambda_i^{(j)}=D_i-\Gamma_i^{(j)}$ on the left gives
\begin{equation}
\partial_t C_i^{(j)}(t,t')+\lambda_i C_i^{(j)}(t,t')-\int_0^t ds \Gamma_i^{(j)}(t,s)C_i^{(j)}(s,t')=\int_0^{t'}ds \Omega_i^{(j)}(t,s)R_i^{(j)}(t',s)\,.
\label{eq:covariance_before_noise_expansion}
\end{equation}
Moving the memory term to the right and substituting \eqref{eq:linear_noise_kernel} gives
\begin{align}
\partial_t C_i^{(j)}(t,t')+\lambda_i C_i^{(j)}(t,t')&=\int_0^t ds \Gamma_i^{(j)}(t,s)C_i^{(j)}(s,t')+\int_0^{t'}ds R_i^{(j)}(t',s)\Delta_i(t,s)\nonumber\\
&\quad+\sum_{\ell\in\partial_i^{\mathrm{in}}\setminus j}J_{i\ell}^2\int_0^{t'}ds R_i^{(j)}(t',s)K_\ell^{(i)}(t,s)\,.
\label{eq:covariance_noise_expanded}
\end{align}
Using $K_\ell^{(i)}=C_\ell^{(i)}$ from \eqref{eq:linear_covariance_from_kernel} and expanding the memory kernel gives the Volterra covariance equation
\begin{align}
\partial_t C_i^{(j)}(t,t')+\lambda_i C_i^{(j)}(t,t')&=\sum_{\ell\in\partial_i^{\mathrm{in}}\setminus j}\int_0^t ds J_{i\ell}\widetilde J_{\ell i}R_\ell^{(i)}(t,s)C_i^{(j)}(s,t')\nonumber\\
&\quad+\int_0^{t'}ds\,R_i^{(j)}(t',s)\Delta_i(t,s)+\sum_{\ell\in\partial_i^{\mathrm{in}}\setminus j}J_{i\ell}^2\int_0^{t'}ds R_i^{(j)}(t',s)C_\ell^{(i)}(t,s)\,.
\label{eq:correlation_Volterra}
\end{align}
The first term on the right is the memory action on the root covariance and contains the reciprocal product $J_{i\ell}\widetilde J_{\ell i}$. The final term is the covariance injected by neighbour fluctuations and contains two factors of the incoming coupling $J_{i\ell}$. These two structures coincide only after specializing to symmetric undirected weights, and even then they multiply different dynamical objects: a response kernel in the memory term and a covariance kernel in the fluctuation-injection term. For the white-noise convention $\Delta_i(t,s)=\sigma^2\delta(t-s)$, the direct bath-noise contribution becomes
\begin{equation}
\int_0^{t'}ds\,R_i^{(j)}(t',s)\Delta_i(t,s)=\sigma^2R_i^{(j)}(t',t)\,,
\label{eq:white_noise_covariance_injection}
\end{equation}
where causality of $R_i^{(j)}(t',t)$ supplies the support restriction; equivalently, this term vanishes unless $t\le t'$.

Equations \eqref{eq:mean_Volterra}, \eqref{eq:response_Volterra_expanded}, and \eqref{eq:correlation_Volterra} are the causal Volterra equations decoded from the Gaussian kernel/source recursion. They recover the linear--Gaussian dynamic-cavity equations of Ref.~\cite{TaraboloDallAsta2025}. The first equation propagates the mean, the second identifies the physical response as the causal inverse of the residual operator, and the third propagates the centered covariance by separating memory acting on the root from noise and neighbour-fluctuation injection.

We close with a direct derivation from the shifted linear cavity equation, used only as an internal consistency check. The branch decomposition \eqref{eq:shifted_neighbour_decomposition} gives, for each incoming neighbour,
\begin{equation}
x_\ell(t)=m_\ell^{(i)}(t)+\int_0^Tds R_\ell^{(i)}(t,s)\widetilde J_{\ell i}x_i(s)+\zeta_{\ell\to i}(t)\,,
\label{eq:direct_neighbour_decomposition}
\end{equation}
with the branch-fluctuation covariance already defined in \eqref{eq:direct_neighbour_fluctuation_covariance}. Substituting \eqref{eq:direct_neighbour_decomposition} into the root constraint in \eqref{eq:linear_cavity_shifted} gives
\begin{equation}
\dot x_i(t)+\lambda_i x_i(t)=h_i(t)+M_i^{(j)}(t)+\int_0^t ds \Gamma_i^{(j)}(t,s)x_i(s)+\eta_i^{(j)}(t)\,,
\label{eq:direct_effective_linear_equation}
\end{equation}
where the effective centered noise is
\begin{equation}
\eta_i^{(j)}(t)\equiv\xi_i(t)+\sum_{\ell\in\partial_i^{\mathrm{in}}\setminus j}J_{i\ell}\zeta_{\ell\to i}(t)\,.
\label{eq:direct_effective_noise_def}
\end{equation}
Its covariance follows from the independence of the local noise and the shifted cavity branches, together with \eqref{eq:direct_neighbour_fluctuation_covariance}:
\begin{align}
\bracket{\eta_i^{(j)}(t)\eta_i^{(j)}(t')}_{\mathrm{eff}\Vert x_i}&=\bracket{\xi_i(t)\xi_i(t')}+\sum_{\ell,\ell'\in\partial_i^{\mathrm{in}}\setminus j}J_{i\ell}J_{i\ell'}
\bracket{\zeta_{\ell\to i}(t)\zeta_{\ell'\to i}(t')}_{\mathrm{br}\Vert x_i}^{(j)}\nonumber\\
&=\Delta_i(t,t')+\sum_{\ell\in\partial_i^{\mathrm{in}}\setminus j}J_{i\ell}^2K_\ell^{(i)}(t,t')=\Omega_i^{(j)}(t,t')\,.
\label{eq:direct_effective_noise}
\end{align}
Taking the mean of \eqref{eq:direct_effective_linear_equation} gives \eqref{eq:mean_Volterra}. Differentiating \eqref{eq:direct_effective_linear_equation} with respect to the local source $h_i(t')$, using the independence of $M_i^{(j)}$, $\Gamma_i^{(j)}$, and the centered effective noise law from the local source $h_i$, gives $\Lambda_i^{(j)}R_i^{(j)}=\mathbb I$ and hence \eqref{eq:response_Volterra}. Finally, the centered form of \eqref{eq:direct_effective_linear_equation} is
\begin{equation}
[\Lambda_i^{(j)}(x_i-m_i^{(j)})](t)=\eta_i^{(j)}(t)\,.
\label{eq:direct_centered_effective_equation}
\end{equation}
Solving this equation with the causal inverse $R_i^{(j)}$ gives $x_i-m_i^{(j)}=R_i^{(j)}\eta_i^{(j)}$, and using \eqref{eq:direct_effective_noise} gives $C_i^{(j)}=R_i^{(j)}\Omega_i^{(j)}R_i^{(j)\dagger}$. Expanding this factorization as in \eqref{eq:covariance_double_convolution}--\eqref{eq:correlation_Volterra} recovers the same covariance equation. Thus the direct cavity calculation and the kernel/source calculation agree, while the kernel/source recursion remains the primary closed object produced by the Gaussian cavity formalism.

The limiting cases separate the roles of the different kernels. In the fully non-reciprocal directed case, $\widetilde J_{\ell i}=0$ for every incoming neighbour, so $\Gamma_i^{(j)}=0$ and the shifted kernels reduce to unshifted incoming branch laws. The directed closure of Sec.~\ref{sec:derivations} is then recovered. In the symmetric undirected specialization, $\widetilde J_{\ell i}=J_{\ell i}=J_{i\ell}$, so the memory contribution becomes proportional to $J_{i\ell}^2R_\ell^{(i)}$, while the neighbour-fluctuation injection remains proportional to $J_{i\ell}^2C_\ell^{(i)}$. In a bidirected graph with non-symmetric weights these two coupling products must remain distinct. For white noise, the bath term in the covariance equation reduces to \eqref{eq:white_noise_covariance_injection}. For zero external field, the explicit drive $h_i$ is removed from the mean equation and from the Gaussian source, but the memory kernel, innovation covariance, and response structure are unchanged.

\section{Causal discretization and population dynamics}
\label{sec:causal_discretization_population_dynamics}
The path-measure equations derived above are continuous-time objects. The role of the time discretization is not to replace them, but to give a finite-time causal representation in which the order of updates is explicit and in which the path laws can be implemented numerically. This is particularly useful on sparse graphs because a cavity message is a probability law over an entire history, rather than a law over an instantaneous variable. The discretized construction below is therefore a causal and algorithmic rederivation of the continuous-time equations of Sec.~\ref{sec:derivations}. The continuous-time theory is recovered only after the finite-time causal construction has been specified.

Let $t_n=n\Delta$, $n=0,\ldots,M$, with $M\Delta=T$. We write
\beeq{
x_i^n\equiv x_i(t_n)\,,\qquad h_i^n\equiv h_i(t_n)\,, \qquad x_i^{0:M}\equiv (x_i^0,\ldots,x_i^M), \qquad h_i^{0:M-1}\equiv (h_i^0,\ldots,h_i^{M-1})\,.
}
With the It\^{o} convention, the general pairwise dynamics in Eq.~\eqref{eq:microscopic_dynamics} is represented by
\begin{equation}
x_i^{n+1}=x_i^n+\Delta\left[-f(x_i^n)+\sum_{j=1}^{N}c_{ij}J_{ij}g(x_i^n,x_j^n)+h_i^n\right]+\zeta_i^n \,,\qquad n=0,\ldots,M-1 \,,
\label{eq:causal_discrete_update}
\end{equation}
where the noise increments satisfy
\begin{equation}
\bracket{\zeta_i^n\zeta_j^m}=\sigma^2\Delta\,\delta_{ij}\delta_{nm}\,.
\label{eq:causal_discrete_noise}
\end{equation}
For $\sigma>0$, the one-step transition density of node $i$, conditioned on the states of its incoming neighbours at time $t_n$, is
\begin{equation}
\mathcal T_i^\Delta\left(x_i^{n+1}\mid x_i^n,\{x_j^n\}_{j\in\partial_i^{\mathrm{in}}};h_i^n\right)=\frac{1}{\sqrt{2\pi\sigma^2\Delta}}\exp\left[-\frac{\left(x_i^{n+1}-x_i^n-\Delta b_i^n\right)^2
}{2\sigma^2\Delta}\right]\,,
\label{eq:causal_transition_kernel}
\end{equation}
with
\begin{equation}
b_i^n\equiv-f(x_i^n)+\sum_{j\in\partial_i^{\mathrm{in}}}J_{ij}g(x_i^n,x_j^n)+h_i^n\,.
\label{eq:causal_discrete_drift}
\end{equation}
In the deterministic limit $\sigma\to0$, the Gaussian transition density in Eq.~\eqref{eq:causal_transition_kernel} is replaced by the corresponding Dirac delta enforcing Eq.~\eqref{eq:causal_discrete_update}. The finite-time path density of the full system is then
\begin{equation}
P_N^\Delta[\mathbf x^{0:M}\mid \mathbf h^{0:M-1}]=p_0(\mathbf x^0)\prod_{n=0}^{M-1}\prod_{i=1}^{N}\mathcal T_i^\Delta\left(x_i^{n+1}\mid x_i^n,\{x_j^n\}_{j\in\partial_i^{\mathrm{in}}};h_i^n\right)\,,
\label{eq:causal_global_path_density}
\end{equation}
where the product form displays the causal ordering explicitly: all variables at time $t_{n+1}$ are generated from the state at time $t_n$. For fixed $\Delta$ and $M$, Eq.~\eqref{eq:causal_global_path_density} is the exact path density of the discretized microscopic model. Approximation enters only when the discretized dynamics is compared with the continuous-time process, or later when finite-memory numerical closures replace complete histories by finite windows. This representation uses transition kernels rather than auxiliary response fields, since the auxiliary-field derivation has already supplied the continuous-time path-measure cavity equations.

The discrete analogue of the conditional path law in Sec.~\ref{sec:derivations} is obtained by holding the history of the removed node fixed and propagating the corresponding cavity branch with that history inserted into the local drift. For a neighbour $\ell\in\partial_i$, define
\begin{equation}
\Theta_{\ell\leftarrow i}^n[x_\ell,x_i]\equiv
\begin{cases}
J_{\ell i}g(x_\ell^n,x_i^n), & \ell\in\partial_i^{\mathrm{out}}\,,\\[1mm]
0, & \ell\notin\partial_i^{\mathrm{out}}\,.
\end{cases}
\label{eq:causal_discrete_feedback_drift}
\end{equation}
For any finite horizon \(R\ge 1\), we denote the corresponding imposed-history branch path law by
\begin{equation}
[dx_\ell^{0:R}]P_\ell^{(i),\Delta}\left[x_\ell^{0:R}\Vert x_i^{0:R-1};h_\ell^{0:R-1}\right]\,.
\label{eq:causal_discrete_conditional_kernel}
\end{equation}
It is the normalized path law of the cavity branch of $\ell$ in the graph where $i$ is removed, with local updates
\begin{equation}
x_\ell^{n+1}=x_\ell^n+\Delta\left[-f(x_\ell^n)+\sum_{m\in\partial_\ell^{\mathrm{in}}\setminus i}J_{\ell m}g(x_\ell^n,x_m^n)+h_\ell^n+\Theta_{\ell\leftarrow i}^n[x_\ell,x_i]\right]+\zeta_\ell^n\,, \quad n=0,\ldots,R-1.
\label{eq:causal_branch_update}
\end{equation}
As in the continuous-time notation $P_\ell^{(i)}[x_\ell\Vert x_i;h_\ell]$, the symbol $\Vert$ marks an imposed-history path law, not a posterior conditioning on the full coupled process. For general pairwise $g(x_\ell,x_i)$, the inserted term in Eq.~\eqref{eq:causal_branch_update} is additive in the local update but may depend on the branch trajectory $x_\ell$. The notation keeps this dependence inside the branch law.

Using the imposed-history path laws in Eq.~\eqref{eq:causal_discrete_conditional_kernel}, the finite-time version of the fixed-graph cavity equation is
\begin{align}
P_i^{(j),\Delta}\left[x_i^{0:M}\mid h_i^{0:M-1}\right]&\simeq p_0(x_i^0)\int\left[\prod_{\ell\in\partial_i\setminus j}[dx_\ell^{0:M-1}]
P_\ell^{(i),\Delta}\left[x_\ell^{0:M-1}\Vert x_i^{0:M-2};h_\ell^{0:M-2}\right]\right]\nonumber\\
&\quad\times\prod_{n=0}^{M-1}\mathcal T_{i\setminus j}^{\Delta}\left(x_i^{n+1}\mid x_i^n,\{x_\ell^n\}_{\ell\in\partial_i^{\mathrm{in}}\setminus j};h_i^n\right)\,.
\label{eq:causal_discrete_cavity_equation}
\end{align}
where $\mathcal T_{i\setminus j}^{\Delta}$ is the transition density in Eq.~\eqref{eq:causal_transition_kernel} with the incoming sum restricted to $\partial_i^{\mathrm{in}}\setminus j$. Equation~\eqref{eq:causal_discrete_cavity_equation} is the finite-time causal representation of Eq.~\eqref{eq:cavity_equation_general_density}: the root path is generated from its incoming branch histories, ... while reciprocal feedback is carried by the imposed-history path laws attached to neighbours that also receive input from the root.

For the interactions of the form
\begin{equation}
g(u,v)=g_{\rm add}(v)\,,
\label{eq:additive_input_kernel_form}
\end{equation}
the feedback insertion in Eq.~\eqref{eq:causal_discrete_feedback_drift} is independent of the branch variable:
\begin{equation}
\Theta_{\ell\leftarrow i}^n[x_\ell,x_i]=J_{\ell i}g_{\rm add}(x_i^n)\,.
\label{eq:additive_input_discrete_feedback}
\end{equation}
In this case the imposed-history branch law can be written as the ordinary source-shifted message
\begin{equation}
P_\ell^{(i),\Delta}\left[x_\ell^{0:M}\Vert x_i^{0:M-1};h_\ell^{0:M-1}\right]=P_\ell^{(i),\Delta}\left[x_\ell^{0:M}\mid h_\ell^{0:M-1}+J_{\ell i}g_{\rm add}(x_i^{0:M-1})\right]\,,
\label{eq:additive_input_source_shift_discrete}
\end{equation}
where $J_{\ell i}g_{\rm add}(x_i^{0:M-1})$ denotes the history whose $n$-th component is $J_{\ell i}g_{\rm add}(x_i^n)$. Thus the source-shifted form is a special case of the imposed-history path-law construction. The general pairwise theory remains Eq.~\eqref{eq:causal_discrete_cavity_equation}, with the branch law driven by $J_{\ell i}g(x_\ell^n,x_i^n)$.

The distinction between directed and bidirected interactions is transparent in this finite-time notation. If the graph is purely directed, so that $\partial_i^{\mathrm{in}}\cap\partial_i^{\mathrm{out}}=\varnothing$, then every incoming neighbour of $i$ is not an outgoing neighbour of $i$. Hence the messages that drive the root are unshifted, while outgoing-only imposed-history path laws integrate to one by normalization. The recursion becomes
\begin{align}
P_i^{(j),\Delta}\left[x_i^{0:M}\mid h_i^{0:M-1}\right]&\simeq p_0(x_i^0)\int\left[
\prod_{\ell\in\partial_i^{\mathrm{in}}\setminus j}[dx_\ell^{0:M-1}]P_\ell^{(i),\Delta}\left[
x_\ell^{0:M-1}\mid h_\ell^{0:M-2}\right]\right]\nonumber\\
&\quad\times\prod_{n=0}^{M-1}\mathcal T_{i\setminus j}^{\Delta}\left(x_i^{n+1}\mid x_i^n,\{x_\ell^n\}_{\ell\in\partial_i^{\mathrm{in}}\setminus j};h_i^n\right)\,.
\label{eq:causal_discrete_directed_cavity}
\end{align}
When reciprocal edges are present, a neighbour in $\partial_i^{\mathrm{in}}\cap\partial_i^{\mathrm{out}}$ both drives the root and is driven by the imposed root history. Its contribution therefore remains an imposed-history path law $P_\ell^{(i),\Delta}[x_\ell\Vert x_i;h_\ell]$. At a fixed time horizon this still gives a causal construction on a tree: to generate the root path up to $t_M$, one recursively generates the branch histories needed one time slice earlier, with imposed histories propagated along reciprocal edges. The resulting finite-depth tree representation is written explicitly in App.~\ref{app:finite_depth_reciprocal_tree}.

The same finite-time equations give a natural population-dynamics representation, with the same ensemble-level hierarchy as the continuous-time cavity equations. On a fixed graph, the unknowns are the path messages $P_i^{(j),\Delta}$. At fixed $\Delta$ and horizon $M$, these messages are elements of a finite-time message space $\mathcal M^\Delta_M$, whose elements are maps $\eta^\Delta$ such that $\eta^\Delta[\cdot\mid h^{0:M-1}]$ is a normalized path probability on histories $x^{0:M}$, possibly with source-history arguments or imposed-history arguments. After averaging over a sparse locally tree-like ensemble, the corresponding object is therefore not immediately a single path law, but a probability law $Q^\Delta$ over path-probability messages. The barycenters of this law give the node-sampled and edge-sampled path probabilities, exactly as in the continuous ensemble construction.

The discrete cavity update in Eq.~\eqref{eq:causal_discrete_cavity_equation} defines an operator $\mathfrak G^\Delta$ which takes a sampled local neighbourhood, its edge types, its couplings, and the incoming branch messages, and returns a new message. In distributional form this can be written schematically as
\begin{equation}
Q^\Delta(d\eta)=\int\mathcal D\omega\mathcal P(d\omega)\int\prod_{a\in\mathcal B(\omega)}Q_a^\Delta(d\eta_a)\delta_{\mathfrak G^\Delta(\omega,\{\eta_a\}_{a\in\mathcal B(\omega)})}(d\eta)\,,
\label{eq:causal_population_message_law}
\end{equation}
where $\omega$ contains the sampled local degrees, edge types, and couplings, while $\mathcal B(\omega)$ is the set of branch messages required by that sampled neighbourhood. The measures $Q_a^\Delta$ may differ for different edge types when reciprocal and unreciprocated edges are both present. Equation~\eqref{eq:causal_population_message_law} is not a new theoretical closure; it is the distributional form of the same finite-time cavity recursion.

For a generic finite-time message law \(Q^\Delta\) on \(\mathcal M^\Delta_M\), the barycenter is
\begin{equation}
P_Q^\Delta[x^{0:M}\mid h^{0:M-1}]\equiv\int Q^\Delta(d\eta)\eta[x^{0:M}\mid h^{0:M-1}]\,.
\label{eq:causal_discrete_barycenter_general}
\end{equation}
If reciprocal edges are present, the same notation is used for barycenters of imposed-history path laws,
\begin{equation}
P_Q^\Delta[y^{0:M}\Vert x^{0:M-1};\widetilde J]\equiv\int Q^\Delta(d\eta)\eta[y^{0:M}\Vert x^{0:M-1};\widetilde J]\,.
\label{eq:causal_discrete_conditional_barycenter_general}
\end{equation}
The finite-time cavity update is multilinear in the incoming branch messages, because the root transition kernel is multiplied by a product of branch path laws and then integrated over branch histories. Consequently, the first moments of the message laws, namely their barycenters, obey closed finite-$\Delta$ equations whenever the corresponding continuous ensemble update closes at the barycentric level. Population dynamics is an empirical numerical representation of these laws $Q^\Delta$, or of their closed barycenters in cases where the barycentric equation is sufficient.

In the fully directed ensemble with joint in-/out-degree distribution $p_{k,\ell}$ and mean degree
\beeq{
c=\sum_{k,\ell}k\,p_{k,\ell}=\sum_{k,\ell}\ell\,p_{k,\ell}\,,
}
there are two natural barycenters. The first is the path law $P_s^\Delta$ of a source node reached by following a uniformly sampled directed edge backwards to its source. Such a source is sampled with the out-degree-biased weight $\ell p_{k,\ell}/c$. The second is the node-uniform path law $P^\Delta$, obtained by sampling the root with weight $p_{k,\ell}$. The source-sampled law satisfies
\begin{align}
P_s^\Delta[x^{0:M}\mid h^{0:M-1}]&=\sum_{k,\ell\ge0}\frac{\ell p_{k,\ell}}{c}\,p_0(x^0)\int\left[\prod_{r=1}^{k}[dx_r^{0:M-1}]P_s^\Delta[x_r^{0:M-1}\mid 0]\right]\int\left[\prod_{r=1}^{k}p_J(dJ_r)\right]\nonumber\\
&\quad\times\prod_{n=0}^{M-1}\mathcal T^\Delta\left(x^{n+1}\mid x^n,\{x_r^n,J_r\}_{r=1}^{k};h^n\right)\,,
\label{eq:causal_directed_source_sampled_path_law}
\end{align}
where \(\mathcal T^\Delta\) is the one-node transition density with drift
\begin{equation}
b^n=-f(x^n)+\sum_{r=1}^{k}J_r g(x^n,x_r^n)+h^n\,.
\label{eq:causal_directed_population_drift}
\end{equation}
The node-uniform law is
\begin{align}
P^\Delta[x^{0:M}\mid h^{0:M-1}]&=\sum_{k,\ell\ge0}p_{k,\ell}\,p_0(x^0)\int\left[\prod_{r=1}^{k}[dx_r^{0:M-1}]P_s^\Delta[x_r^{0:M-1}\mid 0]\right]\int\left[\prod_{r=1}^{k}p_J(dJ_r)\right]\nonumber\\
&\quad\times\prod_{n=0}^{M-1}\mathcal T^\Delta\left(x^{n+1}\mid x^n,\{x_r^n,J_r\}_{r=1}^{k};h^n\right)\,.
\label{eq:causal_directed_node_sampled_path_law}
\end{align}
Equations~\eqref{eq:causal_directed_source_sampled_path_law}--\eqref{eq:causal_directed_node_sampled_path_law} are the finite-time directed sparse-network path-probability equations. They are the discrete causal form of the directed path-probability equation obtained in Sec.~\ref{sec:derivations} and related to \cite{Metz2025}. If the in- and out-degrees are independent, $p_{k,\ell}=p_{\mathrm{in}}(k)p_{\mathrm{out}}(\ell)$, then
\beeq{
\sum_{\ell\ge0}\frac{\ell p_{k,\ell}}{c}=p_{\mathrm{in}}(k)\,,
}
so the source-sampled and node-sampled barycenters coincide, $P_s^\Delta=P^\Delta$. In that case the directed equation reduces to the single self-consistency
\begin{align}
P^\Delta[x^{0:M}\mid h^{0:M-1}]&=\sum_{k\ge0}p_{\mathrm{in}}(k)\,p_0(x^0)\int\left[\prod_{r=1}^{k}[dx_r^{0:M-1}]P^\Delta[x_r^{0:M-1}\mid 0]\right]\int\left[\prod_{r=1}^{k}p_J(dJ_r)\right]\nonumber\\
&\quad\times\prod_{n=0}^{M-1}\mathcal T^\Delta\left(x^{n+1}\mid x^n,\{x_r^n,J_r\}_{r=1}^{k};h^n\right)\,.
\label{eq:causal_directed_factorized_path_law}
\end{align}
This is the simplest setting for population dynamics: the unknown law is a single probability distribution over finite trajectories. A population-dynamics implementation of Eq.~\eqref{eq:causal_directed_factorized_path_law} represents $P^\Delta$ by a large empirical population of histories. One update step samples an indegree $k$, samples $k$ histories from the current population, samples $k$ couplings from $p_J$, generates a new entire history using the causal transition kernel, and replaces one population element by this new history. After convergence of this distributional iteration, path observables are estimated as averages over the empirical population. The update is an algorithmic representation of the path-probability fixed point; no finite-size network realization is generated in the population-dynamics step.

For ensembles containing both unreciprocated directed edges and reciprocal edges, the same hierarchy persists but the message law has more structure. In the schematic finite-time formula below, a root sampled from $p_{k,\ell,b}$ has total in-degree $k$, total out-degree $\ell$, and reciprocal degree $b$. Thus it has $k-b$ purely incoming directed neighbours, $\ell-b$ purely outgoing directed neighbours, and $b$ reciprocal neighbours. Directed incoming neighbours contribute unconditioned source-sampled path laws, while reciprocal neighbours contribute imposed-history path laws driven by the root history. If $P^{\rightarrow,\Delta}_s$ denotes the barycenter of messages sent along purely directed edges and $P^{\leftrightarrow,\Delta}_s[\cdot\Vert x;\widetilde J]$ ddenotes the barycenter of reciprocal imposed-history path laws, then the node-uniform path law has the schematic finite-time form
\begin{align}
P^\Delta[x^{0:M}&\mid h^{0:M-1}]=\sum_{\substack{k,\ell,b\ge 0\\0\le b\le \min(k,\ell)}}
p_{k,\ell,b}\,p_0(x^0)\int \left[\prod_{r=1}^{k-b}[du_r^{0:M-1}]P^{\rightarrow,\Delta}_s[u_r^{0:M-1}\mid 0]\right]\nonumber\\
&\quad\times\int \left[\prod_{r=1}^{k-b}p_J(dJ_r)\right]\int \left[\prod_{s=1}^{b}p_{\leftrightarrow}(dJ_s,d\widetilde J_s)\right]\int \left[\prod_{s=1}^{b}[dy_s^{0:M-1}]P^{\leftrightarrow,\Delta}_s[y_s^{0:M-1}\Vert x^{0:M-2};\widetilde J_s]\right]
\nonumber\\
&\quad\times\prod_{n=0}^{M-1}\mathcal T^\Delta\left(x^{n+1}\mid x^n,\{u_r^n,J_r\}_{r=1}^{k-b},\{y_s^n,J_s\}_{s=1}^{b};h^n\right)\,.
\label{eq:causal_mixed_directed_reciprocal_path_law}
\end{align}
Here $J_s$ is the coupling from reciprocal neighbour $s$ into the root, while $\widetilde J_s$ is the coupling through which the root history is imposed on that neighbour. Equation~\eqref{eq:causal_mixed_directed_reciprocal_path_law} is not a closed scalar population equation for a single unconditioned path law. It is a barycentric representation of a probability law over imposed-history path laws. In the fully directed case this conditional path-law structure collapses to Eq.~\eqref{eq:causal_directed_node_sampled_path_law}; under independent in- and out-degrees it further reduces to Eq.~\eqref{eq:causal_directed_factorized_path_law}. In reciprocal graphs it remains part of the finite-time causal construction.

The conditional path-law structure in Eq.~\eqref{eq:causal_mixed_directed_reciprocal_path_law} becomes explicit in the finite-depth construction for bidirected graphs. To isolate the causal mechanism, consider the zero-noise finite-time dynamics on an undirected regular random graph of degree $c$, with constant couplings and an additive-input kernel $g(u,v)=g_{\rm add}(v)$. This example is used only to make the causal structure of the finite-depth recursion explicit. In this setting the reciprocal ensemble equation is represented by a recursive family of branch laws. Let
\beeq{
W_M^\Delta[y^{0:M-1}\mid h^{0:M-2}]
}
denote the path law of a branch history $y^{0:M-1}$ driven by an imposed source history $h^{0:M-2}$. The loss of one time layer is causal: a branch history ending at $t_{M-1}$ only requires its parent history up to $t_{M-2}$.

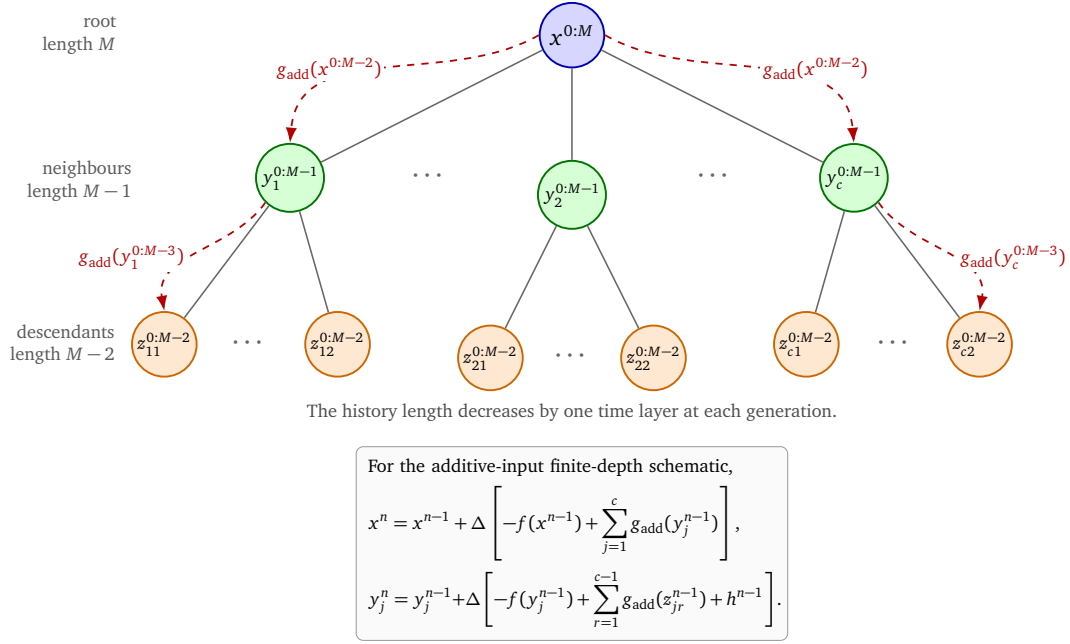
\begin{figure}[t]
\centering
\resizebox{0.94\linewidth}{!}{%
\begin{tikzpicture}[
    font=\small,
    >=Latex,
    root/.style={circle,draw=blue!70!black,fill=blue!16,thick,minimum size=9.5mm,inner sep=1pt},
    branch/.style={circle,draw=green!45!black,fill=green!16,thick,minimum size=9.0mm,inner sep=1pt,font=\scriptsize},
    leaf/.style={circle,draw=orange!80!black,fill=orange!18,thick,minimum size=8.0mm,inner sep=1pt,font=\scriptsize},
    edge/.style={draw=black!60,line width=0.65pt},
    imposed/.style={->,draw=red!70!black,dashed,line width=0.75pt},
    note/.style={draw=black!40,rounded corners=3pt,fill=black!2,align=left,inner sep=5pt,font=\scriptsize},
    levellabel/.style={font=\scriptsize,align=right,text=black!65},
    centerlabel/.style={font=\scriptsize,align=center,text=black!70,fill=white,inner sep=1pt},
    arrowlabel/.style={font=\scriptsize,fill=white,inner sep=1.0pt,text=red!70!black},
    ellipsis/.style={font=\large,text=black!60}
]

\node[levellabel] at (-7.25,0) {root\\length $M$};
\node[levellabel] at (-7.25,-2.15) {neighbours\\length $M-1$};
\node[levellabel] at (-7.50,-4.55) {descendants\\length $M-2$};

\node[root] (x) at (0,0) {$x^{0:M}$};

\node[branch] (y1) at (-4.15,-2.10) {$y_1^{0:M-1}$};
\node[branch] (y2) at (0,-2.35) {$y_2^{0:M-1}$};
\node[branch] (yc) at (4.15,-2.10) {$y_c^{0:M-1}$};

\node[leaf] (z11) at (-6.00,-4.55) {$z_{11}^{0:M-2}$};
\node[leaf] (z12) at (-3.45,-4.55) {$z_{12}^{0:M-2}$};
\node[leaf] (z21) at (-1.20,-4.75) {$z_{21}^{0:M-2}$};
\node[leaf] (z22) at (1.20,-4.75) {$z_{22}^{0:M-2}$};
\node[leaf] (zc1) at (3.45,-4.55) {$z_{c1}^{0:M-2}$};
\node[leaf] (zc2) at (6.00,-4.55) {$z_{c2}^{0:M-2}$};

\draw[edge] (x) -- (y1);
\draw[edge] (x) -- (y2);
\draw[edge] (x) -- (yc);
\draw[edge] (y1) -- (z11);
\draw[edge] (y1) -- (z12);
\draw[edge] (y2) -- (z21);
\draw[edge] (y2) -- (z22);
\draw[edge] (yc) -- (zc1);
\draw[edge] (yc) -- (zc2);

\node[ellipsis] at (-2.1,-2.10) {$\cdots$};
\node[ellipsis] at (2.1,-2.10) {$\cdots$};
\node[ellipsis] at (-4.75,-4.55) {$\cdots$};
\node[ellipsis] at (0,-4.75) {$\cdots$};
\node[ellipsis] at (4.75,-4.55) {$\cdots$};

\draw[imposed] (x.west) to[out=210,in=85]
    node[arrowlabel,pos=.48,left=2pt] {$g_{\rm add}(x^{0:M-2})$} (y1.north);
\draw[imposed] (x.east) to[out=-30,in=95]
    node[arrowlabel,pos=.48,right=2pt] {$g_{\rm add}(x^{0:M-2})$} (yc.north);

\draw[imposed] (y1.south west) to[out=235,in=100]
    node[arrowlabel,pos=.55,left=2pt] {$g_{\rm add}(y_1^{0:M-3})$} (z11.north);
\draw[imposed] (yc.south east) to[out=-55,in=80]
    node[arrowlabel,pos=.55,right=2pt] {$g_{\rm add}(y_c^{0:M-3})$} (zc2.north);

\node[centerlabel] at (0,-5.55) {The history length decreases by one time layer at each generation.};

\node[note,anchor=north,text width=6.0cm] at (0,-6.05) {%
For the additive-input finite-depth schematic,\\[1mm]
$\displaystyle x^n=x^{n-1}
+\Delta\left[-f(x^{n-1})+\sum_{j=1}^{c}g_{\rm add}(y_j^{n-1})\right],$\\[1mm]
$\displaystyle y_j^n=y_j^{n-1}
+\Delta\left[-f(y_j^{n-1})+\sum_{r=1}^{c-1}g_{\rm add}(z_{jr}^{n-1})+h^{n-1}\right].$
};

\end{tikzpicture}%
}
\caption{Finite-depth causal tree for the bidirected population-dynamics construction. The root history is generated from neighbour histories one time layer shorter; each neighbour history is generated from its own descendants with one further time layer removed. The recursion terminates at leaves drawn from the initial distribution. The schematic is drawn for the zero-noise, constant-coupling additive-input example used to make the causal structure explicit. In the general pairwise formulation, ordinary source histories are replaced by the corresponding imposed-history branch path laws.}
\label{fig:finite_depth_bidirected_tree}
\end{figure}

The root path law at horizon $M$ is
\begin{align}
P_M^\Delta[x^{0:M}]&=p_0(x^0)\int\left[\prod_{j=1}^{c}[dy_j^{0:M-1}] W_M^\Delta\left[y_j^{0:M-1}\mid g_{\rm add}(x^{0:M-2})\right]\right]\nonumber\\
&\quad\times\prod_{n=1}^{M}\delta\left[x^n-x^{n-1}-\Delta\left(-f(x^{n-1})+\sum_{j=1}^{c}g_{\rm add}(y_j^{n-1})\right)\right]\,.
\label{eq:bidirected_root_finite_depth_law}
\end{align}
The branch law satisfies the one-layer-shorter recursion
\begin{align}
W_M^\Delta[y^{0:M-1}\mid h^{0:M-2}]&=p_0(y^0)\int\left[\prod_{r=1}^{c-1}[dz_r^{0:M-2}] W_{M-1}^\Delta\left[z_r^{0:M-2}\mid g_{\rm add}(y^{0:M-3})\right]\right]\nonumber\\
&\quad\times\prod_{n=1}^{M-1}\delta\left[y^n-y^{n-1}-\Delta\left(-f(y^{n-1})+\sum_{r=1}^{c-1}g_{\rm add}(z_r^{n-1})+h^{n-1}\right)\right]\,,
\label{eq:bidirected_branch_finite_depth_recursion}
\end{align}
with the initialization
\begin{equation}
W_1^\Delta[y^0]=p_0(y^0)\,.
\label{eq:bidirected_branch_finite_depth_initial}
\end{equation}
The convention in Eq.~\eqref{eq:bidirected_branch_finite_depth_recursion} is that source histories of negative length are absent at the leaves. Equations~\eqref{eq:bidirected_root_finite_depth_law}--\eqref{eq:bidirected_branch_finite_depth_initial} give the arbitrary-horizon version of the finite-depth causal tree construction: the root path of length $M$ is generated from branch histories of length $M-1$, those branch histories are generated from descendants of length $M-2$, and the recursion continues until the leaves are drawn from $p_0$. The object propagated by the algorithm is therefore a conditional branch law, not a single unconditioned trajectory distribution. In the notation of the general pairwise theory, the same construction replaces the source history $g_{\rm add}(x^{0:M-2})$ by the conditional interaction channel $Jg(y^n,x^n)$ inside the branch transition kernel, with the appropriate directed coupling restored. Figure~\ref{fig:finite_depth_bidirected_tree} illustrates this causal loss of one time layer at each generation.

The numerical implication is that population dynamics acts on the object that closes the corresponding ensemble equation. In the directed factorized case, this object is a population of trajectories representing $P^\Delta$. With correlated in- and out-degrees, it is a population representing $P_s^\Delta$, from which $P^\Delta$ is computed by the node-sampled equation. With reciprocal edges, it is in general a population of imposed-history path laws or a sampled finite-depth tree representation of those path laws. The latter construction is heavier because a reciprocal branch is a map from an imposed history to a distribution over branch histories, rather than an unconditioned trajectory law. This increase in representation cost is a consequence of temporal feedback along reciprocal edges; it does not alter the general continuous-time pairwise cavity theory.

\begin{figure}[t!]
\centering
\includegraphics[width=0.92\linewidth]{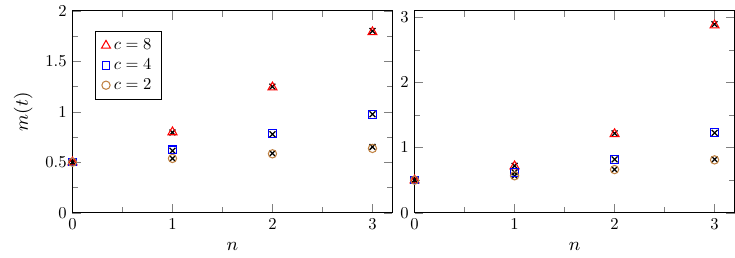}
\caption{Comparison between direct finite-graph simulations and finite-depth population dynamics for undirected random graphs with Poisson degree distribution of mean $c$. Coupling strengths are sampled from a Gaussian distribution with unit mean and unit variance. The time-discretization parameter is $\Delta=0.1$. Crosses denote direct finite-graph simulations of the discretized microscopic dynamics with $N=15000$, while the remaining symbols denote the finite-depth population-dynamics construction with population size $N_{\rm pop}=10^5$. The left panel shows the recurrent neural-network model and the right panel shows the Lotka--Volterra model with immigration rate $\lambda=10^{-2}$. Error bars are not shown.}
\label{fig:bidirected_poisson_comparison}
\end{figure}

\begin{figure}[t!]
\centering
\includegraphics[width=0.92\linewidth]{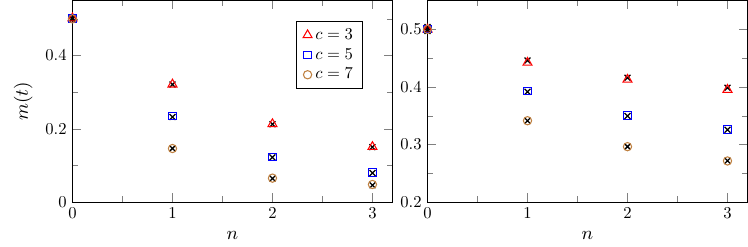}
\caption{Comparison between direct finite-graph simulations and finite-depth population dynamics for undirected regular random graphs with fixed degree $c$. Coupling strengths are sampled from a Gaussian distribution with mean $\langle J\rangle=-1$ and unit variance. The time-discretization parameter is $\Delta=0.1$. Crosses denote direct finite-graph simulations of the discretized microscopic dynamics with $N=15000$, while the remaining symbols denote the finite-depth population-dynamics construction with population size $N_{\rm pop}=10^5$. The left panel shows the recurrent neural-network model and the right panel shows the Lotka--Volterra model with immigration rate $\lambda=10^{-2}$. Error bars are not shown.}
\label{fig:bidirected_rrg_comparison}
\end{figure}

The finite-depth construction can be implemented directly for concrete dynamics on undirected sparse graph ensembles. Figures~\ref{fig:bidirected_poisson_comparison} and \ref{fig:bidirected_rrg_comparison} compare direct finite-graph simulations of the discretized microscopic dynamics with the finite-depth population-dynamics construction. The examples shown are recurrent neural-network dynamics and Lotka--Volterra dynamics, and the plotted observable is the empirical mean $m(t)$. These comparisons illustrate the numerical implementation of the causal tree recursion for bidirected sparse graphs, distinguishing direct finite-graph simulations of the discretized update from the corresponding finite-depth population-dynamics estimates.

The finite-depth construction used in Figs.~\ref{fig:bidirected_poisson_comparison} and \ref{fig:bidirected_rrg_comparison} should be distinguished from the further approximation needed to use population dynamics at long times. For any fixed observation horizon $T=M\Delta$, the causal discretization above gives a finite-time representation of the path law in Eq.~\eqref{eq:causal_discrete_cavity_equation}. A direct simulation of Eq.~\eqref{eq:causal_discrete_update} on a sampled finite graph is likewise an exact simulation of the discretized microscopic dynamics. The difficulty is not the microscopic update, but the fact that an exact population-dynamics implementation of the finite-time cavity equation carries complete histories $x^{0:M}$. As $M$ grows, the represented objects grow with the observation time. The finite-memory closures introduced below address this computational growth by replacing the complete history by a moving finite window. They should therefore be understood as numerical closures of the finite-time population dynamics, not as modifications of the discrete dynamics itself.

We distinguish three such numerical closures. The rolling cavity propagates a moving window while resampling the local marked environment when the window is advanced. The root-quenched rolling cavity retains the root degree and incident reciprocal couplings along the rolling window, while still resampling deeper branch environments from the population. The endpoint mean-corrected rolling cavity uses an annealed moving-window population as a reference barycenter and combines it with endpoint fluctuations generated by a marked-environment proposal.

At a fixed finite horizon the causal depth has a simple interpretation. To generate a root history up to $t_L$, the root requires incoming branch histories up to $t_{L-1}$; each of those branches requires its descendants only up to $t_{L-2}$; and the recursion terminates at the initial law $p_0$. Thus the decreasing length of the histories away from the root is not an approximation, but the causal ordering of the discretized path measure. The complete finite-time population dynamics follows this construction with $L=M$. The rolling cavity fixes instead a memory depth $L$ and propagates particles carrying only the window
\begin{equation}
W_a^n=(x_a^{n-L+1},\ldots,x_a^n)\,.
\label{eq:moving_window_definition}
\end{equation}
A depth-$L$ causal tree is then sampled around this window to generate a new endpoint $x_a^{n+1}$. The oldest states of newly sampled branches are drawn from the empirical population at the oldest time in the window, and the descendant histories are propagated forward using the same imposed-history path laws $P_\ell^{(i),\Delta}[\cdot\Vert\cdot]$ introduced in Eq.~\eqref{eq:causal_discrete_conditional_kernel}. After the endpoint has been generated, the window is shifted,
\begin{equation}
(x_a^{n-L+1},\ldots,x_a^n)
\longmapsto
(x_a^{n-L+2},\ldots,x_a^n,x_a^{n+1})\,.
\label{eq:moving_window_shift}
\end{equation}
This construction turns the finite-time causal tree into a time-stationary numerical procedure, but it also changes the represented law.

\begin{figure}[t]
\centering
\resizebox{\textwidth}{!}{%
\begin{tikzpicture}[
    font=\small,
    >=Latex,
    panel/.style={draw=black!45,rounded corners=4pt,fill=black!2,line width=0.35pt},
    root/.style={circle,draw=black,fill=black!12,thick,minimum size=6.4mm,inner sep=0.5pt,font=\scriptsize},
    dA/.style={circle,draw=blue!60!black,fill=blue!10,minimum size=4.4mm,inner sep=0pt},
    dB/.style={circle,draw=green!45!black,fill=green!11,minimum size=4.0mm,inner sep=0pt},
    dC/.style={circle,draw=orange!80!black,fill=orange!17,minimum size=3.7mm,inner sep=0pt},
    oldA/.style={circle,draw=blue!60!black,fill=blue!8,minimum size=3.8mm,inner sep=0pt,opacity=.72},
    oldB/.style={circle,draw=green!45!black,fill=green!10,minimum size=3.4mm,inner sep=0pt,opacity=.72},
    oldC/.style={circle,draw=orange!80!black,fill=orange!14,minimum size=3.2mm,inner sep=0pt,opacity=.72},
    newA/.style={circle,draw=red!70!black,fill=red!9,minimum size=3.8mm,inner sep=0pt},
    newB/.style={circle,draw=purple!70!black,fill=purple!8,minimum size=3.4mm,inner sep=0pt},
    newC/.style={circle,draw=orange!90!black,fill=orange!15,minimum size=3.2mm,inner sep=0pt},
    exactedge/.style={draw=teal!70!black,line width=0.72pt},
    oldedge/.style={draw=teal!65!black,line width=0.58pt,opacity=.58},
    newedge/.style={draw=red!75!black,line width=0.62pt,dashed},
    outedge/.style={draw=black!28,line width=0.35pt,densely dotted},
    arr/.style={->,draw=black!65,line width=0.65pt},
    win/.style={draw=black!45,rounded corners=2pt,fill=white,align=center,font=\scriptsize,inner sep=2.0pt},
    lab/.style={font=\scriptsize,align=center},
    title/.style={font=\bfseries,anchor=west}
]

\draw[panel] (-0.35,2.05) rectangle (7.45,-5.55);
\draw[panel] (7.95,2.05) rectangle (17.85,-5.55);

\node[title] at (-0.10,1.85) {(a) Exact finite graph};
\node[title] at (8.20,1.85) {(b) Moving-window population closure};

\node[win,text width=2.75cm] (Wi0) at (1.70,1.30)
    {$W_i^n=(x_i^{n-L+1},\ldots,x_i^n)$};

\node[win,text width=3.25cm] (Wi1) at (5.65,1.30)
    {$W_i^{n+1}=(x_i^{n-L+2},\ldots,x_i^{n+1})$};

\draw[arr] (Wi0.east) -- (Wi1.west);

\node[root] (ei) at (0.70,-1.80) {$i$};

\node[dA] (e1a) at (1.90,-0.50) {};
\node[dA] (e1b) at (1.90,-1.80) {};
\node[dA] (e1c) at (1.90,-3.10) {};

\node[dB] (e2a) at (3.25, 0.15) {};
\node[dB] (e2b) at (3.25,-0.90) {};
\node[dB] (e2c) at (3.25,-1.35) {};
\node[dB] (e2d) at (3.25,-1.95) {};
\node[dB] (e2e) at (3.25,-2.55) {};
\node[dB] (e2f) at (3.25,-3.20) {};
\node[dB] (e2g) at (3.25,-4.15) {};

\node[dC] (e3a) at (4.70, 0.55) {};
\node[dC] (e3b) at (4.70, 0.05) {};
\node[dC] (e3c) at (4.70,-0.60) {};
\node[dC] (e3d) at (4.70,-1.10) {};
\node[dC] (e3e) at (4.70,-1.50) {};
\node[dC] (e3f) at (4.70,-2.00) {};
\node[dC] (e3g) at (4.70,-2.40) {};
\node[dC] (e3h) at (4.70,-2.90) {};
\node[dC] (e3i) at (4.70,-3.35) {};
\node[dC] (e3j) at (4.70,-3.85) {};
\node[dC] (e3k) at (4.70,-4.25) {};
\node[dC] (e3l) at (4.70,-4.75) {};

\draw[exactedge] (ei)--(e1a);
\draw[exactedge] (ei)--(e1b);
\draw[exactedge] (ei)--(e1c);

\draw[exactedge] (e1a)--(e2a);
\draw[exactedge] (e1a)--(e2b);

\draw[exactedge] (e1b)--(e2c);
\draw[exactedge] (e1b)--(e2d);
\draw[exactedge] (e1b)--(e2e);

\draw[exactedge] (e1c)--(e2f);
\draw[exactedge] (e1c)--(e2g);

\draw[exactedge] (e2a)--(e3a);
\draw[exactedge] (e2a)--(e3b);

\draw[exactedge] (e2b)--(e3c);
\draw[exactedge] (e2b)--(e3d);

\draw[exactedge] (e2c)--(e3e);
\draw[exactedge] (e2d)--(e3f);
\draw[exactedge] (e2e)--(e3g);

\draw[exactedge] (e2f)--(e3h);
\draw[exactedge] (e2f)--(e3i);

\draw[exactedge] (e2g)--(e3j);
\draw[exactedge] (e2g)--(e3k);
\draw[exactedge] (e2g)--(e3l);

\draw[outedge] (e3d) .. controls (5.55,-0.85) and (5.55,-2.00) .. (e3f);
\draw[outedge] (e3i) .. controls (5.65,-3.35) and (5.65,-4.50) .. (e3l);

\node[lab,anchor=west,text=black!55] at (5.08,-0.20) {rest of graph};

\draw[decorate,decoration={brace,amplitude=3pt},teal!70!black,line width=0.6pt]
  (5.15,-4.95) --
  node[lab,right=4pt,text=teal!60!black] {same $\mathcal E_i$}
  (5.15,0.65);

\node[lab,text width=6.7cm] at (3.55,-5.22)
    {The same rooted marked ball is used\\before and after the window shift.};

\node[win,text width=2.35cm] (Wa0) at (9.75,1.30) {$W_a^n$};
\node[win,text width=2.45cm] (Wa1) at (15.50,1.30) {$W_a^{n+1}$};

\draw[arr] (Wa0.east) --
node[lab,below,text width=2.55cm] {roll window; draw\\a new environment}
(Wa1.west);

\node[root] (oa) at (8.65,-1.80) {$a$};

\node[oldA] (o1a) at (9.55,-0.85) {};
\node[oldA] (o1b) at (9.55,-1.80) {};
\node[oldA] (o1c) at (9.55,-2.75) {};

\node[oldB] (o2a) at (10.45,-0.30) {};
\node[oldB] (o2b) at (10.45,-1.00) {};
\node[oldB] (o2c) at (10.45,-1.55) {};
\node[oldB] (o2d) at (10.45,-2.20) {};
\node[oldB] (o2e) at (10.45,-2.95) {};

\node[oldC] (o3a) at (11.35,-0.05) {};
\node[oldC] (o3b) at (11.35,-0.55) {};
\node[oldC] (o3c) at (11.35,-0.95) {};
\node[oldC] (o3d) at (11.35,-1.45) {};
\node[oldC] (o3e) at (11.35,-2.05) {};
\node[oldC] (o3f) at (11.35,-2.55) {};
\node[oldC] (o3g) at (11.35,-3.15) {};

\draw[oldedge] (oa)--(o1a);
\draw[oldedge] (oa)--(o1b);
\draw[oldedge] (oa)--(o1c);

\draw[oldedge] (o1a)--(o2a);
\draw[oldedge] (o1a)--(o2b);

\draw[oldedge] (o1b)--(o2c);
\draw[oldedge] (o1b)--(o2d);

\draw[oldedge] (o1c)--(o2e);

\draw[oldedge] (o2a)--(o3a);
\draw[oldedge] (o2a)--(o3b);

\draw[oldedge] (o2b)--(o3c);
\draw[oldedge] (o2c)--(o3d);
\draw[oldedge] (o2d)--(o3e);
\draw[oldedge] (o2e)--(o3f);
\draw[oldedge] (o2e)--(o3g);

\node[lab,text=teal!60!black] at (10.20,-3.72)
    {sampled $\mathcal E_a^{(n)}$};

\node[root] (na) at (13.30,-1.80) {$a$};

\node[newA] (n1a) at (14.20,-0.55) {};
\node[newA] (n1b) at (14.20,-1.35) {};
\node[newA] (n1c) at (14.20,-2.20) {};
\node[newA] (n1d) at (14.20,-3.05) {};

\node[newB] (n2a) at (15.10,-0.10) {};
\node[newB] (n2b) at (15.10,-0.85) {};
\node[newB] (n2c) at (15.10,-1.35) {};
\node[newB] (n2d) at (15.10,-1.95) {};
\node[newB] (n2e) at (15.10,-2.55) {};
\node[newB] (n2f) at (15.10,-3.35) {};

\node[newC] (n3a) at (16.00, 0.10) {};
\node[newC] (n3b) at (16.00,-0.35) {};
\node[newC] (n3c) at (16.00,-0.85) {};
\node[newC] (n3d) at (16.00,-1.30) {};
\node[newC] (n3e) at (16.00,-1.85) {};
\node[newC] (n3f) at (16.00,-2.35) {};
\node[newC] (n3g) at (16.00,-2.90) {};
\node[newC] (n3h) at (16.00,-3.40) {};
\node[newC] (n3i) at (16.00,-3.85) {};

\draw[newedge] (na)--(n1a);
\draw[newedge] (na)--(n1b);
\draw[newedge] (na)--(n1c);
\draw[newedge] (na)--(n1d);

\draw[newedge] (n1a)--(n2a);
\draw[newedge] (n1a)--(n2b);

\draw[newedge] (n1b)--(n2c);

\draw[newedge] (n1c)--(n2d);
\draw[newedge] (n1c)--(n2e);

\draw[newedge] (n1d)--(n2f);

\draw[newedge] (n2a)--(n3a);
\draw[newedge] (n2a)--(n3b);

\draw[newedge] (n2b)--(n3c);
\draw[newedge] (n2c)--(n3d);
\draw[newedge] (n2d)--(n3e);

\draw[newedge] (n2e)--(n3f);
\draw[newedge] (n2e)--(n3g);

\draw[newedge] (n2f)--(n3h);
\draw[newedge] (n2f)--(n3i);

\node[lab,text=red!65!black] at (14.85,-4.23)
    {new $\widetilde{\mathcal E}_a^{(n+1)}$};

\draw[arr] (11.70,-1.80) --
node[lab,above,text width=1.2cm] {discard}
(12.65,-1.80);

\draw[exactedge] (8.35,-4.83) -- (8.85,-4.83);
\node[lab,anchor=west] at (8.95,-4.83) {persistent mark};

\draw[newedge] (12.10,-4.83) -- (12.60,-4.83);
\node[lab,anchor=west] at (12.70,-4.83) {fresh draw after rolling};

\end{tikzpicture}%
}
\caption{Graph-level view of the moving-window approximation. In the exact finite graph, the history window of a root $i$ advances while the same rooted marked neighbourhood $\mathcal E_i$ is retained. In the moving-window population closure, the particle label $a$ carries the shifted window, but the local marked environment used to generate the next endpoint is sampled anew. Thus the approximation is not in the one-step update, but in replacing the joint law of a window and its persistent rooted environment by a marginal window law together with an annealed environment draw.}
\label{fig:rolling_marked_environment}
\end{figure}
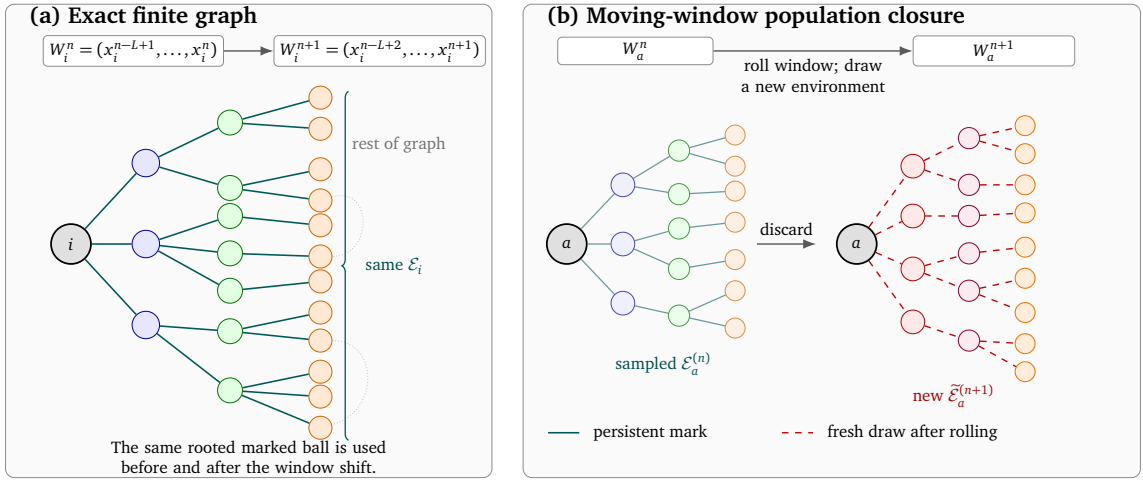

The approximation is shown schematically in Fig.~\ref{fig:rolling_marked_environment}. In the exact finite graph, advancing from one window to the next does not change the marked local environment of the root: the degree, reciprocal couplings, neighbour identities, and deeper branch environments are the same objects at all times. By contrast, the rolling cavity propagates only the marginal law of $W_a^n$. The local tree used to generate the next endpoint is newly sampled when the window advances. In schematic notation, if $\mathcal E$ denotes the rooted marked environment attached to a window, the closure replaces the joint law by an annealed product law of the form
\begin{equation}
\mu_n^L(dW,d\mathcal E)
\approx
\mu_n^L(dW)\,\mathbb P(d\mathcal E)\,.
\label{eq:moving_window_environment_factorization}
\end{equation}
Equation~\eqref{eq:moving_window_environment_factorization} is not a theorem about the exact graph process; it is a compact description of the modelling step induced by resampling the local environment when the window is rolled.

The effect of discarding the persistent marked environment is most transparent at the level of the first two moments of a typical root. Let $X^n$ denote the root state in a finite-memory population closure and write the interaction input as
\begin{equation}
\mathcal I^n=\sum_{\ell=1}^{K}J_{\ell}\,g(X^n,Y_\ell^n)\,,
\label{eq:finite_memory_interaction_input}
\end{equation}
where $K$, the couplings $J_\ell$, and the branch states $Y_\ell^n$ are sampled according to the closure under consideration. The corresponding one-step drift, in the notation of Eq.~\eqref{eq:causal_discrete_drift}, is
\begin{equation}
b^n=-f(X^n)+\mathcal I^n+h^n\,.
\label{eq:finite_memory_typical_drift}
\end{equation}
The scalar update is
\begin{equation}
X^{n+1}=X^n+\Delta b^n+\zeta^n\,.
\label{eq:finite_memory_typical_update}
\end{equation}
Since $\bracket{\zeta^n}=0$ and $\bracket{(\zeta^n)^2}=\sigma^2\Delta$, with the It\^{o} noise increment independent of the time-$n$ state and drift, the first moment $m^n=\mathbb E[X^n]$ obeys
\begin{equation}
m^{n+1}=m^n+\Delta\,\mathbb E[b^n]\,.
\label{eq:finite_memory_mean_balance}
\end{equation}
Similarly, expanding the square of Eq.~\eqref{eq:finite_memory_typical_update} gives
\begin{equation}
q^{n+1}=q^n+2\Delta\mathbb E[X^n b^n]+\Delta^2\mathbb E[(b^n)^2]+\sigma^2\Delta\,,\qquad q^n=\mathbb E[(X^n)^2]\,.
\label{eq:finite_memory_second_moment_balance}
\end{equation}
For the centered variance $v^n=q^n-(m^n)^2$, this becomes
\begin{equation}
v^{n+1}=v^n+2\Delta\operatorname{Cov}(X^n,b^n)+\Delta^2\operatorname{Var}(b^n)
+\sigma^2\Delta\,.
\label{eq:finite_memory_variance_balance}
\end{equation}
These identities are algebraic consequences of Eq.~\eqref{eq:causal_discrete_update} and do not rely on a source-shift form of the interaction. For a general pairwise kernel $(g(x,x')$, the input $\mathcal I^n$ depends explicitly on $X^n$. Hence the relevant object in Eq.~\eqref{eq:finite_memory_variance_balance} is the covariance between the state and the full local drift, not only the covariance with an external input field. Persistent local environments contribute to this covariance through the joint law of $X^n$, $K$, the couplings, and the branch histories. A moving-window closure which resamples the environment can therefore suppress part of the covariance that controls the width of the one-site law, even when the average drift is well approximated.

The root-quenched rolling cavity enriches the moving-window state by attaching the root mark to each population particle. Instead of propagating only $W_a^n$, one propagates
\begin{equation}
(W_a^n,\mathcal E_a^{\rm root})\,,
\qquad
\mathcal E_a^{\rm root}
=
\left(K_a,\{J_{a\ell},J_{\ell a}\}_{\ell=1}^{K_a}\right),
\label{eq:root_marked_window_particle}
\end{equation}
where the degree of the root and its incident reciprocal couplings are held fixed across window shifts. The imposed-history branch laws beyond the first layer are still sampled from the population, so Eq.~\eqref{eq:root_marked_window_particle} is not a full quenched cavity law. It is a root-level marked numerical closure designed to retain the part of the local environment that is most directly tied to the root window. The moment balance in Eq.~\eqref{eq:finite_memory_variance_balance} suggests why such retained marks can affect second moments: they alter the joint law entering $\operatorname{Cov}(X^n,b^n)$, not merely the marginal law of $X^n$.

The following moment-guided projections of the interaction input are diagnostic variants rather than the three finite-memory closures compared below. They interpolate between an annealed moving-window closure and a root-marked closure. Let $\mathcal I_{\rm ann}^n$ denote the input obtained by resampling the root environment as in the annealed moving-window construction, and let $\mathcal I_{\rm q}^n$ denote the input obtained with the root mark in Eq.~\eqref{eq:root_marked_window_particle}. For a chosen coarse variable $Z_n$, define
\begin{equation}
\mathcal I_{\rm hyb}^n=\mathbb E[\mathcal I_{\rm ann}^n\mid Z_n]+\alpha\left(\mathcal I_{\rm q}^n-\mathbb E[\mathcal I_{\rm q}^n\mid Z_n]\right)\,.
\label{eq:conditional_input_residual_projection}
\end{equation}
This construction replaces the conditional mean of the root-marked input by the corresponding annealed conditional mean, while retaining a centered residual of size controlled by $\alpha$. If $Z_n$ is a fine bin of the current state $X^n$, then the residual in Eq.~\eqref{eq:conditional_input_residual_projection} has approximately zero covariance with $X^n$. Such a projection can therefore remove part of the residual input contribution to $\operatorname{Cov}(X^n,b^n)$. This observation does not by itself determine the accuracy of the closure, because the variance balance involves the full drift $b^n=-f(X^n)+\mathcal I^n+h^n$, and for general $g(x,x')$ the interaction input itself has explicit root-state dependence.

\begin{figure}[t!]
\centering
\includegraphics[width=\textwidth]{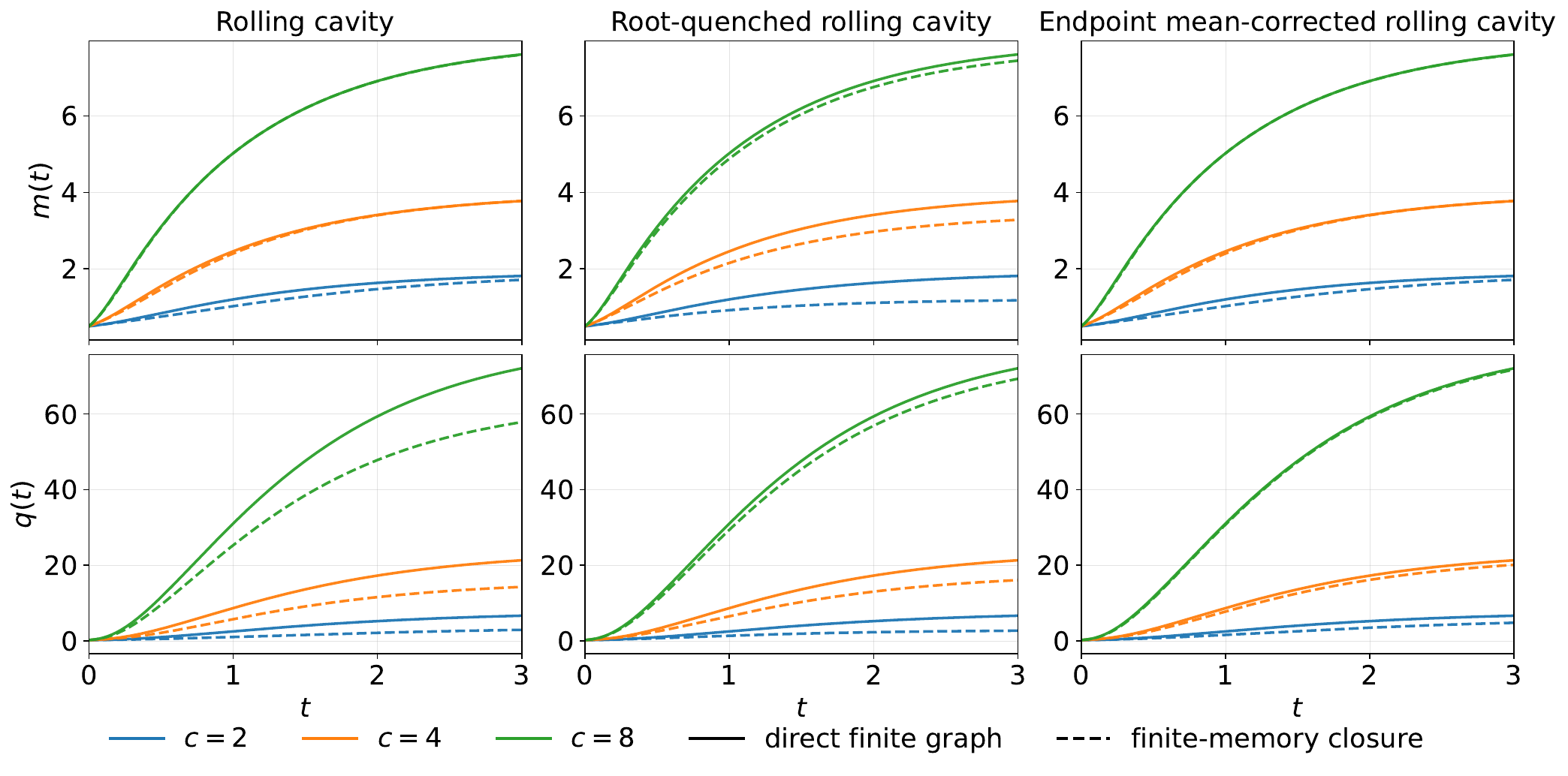}
\caption{Comparison between direct finite-graph simulations and three finite-memory closures for the additive-input RNN on undirected Poisson graphs. The memory depth is fixed to $L=3$. Columns show, from left to right, the rolling cavity, the root-quenched rolling cavity, and the endpoint mean-corrected rolling cavity. The top row shows the mean $m(t)$, and the bottom row shows the second moment $q(t)$. Solid curves are direct finite-graph simulations of Eq.~\eqref{eq:causal_discrete_update}; dashed curves are the corresponding finite-memory closure. Colours denote $c=2,4,8$. The simulations use $\Delta=10^{-2}$, $M=300$, $N=15000$, population size $N_{\rm pop}=10^5$, initial condition $x_i^0=0.5$, symmetric reciprocal couplings $J_{ij}=J_{ji}$ sampled from a Gaussian law with mean $1$ and variance $1$, and $\sigma=0$. Error bars are not shown.}
\label{fig:rnn_finite_memory_closures_L3}
\end{figure}

A coarser projection is obtained by conditioning on no state variable:
\begin{equation}
\mathcal I_{\rm glob}^n=\mathbb E[\mathcal I_{\rm ann}^n]+\alpha\left(\mathcal I_{\rm q}^n-\mathbb E[\mathcal I_{\rm q}^n]\right)\,.
\label{eq:global_input_residual_projection}
\end{equation}
At the input level this gives
\begin{equation}
\operatorname{Cov}(X^n,\mathcal I_{\rm glob}^n)=\alpha\operatorname{Cov}(X^n,\mathcal I_{\rm q}^n)\,,
\label{eq:global_input_residual_covariance}
\end{equation}
since subtracting a constant does not change covariance. Equation~\eqref{eq:global_input_residual_covariance} should be read together with Eq.~\eqref{eq:finite_memory_variance_balance}: the variance evolution depends on $\operatorname{Cov}(X^n,b^n)$, of which the input covariance is only one part. The usefulness of either residual projection is therefore a numerical question tied to the model, the graph ensemble, and the chosen memory depth. These residual projections are diagnostic finite-memory variants; they motivate endpoint-level corrections but are not the three named closures compared in Fig.~\ref{fig:rnn_finite_memory_closures_L3}.

The endpoint mean-corrected rolling cavity is imposed directly at the endpoint distribution. Let $X_{\rm ref}^{n+1}$ be the endpoint generated by an annealed moving-window population, and let $X_{\rm raw}^{n+1}$ be an endpoint generated by a root-marked or residual proposal. The endpoint moment-matched variable is
\begin{equation}
X_{\rm corr}^{n+1}=\mathbb E[X_{\rm ref}^{n+1}]+\beta\left(X_{\rm raw}^{n+1}-\mathbb E[X_{\rm raw}^{n+1}]\right)\,.
\label{eq:endpoint_moment_matched_closure}
\end{equation}
By construction,
\begin{equation}
\mathbb E[X_{\rm corr}^{n+1}]=\mathbb E[X_{\rm ref}^{n+1}]\,,\qquad\operatorname{Var}(X_{\rm corr}^{n+1})=\beta^2\operatorname{Var}(X_{\rm raw}^{n+1})\,.
\label{eq:endpoint_moment_matched_identities}
\end{equation}
The parameter $\beta$ fixes how much of the proposal width is retained; the endpoint mean-corrected closure used in the comparison below takes $\beta=1$, so that the endpoint mean is matched to the annealed reference population while the centered endpoint fluctuations are imported from the marked-environment proposal. More general choices of $\beta$ would impose a variance target explicitly. This affine endpoint matching is not a new exact cavity equation. It is a numerical closure of the finite-memory population dynamics which separates the reference barycenter from the width of a marked-environment proposal. The comparisons below assess how these finite-memory numerical closures behave relative to direct simulations of Eq.~\eqref{eq:causal_discrete_update}.

We now use the additive-input RNN model as a numerical test of the finite-memory closures described above. The comparison is restricted to this source-shift specialization and is used to assess the effect of the three numerical closures at fixed memory depth.

Figure~\ref{fig:rnn_finite_memory_closures_L3} compares these three finite-memory closures against direct finite-graph simulations for the additive-input RNN on the Poisson ensemble. The comparison is made at fixed memory depth $L=3$, so that the effect of changing the closure can be isolated from the effect of changing the temporal window. In this fixed-depth additive-input RNN comparison, the rolling-cavity mean trajectory $m(t)$ lies close to the direct finite-graph curve in the parameter regime shown, while its second moment $q(t)$ lies below the direct finite-graph result. This behaviour is consistent with the environment-resampling step described in Fig.~\ref{fig:rolling_marked_environment}: the marginal window law is propagated, but the persistent marked neighbourhood that contributes to the width of the one-site law is not retained.

The root-quenched rolling cavity changes this behaviour in the direction suggested by the moment-balance argument. Retaining the root mark increases the spread of the population and hence gives a larger $q(t)$, but it also changes the barycentric drift. This is consistent with Eq.~\eqref{eq:finite_memory_variance_balance}: retaining quenched marks changes the joint law entering $\operatorname{Cov}(X^n,b^n)$, but retaining only the root-level mark is not the same as propagating the full quenched branch environment. The endpoint mean-corrected rolling cavity separates these two effects at the level of the generated endpoint. Its mean follows the annealed reference population by construction, while its second moment is determined by the marked-environment endpoint fluctuations. For the parameter regime shown in Fig.~\ref{fig:rnn_finite_memory_closures_L3}, this closure preserves the reference barycenter while giving closer agreement for the second moment than the rolling cavity. The remaining low-connectivity discrepancy is consistent with the interpretation that, for small $c$, root-level persistence alone does not capture all of the quenched branch structure lost by the moving-window approximation.

\section{High-connectivity limits and response channels}
\label{sec:high_connectivity_response_channels}
The sparse cavity equations derived above are finite-connectivity path-measure equations. A high-connectivity or dense limit is a further asymptotic projection of these equations, not a replacement for them. This distinction is important because the existence of a limiting effective path law is not the same as the existence of a closed finite system of equations for a small set of dynamical order parameters such as $m(t)$, $C(t,t')$, and $R(t,t')$. The imposed-history conditional-kernel construction of Sec.~\ref{sec:derivations} remains the general sparse object. In the limit $c\to\infty$, one asks which effective terms survive when many weak inputs are combined: a drift term, a colored-noise term, and, when reciprocal interactions are present, a response or memory term.

We first consider the fully directed case, where the projection can be carried out directly at the level of path probabilities. Let the indegree scale as $k=c\kappa$, and let $\nu_{\mathrm{in}}(\kappa)$ denote the limiting distribution of the rescaled indegree $\kappa$. In this high-connectivity projection, $P[y]$ denotes the incoming-neighbour, or source-sampled, path law entering the directed input. If source nodes of different rescaled degrees have different path laws, this notation should be understood as the corresponding edge-sampled mixture over source types; in homogeneous settings it coincides with the representative path law. The path law $P[x]$ on the left-hand side below is the node-uniform representative law after mixing over the root variable $\kappa$. We assume that incoming branches are independent in the directed locally tree-like limit and that edge couplings are independent of the incoming-neighbour path law. The coupling distribution is scaled as
\begin{equation}
\int p_J(dJ)\,J=\frac{\mu_J}{c}\,, \qquad \int p_J(dJ)\,J^2=\frac{\gamma_J^2}{c}+o(c^{-1})\,.
\label{eq:high_c_coupling_scaling}
\end{equation}
with higher coupling cumulants negligible on the $c^{-1}$ scale. More explicitly, for every order $r\ge 3$, the $r$-th coupling cumulant is assumed to be $o(c^{-1})$, with finite interaction-channel moments under the incoming-neighbour path law on the finite time interval considered. In the directed path-probability equation, a single incoming neighbour contributes, for fixed root path $x$ and response field $\hat x$, a factor of the form
\begin{equation}
A_c[\hat x,x]=\int p_J(dJ)\int[dy] P[y]\exp\left[iJ\int_0^Tdt \hat x(t)g(x(t),y(t))\right]\,.
\label{eq:high_c_single_neighbour_factor}
\end{equation}
Expanding Eq.~\eqref{eq:high_c_single_neighbour_factor} under the scaling \eqref{eq:high_c_coupling_scaling} gives
\begin{align}
A_c[\hat x,x]&=1+\frac{i\mu_J}{c}\int_0^Tdt \hat x(t)M(t\mid x(t))\nonumber\\
&\quad-\frac{\gamma_J^2}{2c}\int_0^Tdt\int_0^Tdt' \hat x(t)\hat x(t')C(t,t'\mid x(t),x(t'))+o(c^{-1})\,,
\label{eq:high_c_single_neighbour_expansion}
\end{align}
where
\begin{equation}
M(t\mid x(t))\equiv\int[dy] P[y] g(x(t),y(t))\,,
\label{eq:high_c_M_def}
\end{equation}
and
\begin{equation}
C(t,t'\mid x(t),x(t'))\equiv\int[dy]\,P[y] g(x(t),y(t))g(x(t'),y(t'))\,.
\label{eq:high_c_C_def}
\end{equation}
The quantity $C(t,t'\mid x(t),x(t'))$ is the second moment of the interaction channel induced by the incoming-neighbour path law; it is not, in general, the centered covariance of the representative trajectory. Since $k=c\kappa$, the product of $k$ such factors has a finite exponential limit. Including the intrinsic Gaussian noise, the resulting directed effective path law can be represented as
\begin{align}
P[x]&=\int_0^\infty d\kappa \nu_{\mathrm{in}}(\kappa)\int[d\omega]P_\kappa[\omega\mid x]\,
p_0(x(0))\nonumber\\
&\quad\times\delta_{(\mathrm F)}\left[\dot x(t)+f(x(t))-h(t)-\mu_J\kappa M(t\mid x(t))-\omega(t)\right]\,,
\label{eq:high_c_directed_path_law}
\end{align}
where $P_\kappa[\omega\mid x]$ is a normalized Gaussian path measure, for each admissible $x$, with zero mean and covariance
\begin{equation}
\bracket{\omega(t)\omega(t')}_{\kappa,x}=\sigma^2\delta(t-t')+\gamma_J^2\kappa C(t,t'\mid x(t),x(t'))\,.
\label{eq:high_c_directed_noise_covariance}
\end{equation}
Equations~\eqref{eq:high_c_directed_path_law}--\eqref{eq:high_c_directed_noise_covariance} are the directed high-connectivity projection of the sparse path-probability equation. They show that the dense limit is naturally formulated as an effective path law for a representative trajectory. For a general pairwise kernel $g(x,x')$, the induced Gaussian field may depend on the root trajectory itself; it is therefore best viewed as a Gaussian kernel $P_\kappa[\omega\mid x]$ conditional on the trajectory $x$, rather than as an externally prescribed additive colored noise independent of $x$.

The dependence on $\nu_{\mathrm{in}}$ is not merely notational. If the rescaled indegrees do not concentrate, the limiting process remains a mixture over $\kappa$. Thus high connectivity does not automatically imply the usual homogeneous fully connected DMFT. The homogeneous fully connected form is recovered only when
\begin{equation}
\nu_{\mathrm{in}}(\kappa)=\delta(\kappa-1)\,,
\label{eq:high_c_homogeneous_degree_limit}
\end{equation}
or under an equivalent condition that removes rescaled degree fluctuations from the effective input statistics. In this sense, the high-connectivity projection retains a memory of degree heterogeneity unless the graph ensemble itself becomes homogeneous on the $k/c$ scale.

The neural-network and Lotka--Volterra kernels illustrate two different projections of the same formulas. For a rate-based recurrent neural network,
\begin{equation}
g(x,x')=\varphi(x')\,,\qquad \varphi(x')=\tanh x'\,,
\label{eq:high_c_NN_kernel}
\end{equation}
the quantities in Eqs.~\eqref{eq:high_c_M_def}--\eqref{eq:high_c_C_def} reduce to
\begin{equation}
M(t)=\int[dy] P[y] \varphi(y(t))\,,
\qquad
C(t,t')=\int[dy] P[y] \varphi(y(t))\varphi(y(t'))\,.
\label{eq:high_c_NN_MC}
\end{equation}
The colored field is then independent of the root trajectory, apart from the self-consistent dependence of $P$ itself. By contrast, Lotka--Volterra dynamics in abundance variables has the multiplicative pairwise kernel
\begin{equation}
g(x,x')=xx'\,.
\label{eq:high_c_LV_abundance_kernel}
\end{equation}
In this case
\begin{equation}
M(t\mid x(t))=x(t)m(t)\,,\qquad m(t)\equiv \int[dy] P[y] y(t)\,,
\label{eq:high_c_LV_M}
\end{equation}
and
\begin{equation}
C(t,t'\mid x(t),x(t'))=x(t)x(t')Q(t,t')\,, \qquad Q(t,t')\equiv \int[dy] P[y] y(t)y(t')\,.
\label{eq:high_c_LV_C}
\end{equation}
Here $Q(t,t')$ is the raw second moment of the abundance variable under the path law $P$. The directed dense LV process therefore has a multiplicative Gaussian channel:
\begin{equation}
\bracket{\omega(t)\omega(t')}_{\kappa,x}=\sigma^2\delta(t-t')+\gamma_J^2\kappa\,x(t)x(t')Q(t,t')\,.
\label{eq:high_c_LV_noise_covariance}
\end{equation}
This state dependence is not a failure of the high-connectivity limit. It is the form taken by the effective path law when the pairwise interaction is multiplicative in the root variable.

The preceding equations also separate two notions of closure. The directed high-connectivity limit gives a self-consistent effective process for $P[x]$. It does not by itself imply that the resulting dynamics can always be reduced to a closed finite set of equations for $m(t)$, $C(t,t')$, and $R(t,t')$. In additive-input or linear models, the effective field often depends on a small number of two-time objects, and one may obtain a closed DMFT system for those objects. For a general nonlinear pairwise kernel, the natural self-consistency may remain at the level of the path law or involve composite dynamical observables. Thus the existence of the dense effective process and the closure of low-dimensional dynamical parameters are distinct questions.

Reciprocal or undirected interactions add a further channel. In the directed calculation, the disorder average produces only the drift and colored-noise contributions shown above. If a link $j\to i$ is accompanied by a feedback link $i\to j$, then the disorder average also couples the insertion of the root into the neighbour equation with the insertion of the neighbour into the root equation. A convenient way to represent this contribution at the level of the dense projection is to allow the reciprocal pair to have a covariance
\begin{equation}
\overline{J_{ij}J_{ji}}-\overline{J_{ij}}\,\overline{J_{ji}} = \frac{\rho_J}{c}+o(c^{-1})\,.
\label{eq:high_c_reciprocal_covariance_scaling}
\end{equation}
The coefficient $\rho_J$ controls the response channel generated by the $O(c^{-1})$ covariance of paired reciprocal couplings. In the formulas below, $\rho_J$ is understood as the effective reciprocal covariance after including the asymptotic density of reciprocal partners among incoming neighbours; equivalently, if that density is kept separate, it multiplies $\rho_J$ in the response-channel strength. The symmetric random case $J_{ij}=J_{ji}$, with nonzero variance at the same scaling, is a particular case in which the response channel is present. The sparse linear--Gaussian bridge of Sec.~\ref{sec:known} already exhibits the finite-connectivity precursor of this term: memory appears only along reciprocal edges, through products of the forward and backward couplings and the response kernels of the neighbouring branches.

For a general pairwise kernel, the corresponding dense response channel is not, in general, the ordinary additive-field response. Let \(P[y,\hat y]\) denote the auxiliary-field representation of the incoming-neighbour/source-sampled path law used in the dense projection, with the same source convention as in the preceding sections. The composite response channel has the schematic form
\begin{equation}
\mathcal R_g(t,s\mid x(t),x(s))\equiv i\int[dy\,d\hat y] P[y,\hat y] g(x(t),y(t)) \hat y(s)\,g(y(s),x(s))\,,
\label{eq:high_c_general_response_channel}
\end{equation}
with the sign convention inherited from the MSRJD weight \eqref{eq:msrjd_weight}, for which physical responses are generated by insertions of $i\hat y$. By causality, this response channel has retarded support in the second time argument. With the convention for \(\rho_J\) stated above, the corresponding contribution to the reciprocal dense effective equation may be written schematically as the retarded drift
\beeq{
\rho_J\kappa\int_0^t ds\,\mathcal R_g(t,s\mid x(t),x(s)).
}
At the same schematic level, the delta constraint in Eq.~\eqref{eq:high_c_directed_path_law} is supplemented, in the reciprocal projection, by subtracting this retarded drift inside the effective residual. The important point is structural: for general $g$, the response channel depends on how the imposed root history enters the neighbour interaction channel, not only on the ordinary response of $y(t)$ to an additive source. Equation~\eqref{eq:high_c_general_response_channel} should therefore be read as a composite response channel in the projected path law, not as a closed equation for the ordinary response kernel $R(t,s)$.

The difference is again transparent in examples. If
\begin{equation}
g(u,v)=g_{\rm add}(v)\,,
\label{eq:high_c_additive_input_again}
\end{equation}
then
\begin{equation}
\mathcal R_g(t,s\mid x(s))=g_{\rm add}(x(s)) R_{g_{\rm add}}(t,s)\,,
\qquad
R_{g_{\rm add}}(t,s)\equiv i\int[dy d\hat y] P[y,\hat y] g_{\rm add}(y(t))\hat y(s)\,.
\label{eq:high_c_additive_response_channel}
\end{equation}
For the linear case $g_{\rm add}(v)=v$, this is the standard response kernel multiplying the imposed history. This is precisely why the reciprocal linear--Gaussian case admits a closed Volterra representation. For abundance-variable Lotka--Volterra, however, $g(u,v)=uv$, and Eq.~\eqref{eq:high_c_general_response_channel} becomes
\begin{equation}
\mathcal R_{\rm LV}(t,s\mid x(t),x(s))=x(t)x(s)i\int[dy d\hat y] P[y,\hat y] y(t)\hat y(s)y(s)\,.
\label{eq:high_c_LV_response_channel}
\end{equation}
The response insertion contains the additional factor $y(s)$. It is therefore not determined by the ordinary additive response $R(t,s)=i\bracket{y(t)\hat y(s)}$ alone. This is an MSR-level manifestation of the closure obstruction in nonlinear symmetric or reciprocal multiplicative systems: the dense effective process may exist, but a closed equation for the usual low-order response sector need not follow.

A change of variables can simplify the representation of a particular model, but it must be interpreted as a change of the entire stochastic problem. In the Lotka--Volterra example, abundance variables $N_i>0$ give a genuinely pairwise kernel $g(N,N')=NN'$. The log-abundance variables $x_i=\log N_i$ move this interaction into an additive-input form, as described in Sec.~\ref{sec:model}. This can be a useful coordinate representation for deriving or comparing additive-input effective processes. However, correlations and responses in the transformed variables are not the same objects as abundance correlations and abundance responses. For example,
\begin{equation}
\bracket{N(t)}=\bracket{e^{x(t)}}\,, \qquad \bracket{N(t)N(t')}=\bracket{e^{x(t)}e^{x(t')}}\,,
\label{eq:high_c_log_variable_observables}
\end{equation}
and an additive source in the $x$-equation corresponds to a different perturbation of the abundance dynamics than an additive source in the $N$-equation. Thus a transformed representation must transform the dynamics, sources, noise, observables, and response functions consistently before it is compared with the abundance-variable formulation.

The dense limit should therefore be viewed as a projection of the sparse path-measure theory onto a smaller set of effective channels. In fully directed high-connectivity limits, this projection can be written for a general pairwise kernel and may produce state-dependent Gaussian noise. In reciprocal or undirected limits, a response channel must also be retained. For additive-input and linear models this channel reduces to familiar response functions; for nonlinear multiplicative models it generally involves composite responses and need not close in the usual $m,C,R$ sector. The general sparse conditional-kernel construction remains valid independently of whether such a low-dimensional dense closure is available.

\section{Conclusions}
\label{sec:conclusions}
We have developed a continuous-time cavity derivation of dynamical mean-field theory for stochastic dynamics on sparse random graphs. The construction is formulated at the level of path measures and starts from a general pairwise interaction kernel $g(x_i,x_j)$. On trees, removing a node separates its neighbouring branches and gives exact recursions for path-probability messages; on locally tree-like sparse random graphs, the same construction gives the corresponding finite-time cavity description in the thermodynamic limit. The resulting theory identifies the sparse dynamical objects that are propagated in the cavity description and that provide the starting point for any reduction to lower-dimensional observables.

The derivation makes explicit how the reciprocity structure of the interaction graph controls the form of the cavity equations. In the fully directed case, outgoing feedback kernels normalize away and the equations reduce to the known sparse directed path-probability closure. When reciprocal edges are present, a neighbouring branch is driven by the imposed history of the receiving node through an inserted drift in the branch equation. This insertion is additive at the level of the branch dynamics, but for a general pairwise kernel it may still depend on the branch trajectory, so the relevant object is a conditional path kernel rather than an unconditioned single-site path law. Source-shift subclasses of the general conditional-kernel construction, including additive-input kernels and the linear interaction as a special case, connect the imposed-history kernel formulation with the familiar response-generating picture. The linear--Gaussian reciprocal model is a solvable specialization in which this connection can be decoded explicitly into closed Volterra equations for means, correlations, and responses, with retarded memory terms generated by reciprocal feedback.

We have also introduced a causal finite-time representation of the path-measure equations. The discretized formulation makes the temporal ordering of the cavity construction explicit and provides the natural representation for numerical solution by population dynamics. In directed ensembles, the algorithm acts on populations of trajectories representing the relevant path laws. In bidirected or reciprocal ensembles, the numerical object is richer: reciprocal feedback requires conditional branch laws, which at finite horizon can be represented by a finite-depth causal tree of histories. This gives a concrete algorithmic realization of the same path-kernel structure derived in continuous time.

The high-connectivity analysis shows how dense effective processes arise as projections of the sparse path-measure theory. In directed high-connectivity limits, the many incoming sparse contributions produce drift and colored-noise channels, with possible dependence on rescaled degree fluctuations and on the state of the representative trajectory. In reciprocal or undirected limits, the dense projection also contains a response channel associated with retarded feedback. Whether these channels reduce to a closed system for a small set of dynamical observables depends on the model class: additive-input and linear models admit familiar reductions, while nonlinear pairwise interactions can require composite response objects or a path-level description.

The main outcome is a sparse-network DMFT framework in which the primary objects are path-probability messages, conditional kernels, and their ensemble laws. Fully directed sparse DMFT is recovered as a special case; reciprocal sparse dynamics requires conditional branch kernels; barycenters and higher-message moments close through the multilinearity of the path update and the independence of incoming branches; and dense DMFT emerges only after additional high-connectivity scaling assumptions. Natural next steps are to develop more efficient numerical representations of reciprocal conditional kernels and to systematize high-connectivity reductions for nonlinear pairwise interactions with nontrivial response structure.

\section*{Acknowledgements}
FLM acknowledges financial support from CNPq (Grants No 402487/2023-0 and \\No
310255/2025-2) and the ICTP through the Associates Programme (2023-2028).

\begin{appendix}
\numberwithin{equation}{section}

\section{$r$-copy path measures for higher moments of message laws}
\label{app:rcopy_moments}
The ensemble construction in Sec.~\ref{sec:derivations} leads first to probability laws over path-probability messages. The barycenter equations used in the main text are obtained by taking the first moment of those laws. In this appendix we describe the corresponding higher moments. The point of view is probabilistic: the $r$-th moment of a message law is a path probability for $r$ copies of the same cavity process evolving in a common quenched cavity environment, with independent dynamical noises conditional on that environment.

Let $\mathcal Q$ be a probability law on the message space $\mathcal M$. A draw $\eta\sim\mathcal Q$ is a normalized path-probability kernel $\eta[\cdot\mid h]$. For $r\ge 1$, define the $r$-copy path measure associated with $\mathcal Q$ by
\beeq{
S_{\mathcal Q}^{(r)}[dx^1,\dots,dx^r\mid h^1,\dots,h^r]\equiv\int \mathcal Q(d\eta)\eta[dx^1\mid h^1]\cdots \eta[dx^r\mid h^r]\,.
\label{eq:S_Q_r_def}
}
Equivalently, one first samples a random message $\eta$ from $\mathcal Q$, and then samples $r$ conditionally independent trajectories $x^1,\dots,x^r$ from that same path law. Hence $S_{\mathcal Q}^{(r)}$ is a normalized path probability on the product path space $\mathcal X^r$:
\beeq{
\int S_{\mathcal Q}^{(r)}[dx^1,\dots,dx^r\mid h^1,\dots,h^r]=\int \mathcal Q(d\eta)\prod_{a=1}^{r}\int \eta[dx^a\mid h^a]=1\,.
\label{eq:S_Q_r_normalization}
}
This object is not the $r$-point correlation function of a single trajectory drawn from the barycenter. It is the $r$-fold product measure of one random path probability, averaged over the law of that random path probability.

For the two edge-sampled message laws introduced in Sec.~\ref{sec:derivations}, we write
\beeq{
S_s^{\rightarrow,(r)}[dx^1,\dots,dx^r\mid h^1,\dots,h^r]\equiv\int \mathcal Q_s^{\rightarrow}(d\eta)\prod_{a=1}^{r}\eta[dx^a\mid h^a]\,,
\label{eq:S_arrow_r_def}
}
and
\beeq{
S_s^{\leftrightarrow,(r)}[dx^1,\dots,dx^r\mid h^1,\dots,h^r]\equiv\int \mathcal Q_s^{\leftrightarrow}(d\eta)\prod_{a=1}^{r}\eta[dx^a\mid h^a]\,.
\label{eq:S_bidir_r_def}
}
For reciprocal messages it is useful to introduce the compact conditional-kernel notation
\beeq{
S_s^{\leftrightarrow,(r)}[dy^1,\dots,dy^r\Vert x^1,\dots,x^r;\tilde J]\equiv S_s^{\leftrightarrow,(r)}\left[dy^1,\dots,dy^r \mid\tilde J g(y^1,x^1),\dots,\tilde J g(y^r,x^r)\right]\,.
\label{eq:S_bidir_r_kernel_def}
}
The same feedback coupling $\tilde J$ appears in all copies because the $r$ copies share the same quenched reciprocal edge. For a general pairwise kernel, the inserted drift $\tilde J g(y^a,x^a)$ may depend on the branch copy $y^a$, so this notation denotes the $r$-copy imposed-history kernel rather than an ordinary source-history evaluation. In the additive-input specialization it reduces to an ordinary source shift in each copy.

We now derive the $r$-copy equations. Consider first the unreciprocated-edge message law $\mathcal Q_s^{\rightarrow}$. In the distributional fixed point, the source degree triple $(k,\ell,b)$ is sampled with weight $(\ell-b)p_{k,\ell,b}/c_{\rightarrow}$. The output message is obtained by applying $\mathfrak{G}_{k-b,b,\mathbf J,\mathbf J^{\leftrightarrow},\tilde{\mathbf J}}$ to $k-b$ unreciprocated incoming messages and $b$ reciprocal incoming messages. To compute the $r$-th moment of the output law, one evaluates the same random output message at $r$ histories $x^1,\dots,x^r$ and fields $h^1,\dots,h^r$. The degree, couplings, and incoming message objects are shared across the $r$ copies, while the local noises $\xi^1,\dots,\xi^r$ are independent conditional on this shared environment. Using the shorthand $\mathbf x^{(r)}=(x^1,\dots,x^r)$ and $\mathbf h^{(r)}=(h^1,\dots,h^r)$, and writing the $r$-copy measures below through their densities with respect to the displayed path measures, this gives
\begin{align}
&S_s^{\rightarrow,(r)}[\mathbf x^{(r)}\mid \mathbf h^{(r)}]=\sum_{\substack{k,\ell,b\ge 0\\0\le b\le \min(k,\ell)}}\frac{(\ell-b)p_{k,\ell,b}}{c_{\rightarrow}}\int\Bigg[\prod_{q=1}^{k-b}p_J(dJ_q)\Bigg]\int\Bigg[\prod_{s=1}^{b}p_{\leftrightarrow}(dJ_s^\leftrightarrow,d\tilde J_s)\Bigg]\nonumber\\
&\quad\times\int\prod_{q=1}^{k-b}\left[\prod_{a=1}^{r}[dx_q^a] S_s^{\rightarrow,(r)}[x_q^1,\dots,x_q^r\mid 0,\dots,0]\right]\int\prod_{s=1}^{b}\left[\prod_{a=1}^{r}[dy_s^a] S_s^{\leftrightarrow,(r)}[y_s^1,\dots,y_s^r\Vert x^1,\dots,x^r;\tilde J_s]\right]\nonumber\\
&\quad\times\prod_{a=1}^{r}\Bigg\{p_0(x^a(0))\int[d\xi^a]P_\sigma[\xi^a]\delta_{(F)}\Bigg[
\dot x^a(t)+f(x^a(t))\nonumber\\
&-h^a(t)-\sum_{q=1}^{k-b}J_q\,g(x^a(t),x_q^a(t))-\sum_{s=1}^{b}J_s^\leftrightarrow\,g(x^a(t),y_s^a(t))-\xi^a(t)\Bigg]\Bigg\}\,.
\label{eq:S_arrow_r_general}
\end{align}
The same construction for the reciprocal-edge message law uses the size-biased weight $b\,p_{k,\ell,b}/c_{\leftrightarrow}$. Since the message is sent along one oriented reciprocal edge, the receiving reciprocal neighbour is removed from the cavity graph. The source therefore has $k-b$ purely incoming neighbours and $b-1$ remaining reciprocal neighbours. Thus
\begin{align}
&S_s^{\leftrightarrow,(r)}[\mathbf x^{(r)}\mid \mathbf h^{(r)}]=\sum_{\substack{k,\ell,b\ge 0\\1\le b\le \min(k,\ell)}}\frac{b\,p_{k,\ell,b}}{c_{\leftrightarrow}}\int\Bigg[\prod_{q=1}^{k-b}p_J(dJ_q)\Bigg]\int\Bigg[\prod_{s=1}^{b-1}p_{\leftrightarrow}(dJ_s^\leftrightarrow,d\tilde J_s)\Bigg]\nonumber\\
&\quad\times\int\prod_{q=1}^{k-b}\left[\prod_{a=1}^{r}[dx_q^a]\,S_s^{\rightarrow,(r)}[x_q^1,\dots,x_q^r\mid 0,\dots,0]\right]\int\prod_{s=1}^{b-1}\left[\prod_{a=1}^{r}[dy_s^a] S_s^{\leftrightarrow,(r)}[y_s^1,\dots,y_s^r\Vert x^1,\dots,x^r;\tilde J_s]\right]\nonumber\\
&\quad\times\prod_{a=1}^{r}\Bigg\{p_0(x^a(0))\int[d\xi^a]P_\sigma[\xi^a]\delta_{(F)}\Bigg[
\dot x^a(t)+f(x^a(t))-h^a(t)\nonumber\\
&-\sum_{q=1}^{k-b}J_q\,g(x^a(t),x_q^a(t))-\sum_{s=1}^{b-1}J_s^\leftrightarrow\,g(x^a(t),y_s^a(t))-\xi^a(t)\Bigg]\Bigg\}\,.
\label{eq:S_bidir_r_general}
\end{align}
The convention is that empty products are equal to one. Equation \eqref{eq:S_arrow_r_general} is present only when $c_{\rightarrow}>0$, and Eq.~\eqref{eq:S_bidir_r_general} only when $c_{\leftrightarrow}>0$.

The factorization mechanism in \eqref{eq:S_arrow_r_general}--\eqref{eq:S_bidir_r_general} is the $r$-copy version of the first-moment argument in the main text. Distinct incoming branches are independent in the locally tree-like limit, so the average factorizes across different branch slots. Within one branch slot, however, the same random incoming message is evaluated $r$ times, once for each copy. Thus a single unreciprocated incoming branch contributes
\beeq{
\int \mathcal Q_s^{\rightarrow}(d\eta)\eta[dx^1\mid 0]\cdots\eta[dx^r\mid 0]=S_s^{\rightarrow,(r)}[dx^1,\dots,dx^r\mid 0,\dots,0]\,,
\label{eq:r_copy_branch_factor_arrow}
}
and a single reciprocal incoming branch contributes
\beeq{
\int \mathcal Q_s^{\leftrightarrow}(d\eta)\eta[dy^1\Vert x^1;\tilde J]\cdots \eta[dy^r\Vert x^r;\tilde J]=S_s^{\leftrightarrow,(r)}[dy^1,\dots,dy^r\Vert x^1,\dots,x^r;\tilde J]\,.
\label{eq:r_copy_branch_factor_bidir}
}
The $r$-copy closure therefore involves $r$-copy measures, not products of $r$ barycenters.

The node-uniform $r$-copy path probability is obtained by sampling the root degree triple from $p_{k,\ell,b}$ rather than from an edge-biased law. Denoting this path measure by $S^{(r)}[\mathbf x^{(r)}\mid\mathbf h^{(r)}]$, one obtains
\begin{align}
&S^{(r)}[\mathbf x^{(r)}\mid \mathbf h^{(r)}]=\sum_{\substack{k,\ell,b\ge 0\\0\le b\le \min(k,\ell)}}p_{k,\ell,b}\int\Bigg[\prod_{q=1}^{k-b}p_J(dJ_q)\Bigg]\int\Bigg[\prod_{s=1}^{b}p_{\leftrightarrow}(dJ_s^\leftrightarrow,d\tilde J_s)\Bigg]\nonumber\\
&\quad\times\int\prod_{q=1}^{k-b}\left[\prod_{a=1}^{r}[dx_q^a] S_s^{\rightarrow,(r)}[x_q^1,\dots,x_q^r\mid 0,\dots,0]\right]\int\prod_{s=1}^{b}\left[\prod_{a=1}^{r}[dy_s^a] S_s^{\leftrightarrow,(r)}[y_s^1,\dots,y_s^r\Vert x^1,\dots,x^r;\tilde J_s]\right]\nonumber\\
&\quad\times\prod_{a=1}^{r}\Bigg\{p_0(x^a(0))\int[d\xi^a]P_\sigma[\xi^a]\delta_{(F)}\Bigg[
\dot x^a(t)+f(x^a(t))\nonumber\\
&-h^a(t)-\sum_{q=1}^{k-b}J_q\,g(x^a(t),x_q^a(t))-\sum_{s=1}^{b}J_s^\leftrightarrow\,g(x^a(t),y_s^a(t))-\xi^a(t)\Bigg]\Bigg\}\,.
\label{eq:S_node_r_general}
\end{align}

Setting $r=1$ in \eqref{eq:S_arrow_r_general}--\eqref{eq:S_node_r_general} recovers the barycenter equations of Sec.~\ref{sec:derivations}. Indeed,
\beeq{
S_s^{\rightarrow,(1)}[dx\mid h]=P_s^{\rightarrow}[dx\mid h]\,,\qquad S_s^{\leftrightarrow,(1)}[dx\mid h]=P_s^{\leftrightarrow}[dx\mid h]\,, \qquad S^{(1)}[dx\mid h]=P[dx\mid h]\,.
\label{eq:r_one_recovery}
}
Thus the main-text sparse-DMFT equations are the first member of the $r$-copy hierarchy.

The fully directed non-reciprocal case is obtained by setting $b=0$, $c_{\leftrightarrow}=0$, $c_{\rightarrow}=c$, and $p_{k,\ell,b}=p_{k,\ell}\delta_{b,0}$. There is then a single source-sampled message law and a single $r$-copy source measure $S_s^{(r)}$, satisfying
\begin{align}
S_s^{(r)}[\mathbf x^{(r)}\mid\mathbf h^{(r)}]&=\sum_{k,\ell\ge 0}\frac{\ell p_{k,\ell}}{c}
\int\Bigg[\prod_{q=1}^{k}p_J(dJ_q)\Bigg]\int\prod_{q=1}^{k}\left[\prod_{a=1}^{r}[dx_q^a]\,
S_s^{(r)}[x_q^1,\dots,x_q^r\mid 0,\dots,0]\right]\nonumber\\
&\quad\times\prod_{a=1}^{r}\Bigg\{p_0(x^a(0))\int[d\xi^a]P_\sigma[\xi^a]\delta_{(F)}\Bigg[
\dot x^a(t)+f(x^a(t))\nonumber\\
&-h^a(t)-\sum_{q=1}^{k}J_q\,g(x^a(t),x_q^a(t))-\xi^a(t)\Bigg]\Bigg\}\,.
\label{eq:S_directed_r_source}
\end{align}
The node-uniform $r$-copy path probability is
\begin{align}
S^{(r)}[\mathbf x^{(r)}\mid\mathbf h^{(r)}]&=\sum_{k,\ell\ge 0}p_{k,\ell}\int\Bigg[\prod_{q=1}^{k}p_J(dJ_q)\Bigg]\int\prod_{q=1}^{k}\left[\prod_{a=1}^{r}[dx_q^a] S_s^{(r)}[x_q^1,\dots,x_q^r\mid 0,\dots,0]\right]\nonumber\\
&\quad\times\prod_{a=1}^{r}\Bigg\{p_0(x^a(0))\int[d\xi^a]P_\sigma[\xi^a]\delta_{(F)}\Bigg[
\dot x^a(t)+f(x^a(t))\nonumber\\
&-h^a(t)-\sum_{q=1}^{k}J_q\,g(x^a(t),x_q^a(t))-\xi^a(t)\Bigg]\Bigg\}\,.
\label{eq:S_directed_r_node}
\end{align}
If $p_{k,\ell}=p_{\mathrm{in}}(k)p_{\mathrm{out}}(\ell)$, then the edge-source and node-uniform weights in the directed case coincide after summing over $\ell$, and the distinction between $S_s^{(r)}$ and $S^{(r)}$ disappears.

In the undirected specialization every edge is reciprocal, so $K_i=L_i=B_i=D_i$, $c_{\rightarrow}=0$, and $c_{\leftrightarrow}=c=\sum_d d p_d$. Writing $S_e^{(r)}\equiv S_s^{\leftrightarrow,(r)}$, the edge-message $r$-copy equation becomes
\begin{align}
S_e^{(r)}[\mathbf x^{(r)}\mid\mathbf h^{(r)}]&=\sum_{d\ge 1}\frac{d p_d}{c}\int\Bigg[\prod_{s=1}^{d-1}p_{\leftrightarrow}(dJ_s,d\tilde J_s)\Bigg]\int\prod_{s=1}^{d-1}\left[\prod_{a=1}^{r}[dy_s^a]S_e^{(r)}[y_s^1,\dots,y_s^r\Vert x^1,\dots,x^r;\tilde J_s]
\right]\nonumber\\
&\quad\times\prod_{a=1}^{r}\Bigg\{p_0(x^a(0))\int[d\xi^a]P_\sigma[\xi^a]\delta_{(F)}\Bigg[
\dot x^a(t)+f(x^a(t))-h^a(t)\nonumber\\
&-\sum_{s=1}^{d-1}J_s\,g(x^a(t),y_s^a(t))-\xi^a(t)\Bigg]\Bigg\}\,.
\label{eq:S_undirected_r_edge}
\end{align}
The node-uniform $r$-copy path probability is
\begin{align}
S^{(r)}[\mathbf x^{(r)}\mid\mathbf h^{(r)}]&=\sum_{d\ge 0}p_d\int\Bigg[\prod_{s=1}^{d}p_{\leftrightarrow}(dJ_s,d\tilde J_s)\Bigg]\int\prod_{s=1}^{d}\left[\prod_{a=1}^{r}[dy_s^a] S_e^{(r)}[y_s^1,\dots,y_s^r\Vert x^1,\dots,x^r;\tilde J_s]\right]\nonumber\\
&\quad\times\prod_{a=1}^{r}\Bigg\{p_0(x^a(0))\int[d\xi^a]P_\sigma[\xi^a]\delta_{(F)}\Bigg[
\dot x^a(t)+f(x^a(t))\nonumber\\
&-h^a(t)-\sum_{s=1}^{d}J_s\,g(x^a(t),y_s^a(t))-\xi^a(t)\Bigg]\Bigg\}\,.
\label{eq:S_undirected_r_node}
\end{align}
For symmetric undirected weights one sets $p_{\leftrightarrow}(dJ,d\tilde J)=p_J(dJ)\delta_J(d\tilde J)$. For bidirected graphs with independent directed weights one sets $p_{\leftrightarrow}(dJ,d\tilde J)=p_J(dJ)p_J(d\tilde J)$. The conditional dependence on the receiver histories $x^1,\dots,x^r$ remains part of the undirected $r$-copy message structure.

The case $r=2$ gives the second moment measure of the random message law. For a generic law $\mathcal Q$, this is
\beeq{
S_{\mathcal Q}^{(2)}[dx,dx'\mid h,h']=\int \mathcal Q(d\eta)\eta[dx\mid h]\eta[dx'\mid h']\,.
\label{eq:S_Q_two_copy}
}
After this two-copy path probability has been defined, one may form the connected measure
\beeq{
V_{\mathcal Q}[dx,dx'\mid h,h']=S_{\mathcal Q}^{(2)}[dx,dx'\mid h,h']-\bar\eta[dx\mid h]\bar\eta[dx'\mid h']\,,\qquad \bar\eta[dx\mid h]\equiv S_{\mathcal Q}^{(1)}[dx\mid h]\,.
\label{eq:V_Q_cov_measure}
}
This connected measure probes fluctuations of the random path-probability message across cavity environments. It is distinct from fluctuations of a trajectory inside the averaged path law $\bar\eta$. For path observables $\Phi$ and $\Psi$,
\beeq{
\int V_{\mathcal Q}[dx,dx'\mid h,h'] \Phi[x]\Psi[x']&=\int \mathcal Q(d\eta)\left(\int\eta[dx\mid h]\Phi[x]\right)\left(\int \eta[dx'\mid h']\Psi[x']\right)\\
&-\left(\int \bar\eta[dx\mid h]\Phi[x]\right)\left(\int \bar\eta[dx'\mid h']\Psi[x']\right)\,.
\label{eq:V_Q_observables}
}
Thus $V_{\mathcal Q}$ measures the variation, across random cavity environments, of quenched path-law averages. If this connected measure vanishes on a sufficiently rich class of observables, then the message law is concentrated in the corresponding sense. The $r$-copy hierarchy therefore provides a direct way to formulate concentration questions without assuming concentration in the derivation of the barycenter equations.

The same interpretation also gives a natural numerical representation. Solving the full law $\mathcal Q$ would require a population whose elements are themselves path probabilities. By contrast, $S^{(r)}_{\mathcal Q}$ can be represented by a population of $r$-tuples of trajectories; in reciprocal settings these tuples are generated within the corresponding imposed-history or finite-depth conditional-kernel construction. A single update samples one degree and one set of quenched couplings, draws incoming $r$-tuples from the corresponding $r$-copy populations, evolves $r$ root trajectories in that shared sampled environment, and uses independent dynamical noises for the $r$ copies. For $r=1$ this reduces to the usual population dynamics for the barycenter path law. For $r=2$ it becomes a pair-population dynamics for the two-copy path measure.

\section{Finite-time reciprocal path-tree recursion}
\label{app:finite_depth_reciprocal_tree}
This appendix records a finite-depth version of the causal tree construction for reciprocal graphs. The purpose is not to introduce a second theory, but to make explicit how the finite-time conditional kernels in Sec.~\ref{sec:causal_discretization_population_dynamics} can be generated recursively. The construction is simplest to write for an undirected $d$-regular tree with a common coupling $J$. Random couplings are included by inserting the corresponding coupling integrations along the branches, and non-regular degree distributions are included by sampling the excess degree at each branch. External source histories are suppressed in this appendix for notational simplicity; they can be inserted additively in the branch and root drifts as in Sec.~\ref{sec:causal_discretization_population_dynamics}.

For a root path $x^{0:M}$, the root update from $t_n$ to $t_{n+1}$ depends on neighbour states at time $t_n$. Hence, to generate $x^{0:M}$, one needs neighbour histories only up to time $t_{M-1}$. Conversely, to generate such a neighbour history $y^{0:M-1}$, the branch update needs the imposed root history only up to time $t_{M-2}$, together with the histories of the other $d-1$ neighbours of the branch up to time $t_{M-2}$. This is the causal reason why the recursion loses one time layer at each depth.

Let $W_M^\Delta(dy^{0:M-1}\Vert x^{0:M-2})$ denote the branch law of a neighbour history $y^{0:M-1}$ conditional on the imposed parent history $x^{0:M-2}$. The base object is
\begin{equation}
W_1^\Delta(dy^0)=p_0(y^0) dy^0\,.
\label{eq:appendix_W_base}
\end{equation}
This base law is independent of an imposed parent history. Equivalently, imposed histories of negative length are absent at the leaves. For $M\ge2$, the recursion is
\begin{align}
W_M^\Delta(dy^{0:M-1}\Vert x^{0:M-2})&=[dy^{0:M-1}]p_0(y^0)\int\left[\prod_{r=1}^{d-1}W_{M-1}^\Delta(dz_r^{0:M-2}\Vert y^{0:M-3})\right]\nonumber\\
&\quad\times\prod_{n=0}^{M-2}\mathcal T_{d-1,\Vert x}^{\Delta}\left(y^{n+1}\mid y^n,\{z_r^n\}_{r=1}^{d-1},x^n\right)\,,
\label{eq:appendix_W_recursion}
\end{align}
where
\begin{equation}
\mathcal T_{d-1,\Vert x}^{\Delta}\left(y^{n+1}\mid y^n,\{z_r^n\}_{r=1}^{d-1},x^n\right)=\frac{1}{\sqrt{2\pi\sigma^2\Delta}}\exp\left[-\frac{\left(y^{n+1}-y^n-\Delta b_{\mathrm{br}}^n\right)^2}{2\sigma^2\Delta}\right]\,,
\label{eq:appendix_branch_transition}
\end{equation}
with branch drift
\begin{equation}
b_{\mathrm{br}}^n=-f(y^n)+Jg(y^n,x^n)+\sum_{r=1}^{d-1}Jg(y^n,z_r^n)\,.
\label{eq:appendix_branch_drift}
\end{equation}
The deterministic limit is again obtained by replacing the Gaussian transition kernel by the corresponding Dirac delta. The root path law at horizon $M$ is then
\begin{align}
P_M^\Delta[x^{0:M}]&=[dx^{0:M}]p_0(x^0)\int\left[\prod_{r=1}^{d}W_M^\Delta(dy_r^{0:M-1}\Vert x^{0:M-2})\right]\nonumber\\
&\quad\times\prod_{n=0}^{M-1}\mathcal T_{d}^{\Delta}\left(x^{n+1}\mid x^n,\{y_r^n\}_{r=1}^{d}\right)\,,
\label{eq:appendix_root_law_from_W}
\end{align}
where $\mathcal T_d^\Delta$ is Eq.~\eqref{eq:causal_transition_kernel} with drift
\begin{equation}
b_{\mathrm{root}}^n=-f(x^n)+\sum_{r=1}^{d}Jg(x^n,y_r^n)\,.
\label{eq:appendix_root_drift}
\end{equation}
Equations~\eqref{eq:appendix_W_recursion}--\eqref{eq:appendix_root_law_from_W} display explicitly why reciprocal graphs do not reduce to a single population of unconditioned trajectories at finite time. The branch object $W_M^\Delta$ is a conditional kernel indexed by an imposed parent history, and its recursive construction requires a tree of histories whose depth grows with the time horizon.

For additive-input kernels $g(u,v)=g_{\rm add}(v)$, the parent contribution $Jg(y^n,x^n)=Jg_{\rm add}(x^n)$ is an ordinary source history for the branch. In that case one may equivalently write
\begin{equation}
W_M^\Delta(dy^{0:M-1}\Vert x^{0:M-2})=W_M^\Delta\left(dy^{0:M-1}\mid Jg_{\rm add}(x^{0:M-2})\right)\,,
\label{eq:appendix_W_additive_input}
\end{equation}
with the source-history interpretation inherited from Eq.~\eqref{eq:additive_input_source_shift_discrete}. For a general pairwise kernel, the same contribution remains inside the branch transition kernel through $Jg(y^n,x^n)$.

\section{Trajectory-population implementation of the directed finite-time equation}
\label{app:directed_population_algorithm}
This appendix records the population-dynamics implementation corresponding to the directed finite-time equations in Sec.~\ref{sec:causal_discretization_population_dynamics}. The algorithm represents a path-probability law by an empirical population of complete histories. It does not generate a finite network. It iterates the distributional path equation directly.

For correlated in- and out-degrees, the primary population represents the source-sampled law $P_s^\Delta$ in Eq.~\eqref{eq:causal_directed_source_sampled_path_law}. Choose a population size $N_{\mathrm{pop}}$, a horizon $M$, and a time step $\Delta$. The empirical population is a collection
\beeq{
\mathcal P_{\mathrm{pop}}=\{x_a^{0:M}:a=1,\ldots,N_{\mathrm{pop}}\}\,.
}
The initial values $x_a^0$ are sampled from $p_0$, or fixed according to the initial condition under study, and the remaining time entries may be initialized arbitrarily or by a preliminary forward evolution.

One single-trajectory update of the source-sampled population is as follows. Select an index $a\in\{1,\ldots,N_{\mathrm{pop}}\}$. Draw a degree pair $(k,\ell)$ from the source-biased law
\begin{equation}
\pi_s(k,\ell)=\frac{\ell p_{k,\ell}}{c}\,.
\label{eq:appendix_source_biased_degree_law}
\end{equation}
Draw indices $a_1,\ldots,a_k$ independently and uniformly from $\{1,\ldots,N_{\mathrm{pop}}\}$, and draw couplings $J_1,\ldots,J_k$ independently from $p_J$. Before evolving the replacement trajectory, set its initial value $x_a^0$ by drawing from $p_0$, or by imposing the fixed initial condition under study. If $\sigma>0$, draw independent Gaussian increments $\zeta^n$ with variance $\sigma^2\Delta$. Then replace the whole trajectory $x_a^{0:M}$ by a new trajectory generated from
\begin{equation}
x_a^{n+1}=x_a^n+\Delta\left[-f(x_a^n)+\sum_{r=1}^{k}J_r g(x_a^n,x_{a_r}^n)+h^n\right]+\zeta^n\,,
\qquad n=0,\ldots,M-1 .
\label{eq:appendix_population_update_general}
\end{equation}
For $\sigma=0$, one sets $\zeta^n=0$. Repeating this single-trajectory replacement iterates the finite-time distributional equation for $P_s^\Delta$. In the case of independent in- and out-degrees, $p_{k,\ell}=p_{\mathrm{in}}(k)p_{\mathrm{out}}(\ell)$, the source-biased law reduces to $p_{\mathrm{in}}(k)$, and the same population represents $P^\Delta=P_s^\Delta$.

Once the source-sampled population has converged, the node-uniform law $P^\Delta$ is obtained by the node-sampled equation. In practice, to generate one node-uniform trajectory, draw $(k,\ell)$ from $p_{k,\ell}$, draw $k$ histories from the converged source-sampled population, draw $k$ couplings from $p_J$, set the initial condition from $p_0$ or from the fixed initial condition under study, and evolve Eq.~\eqref{eq:appendix_population_update_general}. Repeating this sampling procedure gives an empirical representation of $P^\Delta$. If $P_s^\Delta=P^\Delta$, this additional sampling step is unnecessary.

For any path observable $F[x^{0:M}]$, the empirical population used in the estimator should represent the target law under consideration, either source-sampled or node-uniform. The corresponding population estimate is
\begin{equation}
\widehat{\bracket{F}}=\frac{1}{N_{\mathrm{pop}}}\sum_{a=1}^{N_{\mathrm{pop}}}F[x_a^{0:M}]\,.
\label{eq:appendix_population_observable_general}
\end{equation}
In particular,
\begin{equation}
\widehat m(t_n)=\frac{1}{N_{\mathrm{pop}}}\sum_{a=1}^{N_{\mathrm{pop}}}x_a^n\,,\qquad\widehat q(t_n)=\frac{1}{N_{\mathrm{pop}}}\sum_{a=1}^{N_{\mathrm{pop}}}(x_a^n)^2\,.
\label{eq:appendix_population_mq}
\end{equation}
These estimates are numerical outputs of the population representation of the path law. No numerical values are implied by the equations above.

For kernels of the form $g(u,v)=g_{\rm add}(v)$, Eq.~\eqref{eq:appendix_population_update_general} becomes
\begin{equation}
x_a^{n+1}=x_a^n+\Delta\left[-f(x_a^n)+\sum_{r=1}^{k}J_r g_{\rm add}(x_{a_r}^n)+h^n\right]+\zeta^n\,.
\label{eq:appendix_population_update_additive_input}
\end{equation}
This is the form most directly implemented in additive-input examples. For the general pairwise model, the update remains Eq.~\eqref{eq:appendix_population_update_general}.

\end{appendix}

\bibliography{SciPost_Example_BiBTeX_File.bib}

\end{document}